%% file: main.tex
\pdfoutput=1
\documentclass[acmsmall]{acmart}

\usepackage{graphicx}%
\usepackage{multirow}%
 
\usepackage{amsmath,amsfonts,amssymb}%
\usepackage{amsthm}%
\usepackage{mathrsfs}%
\usepackage[title]{appendix}%
\usepackage{xcolor}%
\usepackage{textcomp}%
\usepackage{manyfoot}%
\usepackage{booktabs}%
\usepackage{algorithm}%
\usepackage{algorithmicx}%
\usepackage{algpseudocode}%
\usepackage{listings}%
\usepackage{subfigure}%
\usepackage{tabulary}
\usepackage{arydshln}
\usepackage{hhline}
\usepackage{mathtools}
\usepackage{tabularx}
\usepackage{bm}
\usepackage{threeparttable}
\usepackage{tikz}
\usetikzlibrary{mindmap, shadows}
\usepackage{forest}
\usetikzlibrary{positioning}
\usetikzlibrary{shapes, calc, arrows}
\usetikzlibrary{decorations.markings}
\usepackage{enumitem}

\usepackage[final]{changes} % 使用 draft 以显示修改
\let\added\relax

\newcommand{\myColor}{black}

\AtBeginDocument{%
  \providecommand\BibTeX{{%
    \normalfont B\kern-0.5em{\scshape i\kern-0.25em b}\kern-0.8em\TeX}}}

\setcopyright{acmcopyright}
\copyrightyear{2023}
\acmYear{2023}
\acmDOI{XXXXXXX.XXXXXXX}

\usepackage[justification=justified,skip=3pt]{caption}

\renewcommand{\arraystretch}{0.85}

\begin{document}

\title{Robust Recommender System: A Survey and Future Directions}


\author{Kaike Zhang}
\email{zhangkaike21s@ict.ac.cn}
\affiliation{%
  \institution{Institute of Computing Technology, CAS}
  % \streetaddress{No.6 Kexueyuan South Road Zhongguancun, Haidian District}
  \city{Beijing}
  \country{China}
  \postcode{100190}
}
\affiliation{%
  \institution{University of Chinese Academy of Sciences}
  % \streetaddress{No.1 Yanqihu East Rd}
  \city{Beijing}
  \country{China}
  \postcode{101408}
}

\author{Qi Cao}
\authornote{Corresponding author}
\email{caoqi@ict.ac.cn}
\author{Fei Sun}
\email{sunfei@ict.ac.cn}
\affiliation{%
  \institution{Institute of Computing Technology, CAS}
  % \streetaddress{No.6 Kexueyuan South Road Zhongguancun, Haidian District}
  \city{Beijing}
  \country{China}
  \postcode{100190}
}

\author{Yunfan Wu}
\email{wuyunfan19b@ict.ac.cn}
\author{Shuchang Tao}
\email{taoshuchang18z@ict.ac.cn}
\affiliation{%
  \institution{Institute of Computing Technology, CAS}
  % \streetaddress{No.6 Kexueyuan South Road Zhongguancun, Haidian District}
  \city{Beijing}
  \country{China}
  \postcode{100190}
}
\affiliation{%
  \institution{University of Chinese Academy of Sciences}
  % \streetaddress{No.1 Yanqihu East Rd}
  \city{Beijing}
  \country{China}
  \postcode{101408}
}

\author{Huawei Shen}
\email{shenhuawei@ict.ac.cn}
\author{Xueqi Cheng}
\email{cxq@ict.ac.cn}
\affiliation{%
  \institution{Institute of Computing Technology, CAS}
  % \streetaddress{No.6 Kexueyuan South Road Zhongguancun, Haidian District}
  \city{Beijing}
  \country{China}
  \postcode{100190}
}
\affiliation{%
  \institution{University of Chinese Academy of Sciences}
  % \streetaddress{No.1 Yanqihu East Rd}
  \city{Beijing}
  \country{China}
  \postcode{101408}
}

\renewcommand{\shortauthors}{Kaike Zhang, et al.}

\begin{abstract}
  \input{Section/0-Abstract}

\end{abstract}

% TODO



% TODO
% \received{20 February 2023}
% \received[revised]{12 March 2023}
% \received[accepted]{5 June 2023}

\maketitle

\section{Introduction}\label{1-intro}
\input{Section/1-Introduction}

\section{Definition and Taxonomy}\label{2-pre}
\input{Section/2-Preliminary_2}

\section{Robustness against Malicious Attack}\label{3-Attack}
\input{Section/3-against_Attack}

\section{Robustness against Natural Noise}\label{4-Noise}
\input{Section/4-against_Noise}

\section{Evaluation}\label{5-Eva}
\input{Section/5-Evaluation}

\section{Robustness in Various Recommendation Scenarios}\label{6-Dis}
\input{Section/6-Discussion}

\section{Relationship with Other Trustworthy Properties}\label{7-Rel}

\input{Section/7-Relationship}

\section{Open Issues and Future Directions}\label{8-Fut}

\input{Section/8-Future_Direction}

\section{Conclusion}\label{9-con}
\input{Section/9-Conclusion}

\begin{acks}
This work is funded by the National Key R\&D Program of China (2022YFB3103700, 2022YFB3103701), and the National Natural Science Foundation of China under Grant Nos. 62272125, 62102402, U21B2046. Huawei Shen is also supported by Beijing Academy of Artificial Intelligence (BAAI).
\end{acks}

\bibliographystyle{ACM-Reference-Format}
\bibliography{ref}

\clearpage
\appendix
\pagenumbering{arabic}
\renewcommand{\thepage}{A-\arabic{page}}
\setcounter{page}{1}
\section{APPENDIX}
\input{Section/10-Appendix}

\end{document}

%% file: Section/0-Abstract.tex
With the rapid growth of information, recommender systems have become integral for providing personalized suggestions and overcoming information overload. However, their practical deployment often encounters ``dirty'' data, where noise or malicious information can lead to abnormal recommendations. Research on improving recommender systems' robustness against such dirty data has thus gained significant attention. This survey provides a comprehensive review of recent work on recommender systems' robustness. We first present a taxonomy to organize current techniques for withstanding malicious attacks and natural noise. We then explore state-of-the-art methods in each category, including fraudster detection, adversarial training, certifiable robust training \added{for defending against malicious attacks}, and regularization, purification, self-supervised learning \added{for defending against malicious attacks}. Additionally, we \added{summarize evaluation metrics and commonly used datasets for assessing robustness}. We discuss robustness across varying recommendation scenarios and its interplay with other properties like accuracy, interpretability, privacy, and fairness. Finally, we delve into open issues and future research directions in this emerging field. Our goal is to provide readers with a comprehensive understanding of robust recommender systems and to identify key pathways for future research and development. \added{To facilitate ongoing exploration, we maintain a continuously updated GitHub repository with related research: \url{https://github.com/Kaike-Zhang/Robust-Recommender-System}.}

% With the rapid growth of information, recommender systems have become integral for providing personalized suggestions and overcoming information overload. However, their practical deployment often encounters ``dirty'' data, where noise or malicious information can lead to abnormal recommendations. Research on improving recommender systems' robustness against such dirty data has thus gained significant attention. This survey provides a comprehensive review of recent work on recommender systems' robustness. We first present a taxonomy to organize current techniques for withstanding malicious attacks and natural noise. We then explore state-of-the-art methods in each category, including fraudster detection, adversarial training, certifiable robust training for defending against malicious attacks, and regularization, purification, self-supervised learning for defending against malicious attacks. Additionally, we summarize evaluation metrics and commonly used datasets for assessing robustness. We discuss robustness across varying recommendation scenarios and its interplay with other properties like accuracy, interpretability, privacy, and fairness. Finally, we delve into open issues and future research directions in this emerging field. Our goal is to provide readers with a comprehensive understanding of robust recommender systems and to identify key pathways for future research and development. To facilitate ongoing exploration, we maintain a continuously updated GitHub repository with related research: \url{https://github.com/Kaike-Zhang/Robust-Recommender-System}. 

%% file: Section/1-Introduction.tex
In the era of information overload, recommender systems have emerged as powerful tools for providing personalized suggestions across a wide range of applications~\citep{cheng2016wide, covington2016deep, gomez2015netflix}. From e-commerce platforms like Amazon and Alibaba to streaming services such as YouTube and Netflix, recommender systems have become an important part of the user experience. These systems not only aid users in their decision-making processes but also contribute to revenue growth for businesses. Recommender systems leverage large amounts of data, including past user behavior, demographic information, and contextual data, to generate personalized recommendations for various products, services, and content~\citep{jannach2010recommender}.

\textbf{Importance and Pervasiveness of Robustness in Recommender Systems.} Recommender systems heavily rely on data to provide accurate recommendations to users, using past behavior and preferences to predict future actions~\cite{jannach2010recommender}. Despite their widespread adoption and value in numerous industries, the natural openness of recommender systems means practical implementations often face the challenge of dealing with dirty data~\cite{17_wu2021fight, 54_qin2021world}.
In 2018, the New York Times reported an article about the gray industry derived from YouTube's fake views\footnote{https://www.nytimes.com/interactive/2018/08/11/technology/youtube-fake-view-sellers.html}. In the same year, the BBC reported that false comments on online review websites could affect £23 billion in British customer spending\footnote{https://www.bbc.com/news/technology-43907695}. In 2021, Guardian reported that Facebook employees found more than 30 cases of political manipulation involving 25 countries on the platform\footnote{https://www.theguardian.com/technology/2021/apr/12/facebook-loophole-state-backed-manipulation}.
Noisy or malicious information can lead to inaccuracies in the recommendations provided, resulting in a poor user experience or reduced business value. Consequently, ensuring the robustness of recommender systems is essential, as it measures recommender systems' ability to provide stable recommendations when the data is partially damaged~\cite{58_gunawardana2012evaluating,47_oh2022rank,60_o2004collaborative,15_zhang2017robust,fawzi2017robustness}.

\begin{figure}[t]%
\centering
\includegraphics[width=0.9\textwidth]{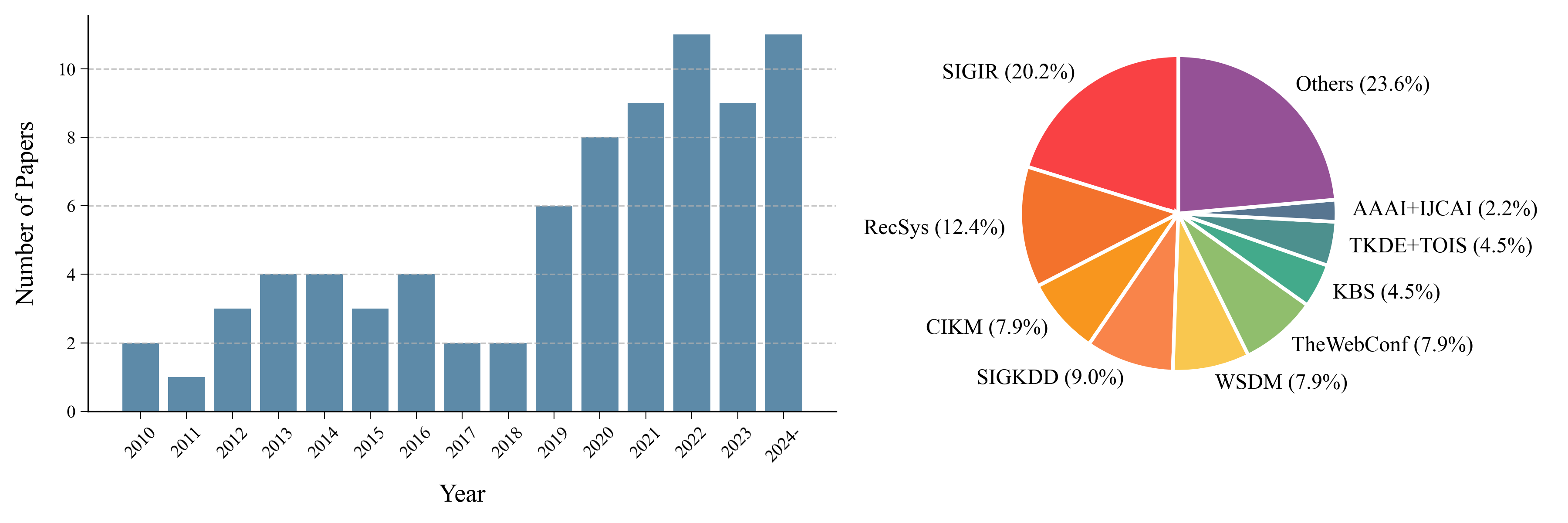}
\caption{\added{The statistics of publications related to robust recommender systems with the publication year and conference/journal.}}\label{pub_statistic}
\end{figure}

In recent years, \added{robustness in} recommender systems has become a focal point of research~\citep{3_anelli2021study, 17_wu2021fight, 27_tian2022learning, 31_gao2022self, 35_wang2022learning, 38_zheng2021multi, 54_qin2021world}. This trend is evident in the growing number of published papers on the topic, as shown in Figure~\ref{pub_statistic}. According to the chart, research interest in \added{this area} has increased significantly since 2019. This surge in attention is also reflected in the inclusion of tutorials and workshops on recommender systems' robustness at various top-tier conferences. For instance, RecSys featured a tutorial on \textit{Adversarial Learning for Recommendation}~\cite{anelli2020adversarial}, dedicated to robustness in recommender systems, in 2020. Similarly, TheWebConf hosted a tutorial on \textit{Trustworthy Recommender Systems} in 2023, highlighting robustness as a key topic. Additionally, several workshops have emphasized the importance of robustness in recommender systems. Specifically, SIGIR’s \textit{Causality in Search and Recommendation}~(2021)~\cite{zhang2021csr}, TheWebConf’s \textit{Decision Making for Information Retrieval and Recommender Systems}~(2023)~\cite{xu2023foreword}, \added{and \textit{The 1st Workshop on Human-Centered Recommender Systems}~(2025)~\cite{zhang20241st}}, along with SIGKDD’s \textit{Workshop on Industrial Recommendation Systems}~(2021)~\cite{xu20212nd}, have identified robustness as a core theme. Therefore, we aim to provide a systematic review of robustness in recommender systems.

\textbf{Difference with Existing Surveys.} 
Numerous surveys on recommender systems have been published recently. Some focus on specific types of recommender systems, such as graph-based recommender systems~\citep{wu2022graph}, knowledge-based recommender systems~\citep{tarus2018knowledge}, deep-learning-based recommender systems~\citep{zhang2019deep}, and reinforcement-learning-based recommender systems~\cite{afsar2022reinforcement}. Others examine broader issues beyond accuracy, including trustworthy recommender systems~\citep{ge2022survey}, debiased recommender systems~\citep{chen2023bias}, explainable recommender systems~\citep{zhang2020explainable}, and evaluation methodologies for recommender systems~\cite{zangerle2022evaluating, alhijawi2022survey}. 
To the best of our knowledge, existing surveys related to robustness primarily focus on a limited subset of recommender systems' robustness, emphasizing adversarial strategies~\citep{deldjoo2021survey, anelli2021adversarial} or the detection of malicious users~\citep{rezaimehr2021survey}. \added{In contrast, our survey provides a comprehensive perspective on robustness in recommender systems, not only considering both adversarial and malicious behaviors but also exploring techniques for mitigating natural noise.} Additionally, while some surveys discuss robustness in Computer Vision and Natural Language Processing~\cite{wang2021measure, zhang2018deep}, they do not specifically address recommender systems. Given the unique characteristics of recommendation tasks, such as the interdependencies between entities and transductive recommendation scenarios, the strategies for enhancing recommender systems' robustness may differ significantly from conventional approaches. Therefore, our survey aims to fill this gap by focusing on robustness in recommender systems and exploring the techniques and strategies specifically tailored to recommendation tasks.

\textbf{Papers Collection.} 
In this survey, we undertook an extensive search across premier conferences and journals, including SIGKDD, SIGIR, Webconf, RecSys, WSDM, CIKM, NeurIPS, ICML, TKDE, TOIS, etc. We employed search keywords such as ``recommender system'', ``recommendation'', ``collaborative filtering'', and ``matrix factorization'', coupled with terms like ``robustness'', ``denoise'', ``defense'', or ``detection'', spanning the period from 2010 to \added{2025}.
To ensure the representative nature of our paper collection, we further screened the publications.
Notably, while a small number of researchers identify \textit{bias} and \textit{sparseness} as issues pertinent to a kind of robustness~\cite{10.1145/3383313.3412261, 38_zheng2021multi, zhang2020deep}, there is a gap with the robustness in general, which measures recommender systems' ability to provide stable recommendations when the data is partially damaged~\cite{58_gunawardana2012evaluating,47_oh2022rank,60_o2004collaborative,15_zhang2017robust,fawzi2017robustness}.
Therefore, in this survey, we consciously exclude papers that solely focus on issues of bias and sparseness to maintain the specificity and relevance of our review.
If you are interested in these properties, there are some relevant surveys that have already been conducted on bias~\citep{chen2023bias} and sparseness~\citep{gope2017survey, abdullah2021eliciting}.

% \textbf{Contributions of this survey.}
% We have organized the papers on the development of robustness in recommender systems to facilitate a swift understanding and entry for researchers into this field, aiming to provide further assistance for the advancement of this area. In summary, this survey offers three significant contributions:
% \begin{itemize}
% \item Comprehensive and systematic review of recommender systems' robustness, including a detailed and complete classification system for different dimensions.
% \item General overviews of the current state-of-the-art methods and the evaluation methods and datasets employed in the field.
% \item Discussions of the robustness of the recommender system in different scenarios, the relationships with other attributes of recommender systems, and the challenges associated with recommender systems' robustness and the anticipated future development trends, offering additional perspectives for researchers in this area.
% \end{itemize}

\textbf{Contributions of this survey.}
We have meticulously arranged a collection of papers regarding the robustness of recommender systems to ensure a quick and efficient understanding for researchers entering into this field. Our goal is to further aid the progress in this area. Briefly, the noteworthy contributions of this survey are as follows:
\begin{itemize}
\item A comprehensive and systematic taxonomy for robustness-enhance methods in recommender systems.
\item An all-encompassing overview of the representative methodologies, as well as evaluation approaches and datasets currently employed in the domain.
\item Detailed discussions encompass various facets: the main consideration of recommender systems' robustness in diverse scenarios, its correlation with other trustworthy properties of recommender systems, as well as open issues coupled with recommender systems' robustness, and trends for future development.
\end{itemize}

The remaining sections of this survey are organized as follows: In Section~\ref{2-pre}, we introduce the concept of robustness in recommender systems, defining what robustness means in this context and outlining a taxonomy for the field. Sections~\ref{3-Attack} and \ref{4-Noise} provide a detailed discussion of state-of-the-art strategies for developing robust recommender systems, addressing both malicious attacks and natural noise, respectively. Section~\ref{5-Eva} explores the metrics and datasets commonly used to evaluate the robustness of recommender systems, \added{and provides a comparative evaluation of representative robustness methods.} In Section~\ref{6-Dis}, we examine robustness considerations across various recommendation scenarios, while Section~\ref{7-Rel} investigates the interrelationship between robustness and other trustworthy properties of recommender systems, such as accuracy, interpretability, privacy, and fairness. Section~\ref{8-Fut} highlights open challenges and potential future directions in the study of robustness in recommender systems. Finally, Section~\ref{9-con} presents a succinct conclusion to this comprehensive survey.

% The remaining sections of this survey are organized as follows: In Section~\ref{2-pre}, we introduce the concept of robustness in the realm of recommender systems, defining what is robustness in recommender systems and discussing the taxonomy within this field. In Sections~\ref{3-Attack} and \ref{4-Noise}, we present a detailed exposition of state-of-the-art strategies for developing robust recommender systems, catering to both malicious attacks and natural noise, respectively. Section~\ref{5-Eva} delves into the metrics and datasets that are commonly employed for evaluating recommender systems' robustness, \added{and provides a comparative evaluation of representative robustness methods.} In Section~\ref{6-Dis}, we discuss the considerations of recommender systems' robustness in various scenarios, while Section~\ref{7-Rel} focuses on the interrelationship between recommender systems' robustness and other trustworthy properties of recommender systems, such as accuracy, interpretability, privacy, and fairness. Moving towards the conclusion, Section~\ref{8-Fut} sheds light on the open issues and potential future directions in the research of recommender systems' robustness. Finally, Section~\ref{9-con} provides a succinct conclusion to this comprehensive survey.

%% file: Section/2-Preliminary_2.tex
% In this section, we provide a formal definition of robustness in the context of recommender systems and outline the taxonomy pertinent to this domain.
% To provide a clear understanding, we first give the formalization of recommendation tasks.
% Suppose we have a set of $M$ users, denoted by $\mathcal{U}$, a set of $N$ items, denoted by $\mathcal{I}$, and a rating set $\mathcal{R}$, containing feedback $r_{u,i}$ of user $u \in \mathcal{U}$ for item $i \in \mathcal{I}$. The collected user-item interactions are represented by $\mathcal{D}$, where $(u, i, r_{u,i}) \in \mathcal{D}$. Our goal is to learn a parametric model $f:\mathcal{U} \times \mathcal{I} \rightarrow \mathcal{R}$ from $\mathcal{D}$ that minimizes the following objective function:
% \begin{equation}
%     \begin{aligned}
%         \mathcal{L}(\mathcal{D}) = \sum_{(u, i, r_{u,i}) \in \mathcal{D}} \psi(f(u,i),r_{u,i}),
%     \end{aligned}
%     \label{eq:org_l}
% \end{equation}
% where $\psi$ is the error function used to measure the distance between the predicted and ground truth labels. In order to facilitate the search for the corresponding meaning of the symbol, we provide the corresponding meaning of the symbol used in this survey in Table~\ref{tab:symbol}.

% In the actual training process, we typically divide the collected data $\mathcal{D}$ into a training set $\mathcal{D}^+$ and a test set $\mathcal{D}^-$. After training on $\mathcal{D}^+$ through $\mathcal{L}(\mathcal{D}^+)$ mentioned in Equation~\ref{eq:org_l}, we obtain an recommender systems model denoted as $f_{\mathcal{D}^+}$. 

In this section, we provide a formal definition of robustness in the context of recommender systems and outline the taxonomy relevant to this domain. \added{To establish a clear foundation, we first formalize recommendation tasks.} Suppose we have a set of $M$ users, denoted by $\mathcal{U}$, and a set of $N$ items, denoted by $\mathcal{I}$. The rating set $\mathcal{R}$ contains feedback $r_{u,i}$ from user $u \in \mathcal{U}$ for item $i \in \mathcal{I}$. The collected user-item interactions are represented by $\mathcal{D}$, where $(u, i, r_{u,i}) \in \mathcal{D}$. Our goal is to learn a parametric model $f:\mathcal{U} \times \mathcal{I} \rightarrow \mathcal{R}$ from $\mathcal{D}$ that minimizes the following objective function:
\begin{equation}
    \begin{aligned}
        \mathcal{L}(\mathcal{D}) = \sum_{(u, i, r_{u,i}) \in \mathcal{D}} \psi(f(u,i),r_{u,i}),
    \end{aligned}
    \label{eq:org_l}
\end{equation}
where $\psi$ is the error function used to measure the \added{difference} between the predicted and ground truth labels. \added{For clarity, we provide a reference table summarizing the symbols used throughout this survey in Appendix Table~\ref{tab:symbol}.} During model training, we typically split the collected data $\mathcal{D}$ into a training set $\mathcal{D}^+$ and a test set $\mathcal{D}^-$. After training on $\mathcal{D}^+$ by optimizing $\mathcal{L}(\mathcal{D}^+)$ in Equation~\ref{eq:org_l}, we obtain a recommender system model, denoted as $f_{\mathcal{D}^+}$.

\subsection{Definition of Robustness in Recommender Systems}
\label{definition_sec}

The term ``robustness'' initially described three essential characteristics of a parameterized model: \textit{efficiency}, \textit{stability}, and \textit{non-breakdown}~\cite{huber1981robust}. 
\textit{Efficiency} refers to \added{maintaining algorithmic efficiency in scenarios that align with model assumptions}, stability demands minimal sensitivity to minor deviations, and non-breakdown ensures performance does not collapse under substantial deviations. \added{With the advancement of deep learning, robustness is generally defined as the ability of a model to maintain stable performance despite variations in relevant factors throughout the entire pipeline. Specifically, robustness can be categorized into three key phases of the machine learning pipeline:}

\begin{itemize}[leftmargin=*]
    \item \added{\textbf{Robustness in Model Training}: The ability to maintain performance despite perturbations or manipulations of training data or gradients, often referred to as poisoning attacks~\cite{jagielski2018manipulating, tian2022comprehensive}. In recommender systems, this involves altering training data (e.g., injecting fake users) to degrade performance or promote/nuke specific items~\cite{tang2020revisiting, christakopoulou2019adversarial}.}
    
    \item \added{\textbf{Robustness in Model Reasoning (Inference)}: The capability of the model to consistently produce stable results for identical or closely related input contexts, also known as test-time perturbations. In recommender systems, this pertains to maintaining recommendation consistency despite slight variations in user or item features, interactions, or environmental conditions~\cite{58_gunawardana2012evaluating}. These variations may arise from attacks or noise introducing misleading recommendations.}
    
    \item \added{\textbf{Robustness in Model Evaluation}: The extent to which performance metrics remain stable under varying evaluation conditions, such as different metrics~\cite{wang2023recad}. This dimension emphasizes the reliability and trustworthiness of evaluation results despite variations in data or metric choices.}
\end{itemize}

\added{Currently, certain recommendation paradigms, such as collaborative filtering, lack inference capabilities for new users and items until the model is explicitly updated. Consequently, they do not naturally support research on robustness in model reasoning (i.e., test-time perturbations). As a result, most robustness research in recommender systems primarily focuses on challenges during the training phase. Therefore, in this survey, we limit our discussion to \textbf{robustness concerning training-phase perturbations}. Additionally, in Section~\ref{5-Eva}, we evaluate the effectiveness of various robustness methods across different types of metrics.} Formally, we define robust recommender systems with respect to training-phase perturbations as follows:

\begin{definition}[$(\epsilon, \varepsilon)$-Robust Recommender Systems]
Given a recommendation model $f$, a dataset $\mathcal{D}$ partitioned into a training set $\mathcal{D}^+$ and a test set $\mathcal{D}^-$, a perturbation magnitude $\epsilon$, and an acceptable performance deviation $\varepsilon$, the recommender system is considered robust if:
\begin{equation}
    |f_{\mathcal{D}^+}(\mathcal{D}^-)-f_{\mathcal{D}^++\Delta}(\mathcal{D}^-)| \leq \varepsilon, \quad \forall \|\Delta\|\leq\epsilon,
\end{equation}
where $f_{\mathcal{D}^+}$ denotes the model trained on clean training data, and $f_{\mathcal{D}++\Delta}$ denotes the model trained on perturbed training data.
\end{definition}

\begin{figure}[t]%
\centering
\includegraphics[width=0.9\textwidth]{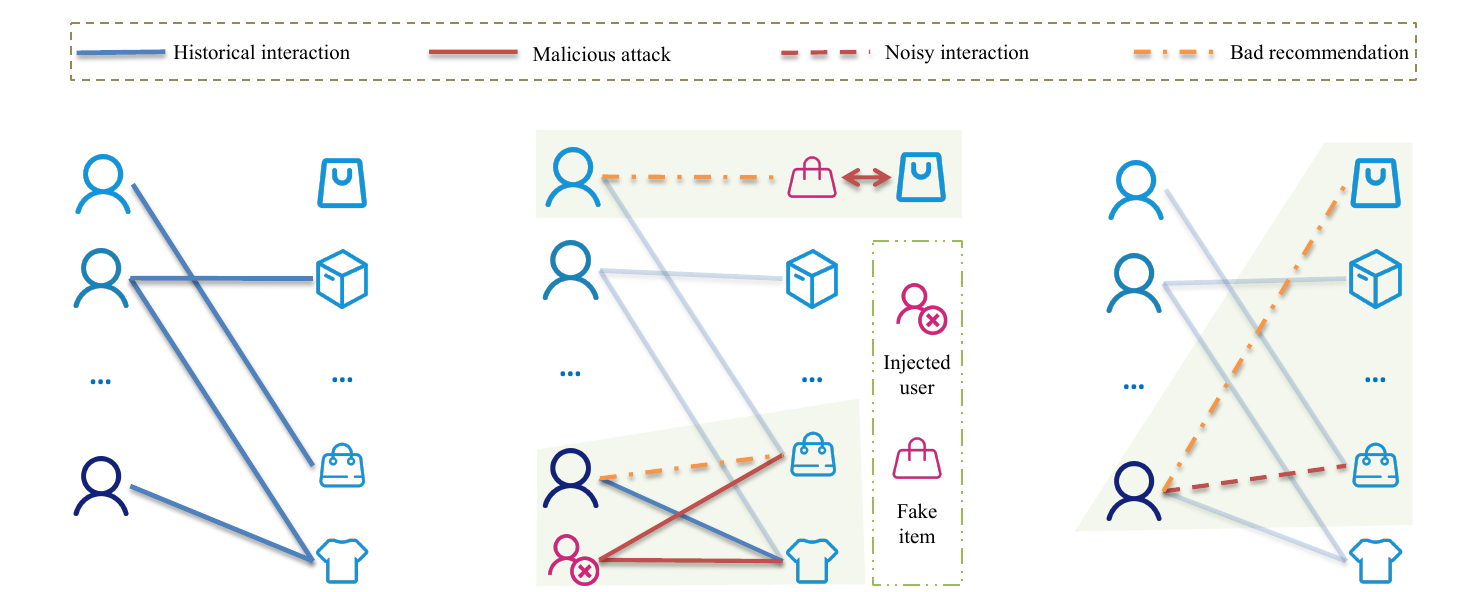}
\vspace{-0.3cm}
\\
\subfigure[User-item interaction graph]{
\includegraphics[width=0.2775\textwidth]{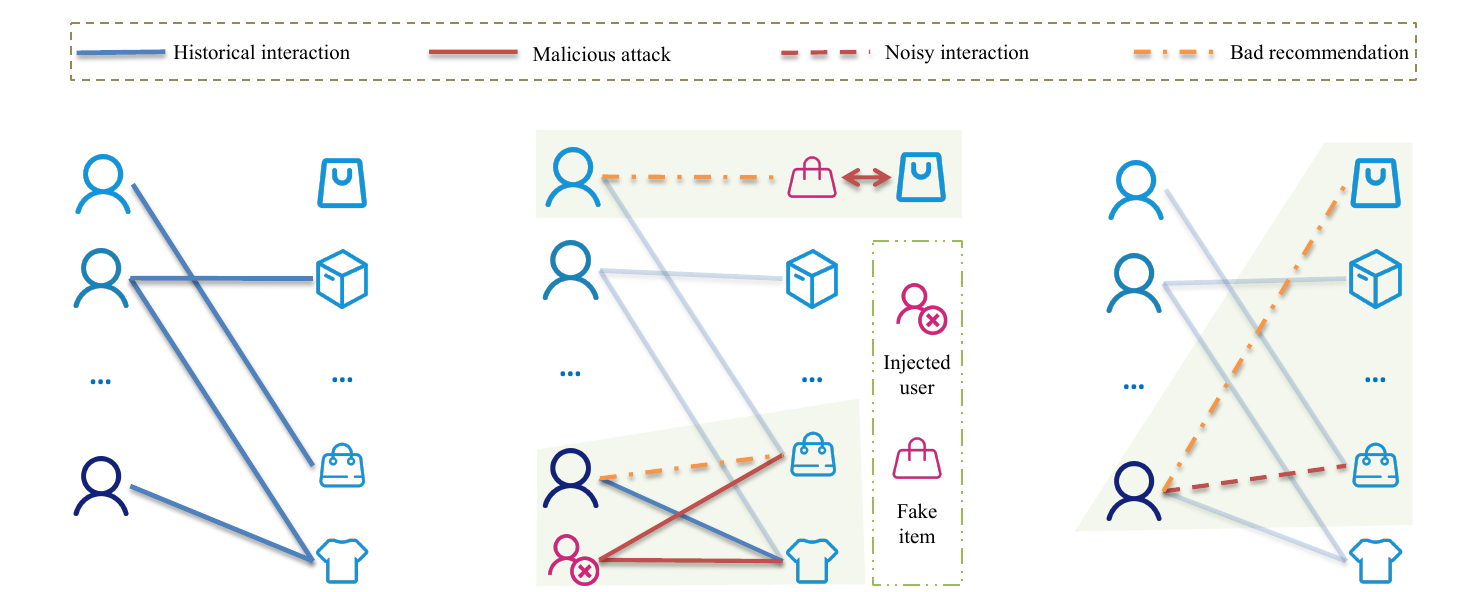}
}
\subfigure[with malicious attack]{
\includegraphics[width=0.3424\textwidth]{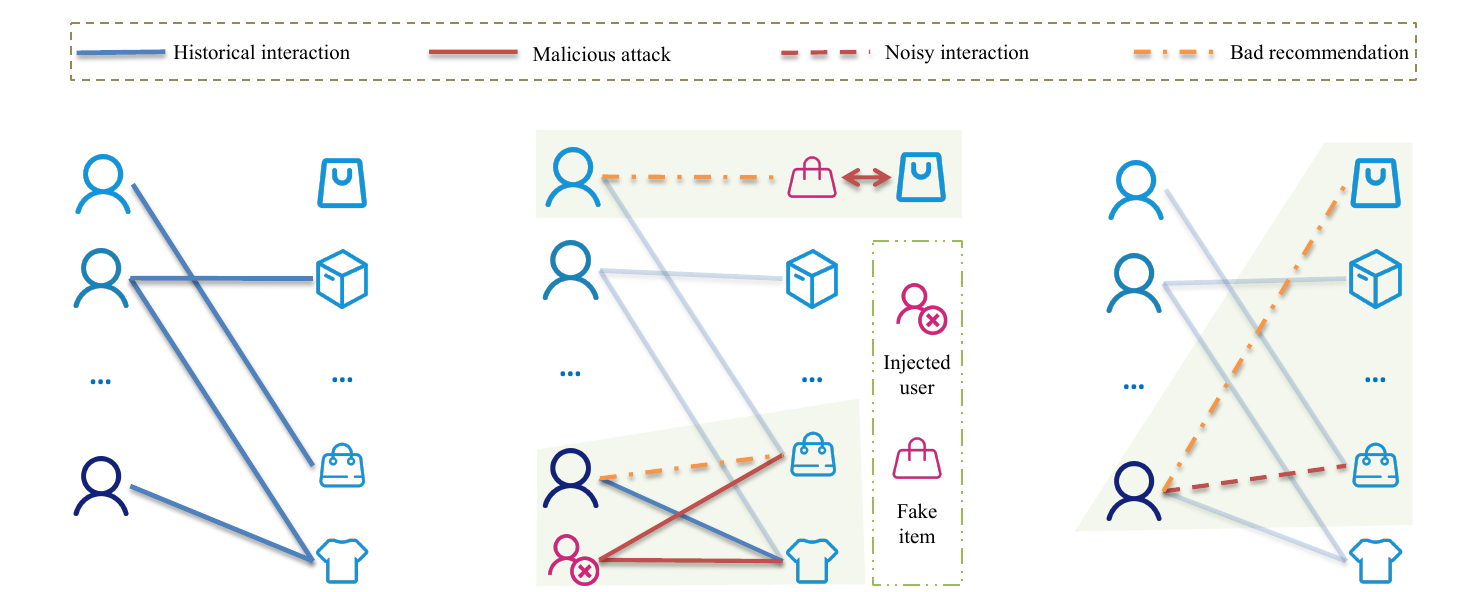}
\label{cat_attack}
}
\subfigure[with noisy interaction]{
\includegraphics[width=0.2801\textwidth]{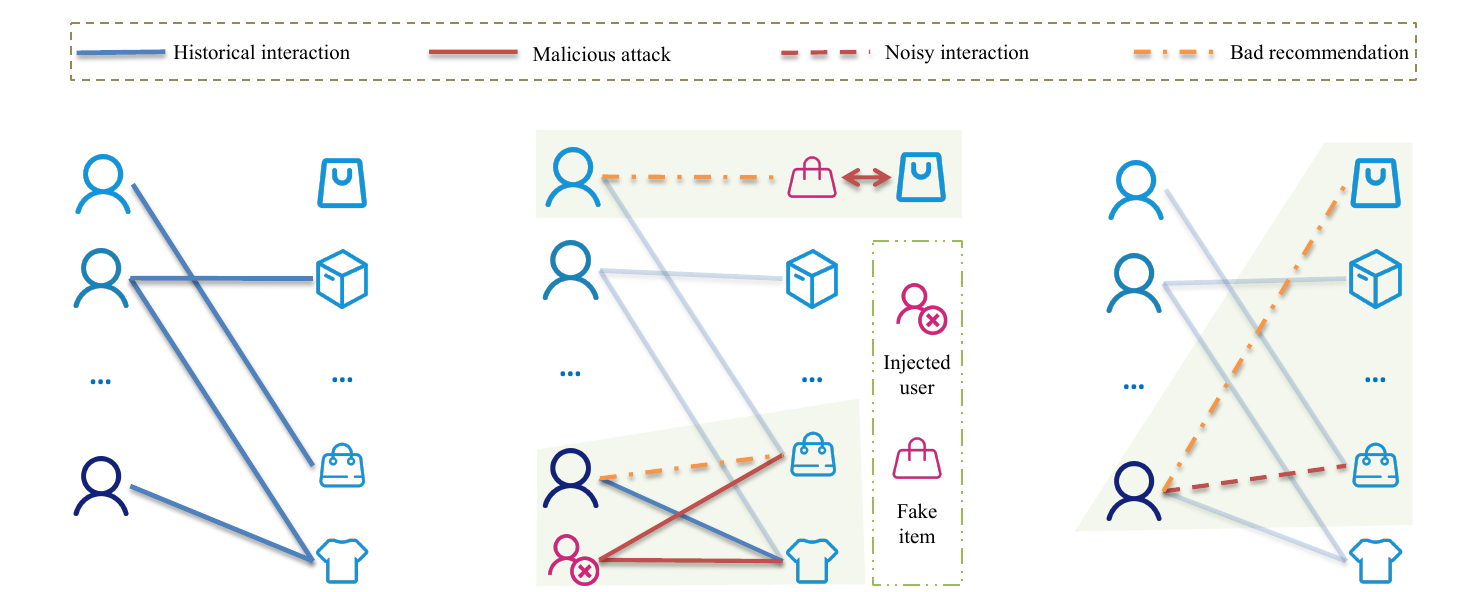}
\label{cat_noise}
}
\caption{User-item interaction graph with malicious attack and natural noise.}
\end{figure}

\subsection{Taxonomy of Robustness in Recommender System}

Recommender systems function as highly interactive platforms, making them vulnerable to various forms of abnormal data. These anomalies may stem from malicious activities, such as injecting fake users and tampering with item information, or from natural noise, which typically arises due to human errors or ambiguities in user behavior.

For malicious attacks, attackers often aim to promote/nuke specific items or degrade the performance of recommender systems. Generally, adversarial scenarios limit attackers' ability to manipulate a user's historical behavior. There are two primary types of attacks in recommendation scenarios, as shown in Figure~\ref{cat_attack}: (1)~\added{\textbf{Item-side information modification}: Attackers alter item side information to artificially boost the popularity of specific items~\cite{cohen2021black}, as depicted at the top of Figure~\ref{cat_attack}.} (2)~\added{\textbf{Fake user injection (shilling attack)}: Attackers introduce fake users to inflate or suppress the exposure of certain items or to degrade overall system performance~\cite{tang2020revisiting, christakopoulou2019adversarial}, as shown at the bottom of Figure~\ref{cat_attack}.}
% \begin{enumerate}[leftmargin=*]
%     \item \added{\textbf{Item-side information modification}: Attackers alter item side information to artificially boost the popularity of specific items~\cite{cohen2021black}, as depicted at the top of Figure~\ref{cat_attack}.}
%     \item \added{\textbf{Fake user injection (shilling attack)}: Attackers introduce fake users to inflate or suppress the exposure of certain items or to degrade overall system performance~\cite{tang2020revisiting, christakopoulou2019adversarial}, as shown at the bottom of Figure~\ref{cat_attack}.}
% \end{enumerate}
Even with restrictions on potential attacks, several defense mechanisms against interaction-level attacks have been proposed~\cite{7_yuan2020exploring, 42_chen2022adversarial, 30_yue2022defending, 9_liu2020certifiable}. \added{A detailed discussion of these defense strategies will be provided in Section~\ref{adv-on-int}.}

Natural noise primarily arises from user-generated factors such as human errors, uncertainty, and ambiguous user behavior~\cite{20_yera2016fuzzy}. For example, noise can manifest in user-item interactions, such as accidental clicks or gift purchases reflecting others' preferences, as illustrated in Figure~\ref{cat_noise}. Additionally, noise may appear in user or item side information, such as incorrect personal details or mislabeled item tags. However, due to limited research addressing these types of noise in recommender systems~\cite{53_liu2021concept}, Figure~\ref{cat_noise} does not include this category.

To enhance the robustness of recommender systems, it is essential to distinguish between different types of abnormal data. Accordingly, we categorize recommender system robustness into two primary types based on the nature of these anomalies:

% \begin{figure}
%     \centering
%     \resizebox{0.81\textwidth}{!}{\input{Tex_Graph/Category2}}
%     \caption{A lookup graph for the reviewed methods on robustness in recommender systems.}
%     \label{fig:category}
% \end{figure}

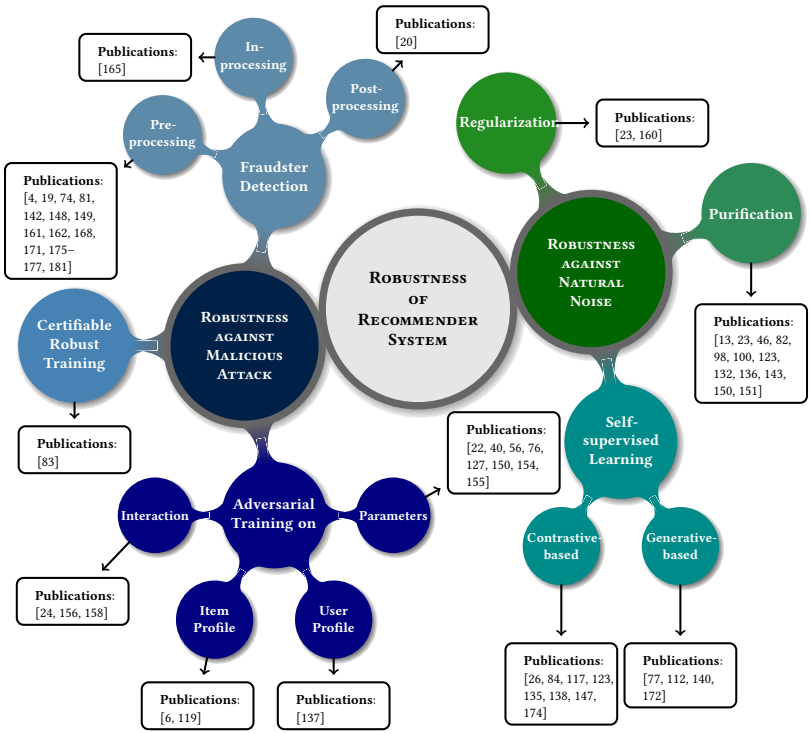
\begin{figure}
    \centering
    \resizebox{0.96\textwidth}{!}{\input{Tex_Graph/Category3}}
    \caption{\added{A lookup graph for the reviewed methods on robustness in recommender systems.}}
    \label{fig:category}
\end{figure}

\begin{itemize}[leftmargin=*]
    \item \added{\textbf{Robustness against Malicious Attacks}: This category focuses on strategies to counteract injected users or falsified item data. Approaches are broadly divided into three types:}
    \begin{itemize}[leftmargin=*]
        \item \added{\textit{Fraudster detection}~\cite{28_wu2012hysad,16_zhang2020gcn,14_cao2020adversarial}: Identifies fraudsters in training data and reduces their influence by either removing them or assigning lower weights.}
        \item \added{\textit{Adversarial training}~\cite{4_he2018adversarial,2_tang2019adversarial,30_yue2022defending}: Enhances model robustness by deliberately introducing small perturbations during training.}
        \item \added{\textit{Certifiable robust training}~\cite{9_liu2020certifiable}: Trains models to function optimally under worst-case perturbations with formal robustness guarantees.}
    \end{itemize}
    
    \item \added{\textbf{Robustness against Natural Noise}: This category addresses methods designed to mitigate the impact of noisy interactions. Techniques can be classified into three types:}
    \begin{itemize}[leftmargin=*]
        \item \added{\textit{Regularization}~\cite{15_zhang2017robust,39_chen2022denoising}: Introduces a regularization term (e.g., R1-norm regularization~\cite{15_zhang2017robust}) in the loss function to reduce the effect of noise.}
        \item \added{\textit{Purification}~\cite{43_wang2021denoising,27_tian2022learning}: Identifies and corrects noise in user-item interactions to enhance performance and robustness.}
        \item \added{\textit{Self-supervised learning}~\cite{46_shenbin2020recvae,35_wang2022learning}: Mitigates noise through generative-based methods (e.g., Denoising Autoencoder~\cite{8_li2015deep}) and contrastive-based approaches (e.g., contrastive learning~\cite{27_tian2022learning}).}
    \end{itemize}
\end{itemize}

The taxonomy of publications on recommender system robustness is presented in Figure~\ref{fig:category}.

%% file: Tex_Graph/Category3.tex
\definecolor{LayerOne}{HTML}{1B4332}  
\definecolor{LayerTwo}{HTML}{2D6A4F}   
\definecolor{LayerThree}{HTML}{D3F8E2}
% \definecolor{LayerThree}{HTML}{40916C} 

% \definecolor{LayerOne}{HTML}{253D36}  
% \definecolor{LayerTwo}{HTML}{375D52}   
% \definecolor{LayerThree}{HTML}{5D8378} 

\tikzset{
  basic/.style = {draw, text width=2.5cm, font=\sffamily, rectangle, align=center, text=white},
  root/.style = {basic, rounded corners=6pt, thin, text width=15em, fill=LayerOne, font=\Large\scshape, minimum height=4em},
  level 2/.style = {basic, rounded corners=6pt, thin, fill=LayerTwo, text width=10em, minimum height=4em},
  level 3/.style = {basic, thin, fill=LayerThree, text width=10.5em, align=left, text=black},
  wide 2/.style = {basic, rounded corners=6pt, thin, fill=LayerTwo, text width=11em, minimum height=4em},
  nonwide 3/.style = {basic, thin, fill=LayerThree, text width=7em, align=left, text=black},
}

\begin{tikzpicture}[
  level 1/.style={sibling distance=44mm},
  edge from parent/.style={->, draw, thick, color=black},
  >=latex
]

% Root node for the first tree
\node[root] (a) {\textbf{Robustness against Malicious Attack}}
  child {node[wide 2, yshift=-0.5cm] (a1) {\textbf{Fraudster Detection} \\ Sec.~\ref{sec:detetct}}}
  child {node[wide 2, yshift=-0.5cm] (a2) {\textbf{Adversarial Training} \\ Sec.~\ref{sec:adv}}}
  child {node[level 2, yshift=-0.5cm] (a3) {\textbf{Certifiable \\ Robust Training} \\ Sec.~\ref{sec:cert}}};

% Second level nodes under Fraudster Detection
\begin{scope}[every node/.style={level 3}]
  \node [below of = a1, xshift=10pt, yshift=-1.2cm] (a11) {\textbf{Pre-processing} Sec.~\ref{sec:3-1-1-pre} \\ \cite{28_wu2012hysad, 61_zhang2012meta, 50_lee2012shilling, 45_zou2013belief, 51_cao2013shilling, 25_zhang2014hht, 63_zhang2014detection, 26_zhou2014detection, 10_zhang2015catch, 64_zhou2015shilling, 11_yang2016re, 52_zhou2016svm, 12_yang2017spotting, 13_aktukmak2019quick, 49_liu2020recommending, zhang2024llm4dec}};
  \node [below of = a11, yshift=-0.9cm] (a12) {\textbf{In-processing} \\ Sec.~\ref{sec:3-1-2-in} \\ \cite{16_zhang2020gcn, zhang2024lorec}};
  \node [below of = a12, yshift=-0.5cm] (a13) {\textbf{Post-processing} Sec.~\ref{sec:3-1-3-post} \\ \cite{14_cao2020adversarial}};
\end{scope}

% Second level nodes under Adversarial Training
\begin{scope}[every node/.style={level 3}]
  \node [below of = a2, xshift=10pt, yshift=-1cm] (a21) {on \textbf{Parameters} \\ Sec.~\ref{sec:3-2-1-para} \\ \cite{4_he2018adversarial,1_du2018enhancing,5_yuan2019adversarial,6_yuan2019adversarial,24_chen2019adversarial,66_tran2019adversarial,32_li2020adversarial,40_ye2023towards, chen2023adversarial, zhang2024improving, zhangunderstanding}};
  \node [below of = a21, yshift=-0.75cm] (a22) {on \textbf{User Profile} \\ Sec.~\ref{sec:3-2-2-user} \\ \cite{17_wu2021fight}};
  \node [below of = a22, yshift=-0.5cm] (a23) {on \textbf{Item Profile} \\ Sec.~\ref{sec:adv_item} \\ \cite{2_tang2019adversarial,3_anelli2021study}};
  \node [below of = a23, yshift=-0.5cm] (a24) {on \textbf{Interaction} \\ Sec.~\ref{adv-on-int} \\ \cite{7_yuan2020exploring, 42_chen2022adversarial, 30_yue2022defending}};
\end{scope}

\begin{scope}[every node/.style={nonwide 3}]
  \node [below of = a3, xshift=15pt, yshift=-1cm] (a31) {Sec.~\ref{sec:cert} \\ \cite{9_liu2020certifiable}};
\end{scope}

% Lines from the root to level 2 nodes
\foreach \value in {1,2,3}
  \draw[->] (a1.west) ++(0.2,-0.68) |- (a1\value.west);
  
\foreach \value in {1,...,4}
  \draw[->] (a2.west) ++(0.2,-0.68) |- (a2\value.west);

\foreach \value in {1}
  \draw[->] (a3.west) ++(0.2,-0.68) |- (a3\value.west);

% Root node for the second tree
\node[root, xshift=13cm] (n) {\textbf{Robustness against Natural Noise}}
  child {node[level 2, yshift=-0.5cm] (n1) {\textbf{Regularization} \\ Sec.~\ref{sec:reg}}}
  child {node[wide 2, yshift=-0.5cm] (n2) {\textbf{Purification} \\ Sec.~\ref{sec:pur}}}
  child {node[wide 2, yshift=-0.5cm] (n3) {\textbf{Self-supervised Learning} \\ Sec.~\ref{sec:ssl}}};

\begin{scope}[every node/.style={nonwide 3}]
  \node [below of = n1, xshift=15pt, yshift=-1cm] (n11) {\cite{15_zhang2017robust,39_chen2022denoising}};
\end{scope}

\begin{scope}[every node/.style={nonwide 3}]
  \node [below of = n2, xshift=8pt, yshift=-1.5cm] (n21) {\cite{71_pham2013preference, 20_yera2016fuzzy, 43_wang2021denoising, 44_bian2021denoising, 53_liu2021concept, 54_qin2021world, 56_wang2021implicit, 27_tian2022learning, 31_gao2022self, 39_chen2022denoising, 70_xie2022denoising, 40_ye2023towards, quan2023robust, wang2023efficient, xia2024neural, he2024double, zhang2025personalized}};
\end{scope}

% Second level nodes under Self-supervised Learning
\begin{scope}[every node/.style={level 3}]
  \node [below of = n3, xshift=10pt, yshift=-0.8cm] (n31) {\textbf{Generative-based} \\ Sec.~\ref{sec:4-3-1-gen} \\ \cite{8_li2015deep,21_wu2016collaborative,38_zheng2021multi,46_shenbin2020recvae, li2024recdiff, jiang2024diffkg}};
  \node [below of = n31, yshift=-1.5cm] (n32) {\textbf{Contrastive-based} \\ Sec.~\ref{sec:4-3-2-con} \\ \cite{67_zhou2020s3, 55_strub2015collaborative, 23_liu2021contrastive, 36_wu2021self, 27_tian2022learning, 34_yang2022knowledge, 35_wang2022learning, 37_chen2022intent, wang2023knowledge, jiang2023adaptive, sun2024self, wang2023denoised}};
\end{scope}

\foreach \value in {1}
  \draw[->] (n1.west) ++(0.2,-0.68) |- (n1\value.west);

\foreach \value in {1}
  \draw[->] (n2.west) ++(0.2,-0.68) |- (n2\value.west);

% Lines from the root to level 2 nodes
\foreach \value in {1,2}
  \draw[->] (n3.west) ++(0.2,-0.68) |- (n3\value.west);

\end{tikzpicture}

%% file: Section/3-against_Attack.tex
In this section, we introduce various representative methods across three key categories---fraudster detection, adversarial training, and certifiable robust training---to enhance the robustness of recommender systems against malicious attacks.

\subsection{Fraudster Detection}
\label{sec:detetct}
\input{Section/3-subsection/1-Detection}

\subsection{Adversarial \added{Training}}
\label{sec:adv}
\input{Section/3-subsection/2-ADV_Training}

\subsection{Certifiable Robust Training}
\label{sec:cert}
\input{Section/3-subsection/3-Certifiable}

%% file: Section/3-subsection/1-Detection.tex
% The detection of fraudsters is paramount for ensuring the robustness of recommender systems, especially against malicious attacks. Such attacks often involve fraudsters providing misleading feedback to manipulate the system~\cite{tang2020revisiting, christakopoulou2019adversarial}. Thus, fraudster detection aims to identify and eliminate these users to maintain reliable recommendations. Fraudster detection approaches can be categorized into three types, corresponding to when detection occurs: pre-processing, in-processing, and post-processing detection (Figure \ref{fig:detection}). Pre-processing detection~\cite{12_yang2017spotting, 52_zhou2016svm} aims to find fraudsters in the original data before training, eliminating their impact from the source. In-processing detection~\cite{16_zhang2020gcn} further leverages model feedback during training to detect and weaken fraudster impact. Post-processing detection~\cite{14_cao2020adversarial}  identifies and corrects bad recommendations caused by fraudsters. Each stage provides tailored detection strategies. Pre-processing is the mainstream approach, while in-processing and post-processing are more recent.

\begin{figure}[t]%
\centering
\subfigure[Fraudster detection in recommender systems]{
\includegraphics[width=0.4577\textwidth]{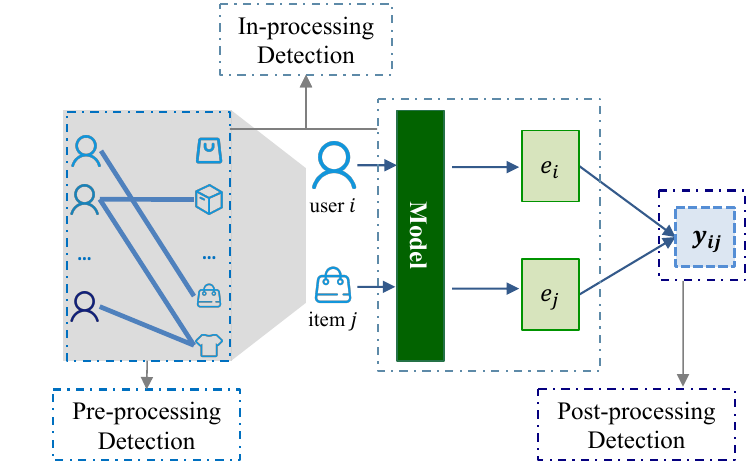}
\label{fig:detection}
}
\subfigure[Adversarial learning in recommender systems]{
\includegraphics[width=0.4423\textwidth]{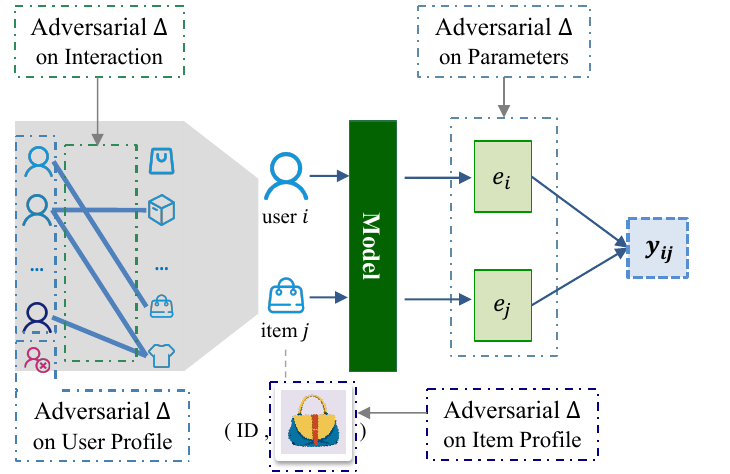}
\label{fig:adversarial}
}
\caption{Fraudster detection \& Adversarial learning}
\end{figure}

Detecting fraudsters is crucial for ensuring the robustness of recommender systems, particularly against malicious attacks. Such attacks often involve fraudsters providing misleading feedback to manipulate the system~\cite{tang2020revisiting, christakopoulou2019adversarial}. Therefore, fraudster detection aims to identify and eliminate these users to maintain reliable recommendations. Fraudster detection approaches can be categorized into three types based on the detection stage: pre-processing, in-processing, and post-processing detection (Figure~\ref{fig:detection}). 
\begin{itemize}[leftmargin=*]
    \item \added{\textbf{Pre-processing detection}~\cite{12_yang2017spotting, 52_zhou2016svm}: Identifies fraudsters in the original data before training, preventing their influence from the outset.}
    \item \added{\textbf{In-processing detection}~\cite{16_zhang2020gcn}: Utilizes model feedback during training to detect and mitigate fraudster impact dynamically.}
    \item \added{\textbf{Post-processing detection}~\cite{14_cao2020adversarial}: Identifies and corrects poor recommendations resulting from fraudsters after model training.}
\end{itemize}
Each stage offers distinct detection strategies. While pre-processing is the most widely used approach, in-processing and post-processing techniques have gained attention in recent research.

\subsubsection{Pre-processing Detection}
\label{sec:3-1-1-pre}

\added{Pre-processing detection refers to identifying and mitigating fraudsters in recommender systems before model training, ensuring a reliable and accurate training process. It consists of two stages,} of which the first stage is \textbf{Feature Extraction}, taking some statistics as the characteristics of each user; the second stage is \textbf{Detection}, through the features extracted in the first stage to detect fraudsters.

\input{Table/features}

\textbf{Feature Extraction}.  
Most pre-processing detection methods rely on feature engineering to incorporate prior knowledge. Early studies~\cite{61_zhang2012meta, 25_zhang2014hht, 26_zhou2014detection} focused on developing \added{user-centric features}. For example, Zhou et al.~\cite{26_zhou2014detection, 64_zhou2015shilling} introduced two indicators based on user attributes: Degree of Similarity with Top Neighbors (DegSim) and Rating Deviation from Mean Agreement (RDMA)~\cite{chirita2005preventing}. DegSim quantifies a user's similarity to their top $k$ nearest neighbors using the average Pearson correlation:
\begin{equation}
    \begin{aligned}
        \textit{DegSim}_u = \frac{\sum_{v \in N_{u}@k}W_{u,v}}{k},
    \end{aligned}
\end{equation}
where $W_{u,v}$ is the Pearson correlation between user $u$ and user $v$, and $N_{u}@k$ represents the $k$ nearest neighbors.  
RDMA measures \added{a user’s rating deviation from the mean agreement of other users on target items, incorporating inverse rating frequency:}
\begin{equation}
    \begin{aligned}
        \textit{RDMA}_u = \frac{\sum_{i \in \mathcal{I}_{u}}|r_{u,i}-\overline{r_i}|/|\mathcal{R}_i|}{|\mathcal{I}_{u}|},
    \end{aligned}
\end{equation}
where $\mathcal{I}_{u}$ is the set of items rated by user $u$, $\overline{r_i}$ is the average rating for item $i$, and $|\mathcal{R}_i|$ is the number of interactions for item $i$. Typically, attackers exhibit higher RDMA values and lower DegSim values~\cite{26_zhou2014detection}.  
Additional features, such as Weighted Deviation from Mean Agreement (WDMA)~\cite{12_yang2017spotting, burke2006classification} and Mean Variance (MeanVar)~\cite{28_wu2012hysad, 11_yang2016re}, further aid fraudster detection, particularly for filler items\footnote{Filler items refer to randomly rated items in an injected user’s profile in early non-optimized shilling attacks~\cite{gunes2014shilling}. A broader definition includes any items in a user’s profile that are neither the highest (push attack) nor the lowest (nuke attack)~\cite{burke2006classification}, which this survey adopts.}. Table~\ref{tab:features} summarizes representative generic features used in pre-processing detection.

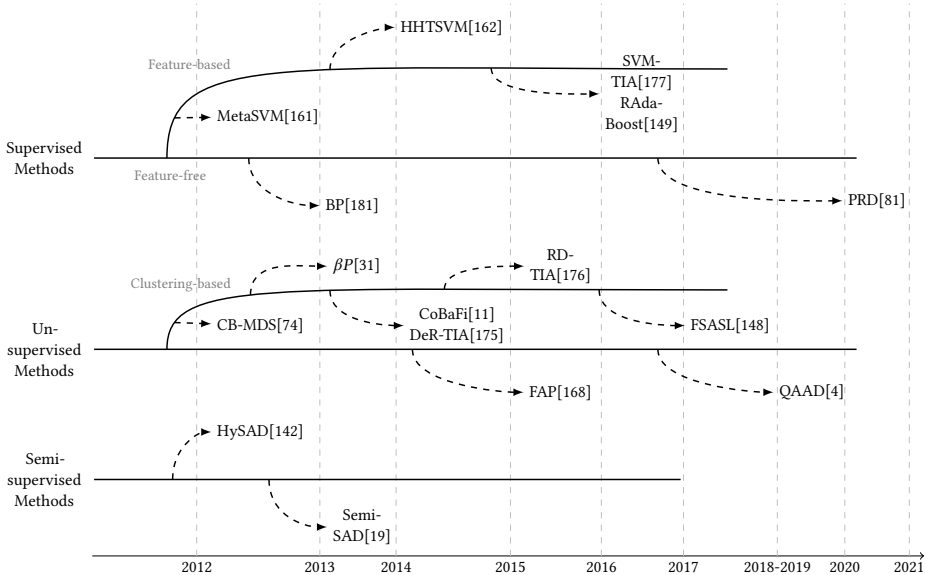
\begin{figure}[t]
    \centering
    \resizebox{0.95\textwidth}{!}{\input{Tex_Graph/1-Detection}}
    \caption{\added{Main development trajectory of pre-processing detection methods.}}
    \label{fig:detection_his}
\end{figure}

\textbf{Detection.}  
Detection methods for pre-processing can be broadly categorized into supervised, unsupervised, and semi-supervised learning approaches. To illustrate the evolution of these methods, Figure~\ref{fig:detection_his} \added{presents their development trajectory}.

\textit{Supervised learning}.  
Supervised fraudster detection relies on labeled datasets to train models for identifying fraudulent users. These datasets are typically constructed by simulating various attack methods and training models to recognize malicious activities. With feature extraction, classification models use feature vectors $\bm{x} = \{x_1, x_2, \dots, x_n\}$ as follows:
\begin{equation}
    \begin{aligned}
        y = f_c(x)
    \end{aligned}
\end{equation}
where $f_c$ can be an SVM-based classifier~\cite{61_zhang2012meta, 25_zhang2014hht, 52_zhou2016svm} (MetaSVM, HHTSVM, SVM-TIA) or an ensemble method such as AdaBoost~\cite{11_yang2016re} (RAdaBoost).  Additionally, some feature-free methods rely on interaction-based approaches. BP~\cite{45_zou2013belief} models the probability distribution of fraudsters and target items based on interactions, using Belief Propagation for inference. Inspired by Bayesian Personalized Ranking (BPR)~\cite{rendle2012bpr}, PRD~\cite{49_liu2020recommending} detects fraudsters by analyzing pairwise ranking distributions, mitigating sample imbalance issues without relying on pre-selected features. \added{Furthermore, LLM4Dec~\cite{zhang2024llm4dec} leverages open-world knowledge of large language models (LLMs) to enhance pre-processing detection using only user interaction data.} Such feature-free approaches improve adaptability by reducing dependence on predefined features, offering resilience against evolving malicious tactics.

\textit{Unsupervised learning}.  
Unsupervised fraudster detection does not require labeled data; instead, it identifies anomalies or patterns in user behavior. Initial clustering-based methods aim to distinguish fraudulent users from genuine ones based on feature distributions. As highlighted by Beutel et al.~\cite{29_beutel2014cobafi}~(CoBaFi), detectable spammers often form their distinct clusters. 

Clustering-based approaches are prevalent. CB-MDS~\cite{50_lee2012shilling} applies hierarchical clustering, while DeR-TIA and RD-TIA~\cite{26_zhou2014detection, 64_zhou2015shilling} use k-means. Some methods assume specific cluster properties. $\beta$P~\cite{33_chung2013betap} employs the $\beta$ distribution to detect attackers with pre-selected features, leveraging statistical properties to identify suspicious patterns. FSASL~\cite{12_yang2017spotting} capitalizes on the tendency of malicious users to mimic genuine users, resulting in higher-density clusters. Similar density-based clustering methods, such as DBSCAN and LOF~\cite{breunig2000lof}, effectively identify densely connected groups of fraudulent users.

Beyond clustering, \added{other strategies exploit the inherent structure of user-item interactions}. FAP~\cite{10_zhang2015catch} \added{detects anomalies by analyzing the graph structure of recommendations.} QAAD~\cite{13_aktukmak2019quick} introduces a probabilistic factorization model to estimate rating likelihoods, \added{assuming that genuine users exhibit higher likelihoods than fraudsters, whose fabricated profiles often show inconsistencies.}

\textit{Semi-supervised learning}.  
Semi-supervised learning leverages a small labeled dataset to train an initial classifier, then iteratively improves it by incorporating unlabeled data~\cite{28_wu2012hysad}. This approach maximizes the utility of limited labeled data while utilizing large volumes of unlabeled data to enhance detection performance. Notable examples include HySAD~\cite{28_wu2012hysad} and Semi-SAD~\cite{51_cao2013shilling}, which initially train a basic classifier on limited labels and refine it using the Expectation-Maximization (EM) algorithm. By learning from unlabeled samples, the classifier progressively improves its accuracy. Other semi-supervised methods, such as self-training, co-training, and multi-view learning~\cite{zhu2009introduction}, could be explored for fraudster detection in the future.

\textit{Conclusion}.  
Each pre-processing detection approach offers unique advantages but also presents challenges. Supervised methods struggle with data imbalance, \added{as genuine users far outnumber fraudsters. Unsupervised techniques, lacking explicit guidance, may fail to distinguish fraudsters who closely resemble genuine users.} Despite these challenges, solutions from other domains can be adapted. For instance, graph-based anomaly detection algorithms~\cite{hooi2016fraudar, hu2016embedding, morales2021selective} leverage the network structure formed by users and items to identify fraudsters. Advanced machine learning techniques, including deep learning~\cite{lecun2015deep} and reinforcement learning~\cite{sutton2018reinforcement}, could further enhance detection by capturing complex behavioral patterns.

\subsubsection{In-processing Detection}
\label{sec:3-1-2-in}
In-processing detection refers to identifying and mitigating fraudulent activities in recommender systems during model training. Unlike pre-processing detection, this approach leverages both the initial data and \added{real-time training information} from the training process, capturing complex and latent relationships among users and items.

In-processing detection~\cite{16_zhang2020gcn,zhang2024lorec} can be formulated as two tasks:

\textbf{Fraudster Detection}:  
\added{Leveraging information from the training process enables more accurate fraudster detection. GraphRfi~\cite{16_zhang2020gcn} incorporates principles from cognitive psychology, assuming that genuine users exhibit coherent and predictable behavior. If a user’s actual behavior significantly deviates from their predicted behavior, they are likely a fraudster. LoRec~\cite{zhang2024lorec} enhances detection by integrating item embeddings learned by the recommender system with user embeddings encoded through a sequential model, effectively modeling fraudster behavior patterns. Additionally, LoRec improves generalization by leveraging large language models (LLMs) to extend detection capabilities beyond specific attack patterns, enhancing robustness against unknown threats.} \added{The fraudster detection task is formalized as:}
\added{
\begin{equation}
    \begin{aligned}
        \mathcal{L}_{\mathrm{detection}} = \mathbb{E}_{u} \left [ -\log \mathbb{P} \left [y=y_u \vert a_u \right ] \right ],
    \end{aligned}
\end{equation}
}
\added{where $\mathbb{P}[y=y_u|a_u]$ represents the probability of user $u$ being classified as a fraudster ($y_u=0$) or a genuine user ($y_u=1$), and $a_u$ denotes auxiliary information obtained during the training process.}

\added{\textbf{Robust Recommendation}:  
By incorporating fraudster probability values from the detection phase, GraphRfi~\cite{16_zhang2020gcn} and LoRec~\cite{zhang2024lorec} dynamically adjust the training weight of each user to mitigate attack impacts. This is achieved through the following objective function:}
\added{
\begin{equation}
    \begin{aligned}
        \mathcal{L}_{\mathrm{rec}} = \mathbb{E}_{u} \left [ \sum_{i \in \mathcal{I}} \mathbb{P}\left [ y=1|a_u \right ] \cdot \mathcal{L}(u, i) \right ],
    \end{aligned}
\end{equation}}
\added{where $\mathcal{L}(u, i)$ represents the loss associated with user $u$ and item $i$.}

\subsubsection{Post-processing Detection}
\label{sec:3-1-3-post}
Post-processing detection \added{occurs after the recommender system has been trained}. Its primary objective is to filter out low-quality recommendations influenced by fraudsters, ensuring that the system provides accurate and reliable suggestions to genuine users.

A representative post-processing detection method was proposed by \citet{14_cao2020adversarial}. In the reinforcement learning (RL)-based recommendation setting, they introduce a two-part detection model. The first component is a Gated Recurrent Unit (GRU) encoder, which encodes the action sequences—i.e., the recommendation lists generated by the RL agent—into a low-dimensional feature vector. The second component is an attention-based decoder with a classifier that differentiates between high-quality and low-quality recommendations. This method effectively identifies poor-quality recommendations resulting from fraudulent activities. The RL-based framework enables the model to continuously refine its decision-making process by learning from previous filtering results, thereby enhancing the overall robustness of the recommender system’s output.

\subsubsection{Discussion of Fraudster Detection}
In the context of fraudster detection in recommender systems, there are three principal strategies: pre-processing, in-processing, and post-processing detection. Pre-processing detection \added{operates before model training}, offering the advantage of computational efficiency during training~\cite{50_lee2012shilling, 13_aktukmak2019quick}. In-processing detection integrates model insights throughout training to achieve improved accuracy~\cite{16_zhang2020gcn}. Meanwhile, post-processing detection aims to filter out poor recommendations for subsequent items~\cite{14_cao2020adversarial}, though its performance may be compromised when filtering entire recommendation lists.

\added{The trade-off between precision and recall is a crucial challenge in fraudster detection.} \added{Lowering the classification threshold may increase fraudster detection rates but risks misidentifying genuine users, negatively affecting user experience and financial outcomes. Thus, future detection methods should prioritize maintaining high precision while improving recall.} Exploring advanced deep learning techniques, such as graph neural networks~\cite{wu2020comprehensive}, appears promising for identifying complex patterns and relationships between users and items, which could further enhance detection.

% \subsubsection{Discussion of Fraudster Detection}
% In the context of fraudster detection within recommendation systems, there are three principal detection strategies: pre-processing, in-processing, and post-processing detection. Pre-processing detection operates outside of model training, offering the advantage of computational efficiency during the training phase~\cite{50_lee2012shilling, 13_aktukmak2019quick}. In-processing detection integrates model insights throughout training to achieve better accuracy~\cite{16_zhang2020gcn}. Meanwhile, post-processing detection aims to filter out bad recommendations~\cite{14_cao2020adversarial}. 

% \added{The trade-off between precision and recall is a crucial challenge in fraudster detection.} \added{Lowering the classification threshold may increase fraudster detection rates but risks misidentifying genuine users, negatively affecting user experience and financial outcomes. Thus, future detection methods should prioritize maintaining high precision while improving recall.} Advancements in deep learning offer promising directions. In particular, graph neural networks (GNNs)~\cite{wu2020comprehensive} have demonstrated potential in capturing complex relationships between users and items, which could further enhance fraudster detection capabilities.

%% file: Table/features.tex
\begin{table}[t]
    \centering
    \renewcommand{\arraystretch}{1.15}
    \caption{Representative generic features in fraudster detection}
    \resizebox{\linewidth}{!}{
    \begin{tabular}{llp{0.8\textwidth}p{0.15\textwidth}}
         \toprule
         \textbf{Feature}       &\textbf{Formula}   &\multirow{1}{0.6\textwidth}{\centering \textbf{Description}}    &\multirow{1}{0.15\textwidth}{\centering \textbf{Publications}}\\
         \midrule
          \multirow{1}*[-0.5em]{\textbf{DegSim}}            & \multirow{1}*[-0.5em]{$\frac{\sum_{v \in N_{u}@k}W_{u,v}}{k}$ }                                         & \textbf{Degree Similarity Neighbors (DegSim)}: Average Pearson Correlation of a profile’s top $k$ neighbors.      & \cite{61_zhang2012meta, 51_cao2013shilling, 63_zhang2014detection, 26_zhou2014detection, 64_zhou2015shilling} \\
         \multirow{1}*[-0.5em]{\textbf{RDMA}}               & \multirow{1}*[-0.5em]{$\sum_{i \in \mathcal{I}_u}\frac{\lvert r_{u,i} - \overline{r_i} \rvert / \lvert \mathcal{R}_i \rvert}{\lvert \mathcal{I}_u \rvert}$}    & \textbf{Rating Deviation Mean Agreement (RDMA)}: Identifies attackers by examining average deviation and item ratings.   & \cite{61_zhang2012meta, 51_cao2013shilling, 63_zhang2014detection, 26_zhou2014detection, 64_zhou2015shilling, 11_yang2016re} \\
         \multirow{1}*[-0.5em]{\textbf{WDA}}                & \multirow{1}*[-0.5em]{$\sum_{i=0}^{N_u}\frac{\lvert r_{u,i} - \overline{r_i} \rvert}{\lvert \mathcal{R}_i \rvert}$}           & \textbf{Weighted Degree Agreement (WDA)}: Utilizes the RDMA numerator without considering profile rating count.           & \multirow{1}*[-0.5em]{\cite{61_zhang2012meta, 11_yang2016re, 12_yang2017spotting}} \\
          \multirow{1}*[-0.5em]{\textbf{WDMA}}              & \multirow{1}*[-0.5em]{$\sum_{i \in \mathcal{I}_u}\frac{\lvert r_{u,i} - \overline{r_i} \rvert / \lvert \mathcal{R}_i \rvert^2}{\lvert \mathcal{I}_u \rvert}$}  & \textbf{Weighted Deviation Mean Agreement (WDMA)}: Based on RDMA, assigns higher weight to sparse item ratings. & \cite{28_wu2012hysad, 61_zhang2012meta, 11_yang2016re, 12_yang2017spotting}\\
          \multirow{1}*[-0.5em]{\textbf{MeanVar}}           & \multirow{1}*[-0.5em]{$\frac{\sum_{r_{u, i} \in \mathcal{R}_u - \mathcal{R}_{u, \mathrm{max}}}(r_{u, i} - \bar{r_u})^2}{\lvert \mathcal{R}_u - \mathcal{R}_{u, \mathrm{max}} \rvert}$}       & \textbf{Mean-Variance (MeanVar)}: Computes variance between filler items and overall average for attack detection. & \multirow{1}*[-0.5em]{\cite{28_wu2012hysad, 11_yang2016re, 12_yang2017spotting}}  \\
          \multirow{1}*[-0.5em]{\textbf{LengthVar}}         & \multirow{1}*[0em]{$\frac{ \left \lvert \lvert \mathcal{R}_u \rvert - \overline{\lvert \mathcal{R}_u \rvert} \right \rvert}{\sum_{v \in \mathcal{U} (\lvert \mathcal{R}_v \rvert - \lvert \overline{\mathcal{R}_u} \rvert)^2}}$}     & \textbf{Length-Variance (LengthVar)}: Uses profile rating count to distinguish between genuine and attack profiles.     & \cite{61_zhang2012meta, 51_cao2013shilling, 11_yang2016re, 12_yang2017spotting}\\
         \bottomrule
    \end{tabular}
    }
    \label{tab:features}
\end{table}

%% file: Tex_Graph/1-Detection.tex
\begin{tikzpicture}[node distance=2cm, auto]
\footnotesize

    \draw[->] (0.82, -6.7) -- (15.4, -6.7);
    \foreach \x/\year in {2.6/2012, 4.7/2013, 6/2014, 7.95/2015, 9.5/2016, 10.9/2017, 12.5/2018-2019, 13.65/2020, 14.55/2021\added{-2024}}
    \draw[dashed, lightgray] (\x, -6.7) -- (\x, 2.6) node[below=9.3cm, black] {\year};
    
    \node[text width=1.5cm, align=center] (1) {Supervised \\ Methods};
    \node[text width=1.5cm, align=center, below=2.3cm of 1] (2) {Un-supervised \\ Methods};
    \node[text width=1.5cm, align=center, below=1.1cm of 2] (3) {Semi-supervised \\ Methods};
    
    \node (45) [right=3.85cm of 1, yshift=-0.8cm]{BP\cite{45_zou2013belief}};
    \node (61) [right=2cm of 1, yshift=0.7cm]{MetaSVM\cite{61_zhang2012meta}};
    \node (25) [right=1.2cm of 61, yshift=1.5cm]{HHTSVM\cite{25_zhang2014hht}};
    \node (52) [right=1.55cm of 25, yshift=-1.12cm, align=center]{SVM-\\TIA\cite{52_zhou2016svm}\\RAda-\\Boost\cite{11_yang2016re}};
    % \node (11) [below=0.8cm of 52, align=center]{RAda-\\Boost\cite{11_yang2016re}};
    \node (49) [right=2.7cm of 52, yshift=-1.8cm]{PRD\cite{49_liu2020recommending}};
    \node (new1) [right=3.2cm of 52, yshift=-0.2cm, align=center]{\added{LLM4Dec}\\ \added{\cite{zhang2024llm4dec}}};

    \node (ept1) [right=13cm of 1] {};
    \draw[line width=0.25mm] (1) to node[pos=0.1, below=0.1cm, font=\scriptsize, text=gray] {Feature-free} (ept1);
    % \draw[dashed, line width=0.25mm] ($(1)!.9!(ept1)$) to (ept1);
    \node (ept2) [left=1cm of ept1, yshift=1.5cm] {};
    \draw[line width=0.25mm] ($(1)!.15!(ept1)$) .. controls +(up:1.8cm) and +(left:8cm) .. node[pos=0.25, above=0.4cm, font=\scriptsize, text=gray] {Feature-based}  (ept2) coordinate[pos=0.15] (EPT2_M1) coordinate[pos=0.5] (EPT2_M2) coordinate[pos=0.7] (EPT2_M3) coordinate[pos=0.95] (EPT2_M4);
    
    \node (50) [right=2cm of 2, yshift=0.4cm]{CB-MDS\cite{50_lee2012shilling}};
    \node (33) [right=0.3cm of 50, yshift=1cm]{$\beta P$\cite{33_chung2013betap}};
    \node (29) [right=0.3cm of 33, align=center, yshift=-1cm]{CoBaFi\cite{29_beutel2014cobafi}\\DeR-TIA\cite{26_zhou2014detection}};
    \node (64) [right=2.35cm of 33, align=center]{RD-\\TIA\cite{64_zhou2015shilling}};
    \node (12) [right=1.5cm of 64,yshift=-1cm]{FSASL\cite{12_yang2017spotting}};
    \node (10) [below=1.5cm of 64] {FAP\cite{10_zhang2015catch}};
    \node (13) [right=3cm of 10] {QAAD\cite{13_aktukmak2019quick}};
    
    \node (ept3) [right=13cm of 2] {};
    \draw[line width=0.25mm] (2) to (ept3);
    \node (ept4) [left=2cm of ept3, yshift=1cm] {};
    \draw[line width=0.25mm] ($(2)!.15!(ept3)$) .. controls +(up:1.2cm) and +(left:8cm) .. node[pos=0.2, above=0.3cm, font=\scriptsize, text=gray] {Clustering-based}  (ept4) coordinate[pos=0.15] (EPT4_M1) coordinate[pos=0.45] (EPT4_M2) coordinate[pos=0.6] (EPT4_M3) coordinate[pos=0.75] (EPT4_M4) coordinate[pos=0.9] (EPT4_M5);

    \node (28) [right=2cm of 3, yshift=0.8cm]{HySAD\cite{28_wu2012hysad}};
    \node (51) [right=0.3cm of 28, align=center, yshift=-1.6cm]{Semi-\\SAD\cite{51_cao2013shilling}};

    \node (ept5) [right=10cm of 3] {};
    \draw[line width=0.25mm] (3) to (ept5);

    \draw[dashed, ->, >=latex, line width=0.25mm] (EPT2_M1) to (61);
    \draw[dashed, ->, >=latex, line width=0.25mm] (EPT2_M2) .. controls +(up:0.6cm) and +(left:1cm) .. (25);
    \draw[dashed, ->, >=latex, line width=0.25mm] (EPT2_M3) .. controls +(down:0.5cm) and +(left:1cm) .. (52);
    \draw[dashed, ->, >=latex, line width=0.25mm] ($(1)!.25!(ept1)$) .. controls +(down:0.6cm) and +(left:1cm) ..  (45);
    % \draw[dashed, ->, >=latex, line width=0.25mm] ($(1)!.45!(ept1)$) .. controls +(down:0.8cm) and +(left:1cm) .. (11);
    \draw[dashed, ->, >=latex, line width=0.25mm] ($(1)!.75!(ept1)$) .. controls +(down:0.8cm) and +(left:1cm) ..  (49);

    \draw[dashed, ->, >=latex, line width=0.25mm] ($(1)!.85!(ept1)$) .. controls +(up:0.5cm) and +(left:1.2cm) ..  node[pos=0.3, above=0.2cm, font=\scriptsize, text=gray] {\added{LLM-empowered}} (new1);
    
    % \draw[->, >=latex, line width=0.25mm] (1) to node[pos=0.3, above, align=center, font=\scriptsize, text=gray] {Feature\\-based}  node[pos=0.75, above, align=center, font=\scriptsize, text=gray] {Probabilistic-based} (45);
    % \draw[->, >=latex, line width=0.25mm] ($(1)!.45!(45)$) .. controls +(up:0.5cm) and +(left:1cm) .. node[pos=0.5, above=0.3cm, font=\scriptsize, text=gray] {SVM-based} (61);
    % \draw[->, >=latex, line width=0.25mm] ($(1)!.45!(45)$) .. controls +(down:1cm) and +(left:3cm) .. node[pos=0.5, above, font=\scriptsize, text=gray] {Others} (11);
    % \draw[->, >=latex, line width=0.25mm] ($(1)!.25!(45)$) .. controls +(down:2cm) and +(left:3cm) .. node[pos=0.5, above, font=\scriptsize, text=gray] {Feature-free} (49);
    % \draw[dashed, line width=0.25mm] (61) to (25);
    % \draw[dashed, line width=0.25mm] (25) to (52);

    \draw[dashed, ->, >=latex, line width=0.25mm] (EPT4_M1) to (50);
    \draw[dashed, ->, >=latex, line width=0.25mm] (EPT4_M2) .. controls +(up:0.6cm) and +(left:1cm) .. (33);
    \draw[dashed, ->, >=latex, line width=0.25mm] (EPT4_M3) .. controls +(down:0.6cm) and +(left:1cm) .. (29);
    \draw[dashed, ->, >=latex, line width=0.25mm] (EPT4_M4) .. controls +(up:0.4cm) and +(left:1cm) .. (64);
    \draw[dashed, ->, >=latex, line width=0.25mm] (EPT4_M5) .. controls +(down:0.6cm) and +(left:1cm) .. (12);
    \draw[dashed, ->, >=latex, line width=0.25mm] ($(2)!.45!(ept3)$) .. controls +(down:0.6cm) and +(left:1cm) .. node[pos=0.8, above=0.2cm, font=\scriptsize, text=gray] {\added{Graph-based}} (10);
    \draw[dashed, ->, >=latex, line width=0.25mm] ($(2)!.75!(ept3)$) .. controls +(down:0.6cm) and +(left:1cm) .. node[pos=0.8, above=0.2cm, font=\scriptsize, text=gray] {\added{Probabilistic Factorization}} (13);
    
    % \draw[->, >=latex, line width=0.25mm] (2) to node[pos=0.6, above, align=center, font=\scriptsize, text=gray] {Clustering-based} (50);
    % \draw[->, >=latex, line width=0.25mm] ($(2)!.35!(50)$) .. controls +(down:1.5cm) and +(left:3cm) .. node[pos=0.3, above=0.1cm, font=\scriptsize, text=gray] {Others} (13) coordinate[pos=0.5] (M);
    % \draw[->, >=latex, line width=0.25mm] (M) .. controls +(up:0.2cm) and +(left:2cm) .. (10);
    % \draw[->, >=latex, line width=0.25mm] (50) to (33);
    % \draw[->, >=latex, line width=0.25mm] (33) to (29);
    % \draw[->, >=latex, line width=0.25mm] (29) to (64);
    % \draw[->, >=latex, line width=0.25mm] (64) to (12);

    \draw[dashed, ->, >=latex, line width=0.25mm] ($(3)!.2!(ept5)$) .. controls +(up:0.6cm) and +(left:1cm) .. (28);
    \draw[dashed, ->, >=latex, line width=0.25mm] ($(3)!.35!(ept5)$) .. controls +(down:0.6cm) and +(left:1cm) .. (51);
    % \draw[->, >=latex, line width=0.25mm] (3) to (28);
    % \draw[->, >=latex, line width=0.25mm] (28) to (51);

\end{tikzpicture}

%% file: Section/3-subsection/2-ADV_Training.tex
Adversarial training has emerged as a promising approach to enhance the robustness of deep models against various malicious attacks~\cite{goodfellow2014explaining, yuan2019adversarial}. This technique introduces small, carefully crafted perturbations to input data during training, improving the model’s resilience to adversarial samples~\cite{tang2020revisiting, christakopoulou2019adversarial}. By learning from these adversarial samples, recommender systems can better generalize to unseen or manipulated data, thereby enhancing overall robustness and reliability.

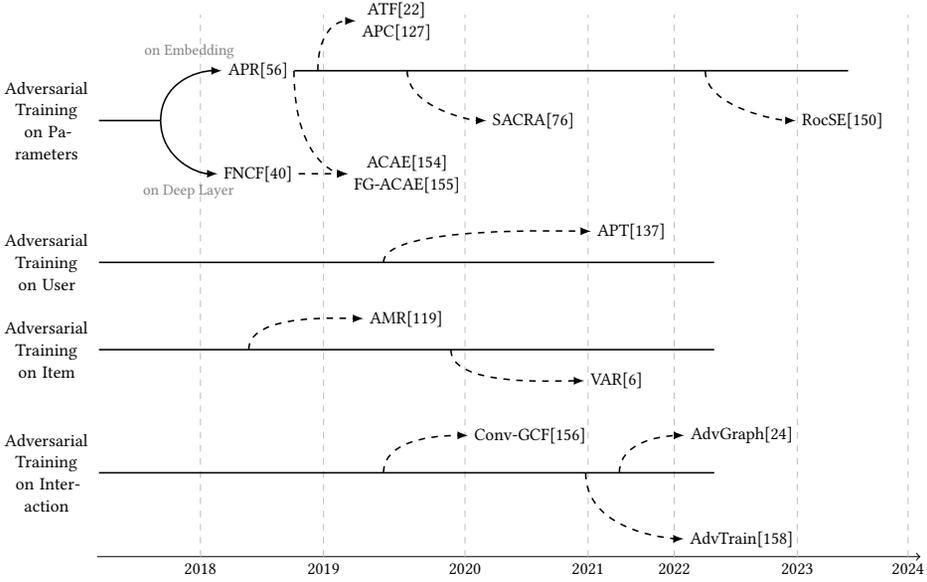
\begin{figure}[t]
    \centering
    \resizebox{0.95\textwidth}{!}{\input{Tex_Graph/2-Adversarial}}
    \caption{\added{Main development trajectory of adversarial training methods.}}
    \label{fig:adv_his}
\end{figure}

Adversarial \added{training} in recommender systems can be categorized based on the nature of the perturbations introduced. Specifically, it can target model parameters, user profiles, item profiles, or user-item interactions. These categories are graphically illustrated in Figure~\ref{fig:adversarial}. To provide a comprehensive view of the evolution of each category, we depict their development trajectories in Figure~\ref{fig:adv_his}.In the following subsections, we delve into each category, providing a comprehensive overview of the representative works.

\subsubsection{Adversarial Perturbation on Parameters}
\label{sec:3-2-1-para}

Most input features in recommender systems, such as user and item IDs or categorical attributes, are discrete. As a result, even minor perturbations can significantly alter their semantics~\cite{4_he2018adversarial}. This discreteness poses challenges for adversarial training, which typically enhances robustness by perturbing input data, as commonly done in computer vision~\cite{goodfellow2014explaining}. To address this limitation, He et al.~\cite{4_he2018adversarial} proposed Adversarial Personalized Ranking (APR), \added{which introduces perturbations at the parameter level to simulate the effects of malicious attacks.} The objective function is formulated as:
\begin{equation}
    \begin{aligned}
        \Theta^* = \arg \min_{\Theta} \max_{\Delta, \left \Vert \Delta \right \Vert \leq \epsilon} \mathcal{L}(\mathcal{D}|\Theta) + \lambda \mathcal{L}(\mathcal{D}|\Theta + \Delta),
    \end{aligned}
\end{equation}
where $\mathcal{L}$ represents the personalized ranking loss function, $\Theta$ denotes the model parameters, $\Delta$ represents perturbations on model parameters, and $\epsilon \geq 0$ controls the \added{perturbation magnitude}.

Subsequent research~\cite{24_chen2019adversarial, 66_tran2019adversarial, 32_li2020adversarial, 40_ye2023towards} has further developed adversarial training in recommender systems, particularly focusing on parameter perturbations. ATF~\cite{24_chen2019adversarial} integrates the Pairwise Interaction Tensor Factorization~\cite{rendle2010pairwise} into APR, enhancing context-aware recommendations~\cite{baltrunas2011matrix}. Similarly, APC~\cite{66_tran2019adversarial} and SACRA~\cite{32_li2020adversarial} build upon APR in their models to improve robustness. \added{RocSE~\cite{40_ye2023towards} further considers the structure of the interaction graph, while SharpCF~\cite{chen2023adversarial} reduces computational complexity by transforming adversarial training into Sharpness-Aware Training.}

\added{Notably, VAT~\cite{zhang2024improving} and PamaCF~\cite{zhangunderstanding} introduce differentiated perturbations to enhance adversarial training. VAT explores user vulnerability and employs vulnerability-aware adversarial training to simultaneously mitigate performance degradation and strengthen defense capabilities. PamaCF provides a theoretical analysis explaining how adversarial training improves both robustness and overall performance in recommender systems. Leveraging these insights, PamaCF further optimizes adversarial collaborative filtering performance.}

Taking a different approach, FNCF~\cite{1_du2018enhancing} investigates the effects of parameter perturbations across various layers of neural network-based recommender systems, shedding light on how deeper layers respond to perturbations. Bridging different techniques, ACAE~\cite{15_zhang2017robust} and FG-ACAE~\cite{6_yuan2019adversarial} combine principles from APR~\cite{4_he2018adversarial} and FNCF~\cite{1_du2018enhancing}, addressing perturbations in both user/item embeddings and deeper model layers.

\subsubsection{Adversarial Perturbation on User Profile}
\label{sec:3-2-2-user}

Adversarial training methods focusing on user profiles take a defensive stance against attacks, aiming to protect recommender systems from various adversarial threats. A straightforward approach involves injecting specific users during training to enhance the model’s robustness against potential malicious users. The objective function is formulated as:
\begin{equation}
    \begin{aligned}
        & \Theta^* = \arg \min_{\Theta} \max_{\Delta, \left \Vert \Delta \right \Vert  \leq \epsilon} \mathcal{L}(\mathcal{D}|\Theta) + \lambda \mathcal{L}(\mathcal{D}+\Delta|\Theta), \\
        & \text{where} ~~\mathcal{D}+\Delta \coloneqq (\mathcal{U} \cup \mathcal{U}_{\mathrm{inj}}, \mathcal{I}, \mathcal{R} \cup \mathcal{R}_{\mathrm{inj}}).
    \end{aligned}
\end{equation}
Here, $\mathcal{U}_{\mathrm{inj}}$ denotes the set of injected users, while $\mathcal{R}_{\mathrm{inj}}$ is the interactions associated with them.

Although injecting adversarial users can strengthen the model’s robustness, it does not necessarily mitigate the impact of pre-existing malicious users within the training data. This raises a critical question: How can we counteract the effects of already poisoned training data?
To address this, \citet{17_wu2021fight} proposed Adversarial Poisoning Training (APT), which follows a ``fight fire with fire'' strategy. Instead of simply removing malicious users, APT introduces empirical risk minimization (ERM) users, whose profiles are designed to counterbalance the effects of adversarial users. The corresponding objective function is:
\begin{equation}
    \begin{aligned}
        \Theta^* = \arg \min_{\Theta} \min_{\mathcal{D}_{\mathrm{ERM}},|\mathcal{D}_{\mathrm{ERM}}|=n} \mathcal{L}(\mathcal{D} \cup \mathcal{D}_{\mathrm{ERM}} | \Theta),
    \end{aligned}
\end{equation}
where $\mathcal{D}_{\mathrm{ERM}}$ represents the set of user profiles that minimize empirical risk, and $n$ specifies the number of ERM users introduced. By adding ERM users to the training dataset, APT improves the recommender system’s robustness, effectively mitigating the effects of malicious users.

\subsubsection{Adversarial Perturbation on Item Profile}
\label{sec:adv_item}
Content-based recommendation leverages rich side information to improve recommendation accuracy. However, this approach is vulnerable to external threats, where malicious entities may manipulate item side information to unfairly promote certain products or distort recommendations. To counteract such adversarial actions, researchers have introduced adversarial perturbations in handling item side information, either in the original data domain~\cite{3_anelli2021study} or in the latent feature space~\cite{2_tang2019adversarial}. The objective function is:
\begin{equation}
    \begin{aligned}
        & \Theta^* = \arg \min_{\Theta} \max_{\Delta, ||\Delta|| \leq \epsilon} \mathcal{L}(\mathcal{D}|\Theta) + \lambda \mathcal{L}(\mathcal{D}+\Delta|\Theta), \\
        & \text{where} ~~\mathcal{D}+\Delta \coloneqq (\mathcal{U}, \mathcal{I}+\Delta, \mathcal{R}),
    \end{aligned}
\end{equation}
where $\mathcal{I}+\Delta$ represents perturbations applied either in the original feature space or the latent space of item side information. The choice between these approaches depends on the primary objective—whether to enhance robust feature extraction\footnote{Feature Extraction: the process of obtaining representations of item side information~\cite{3_anelli2021study}.} or to strengthen recommender systems against adversarial vulnerabilities.

\subsubsection{Adversarial Perturbation on Interaction}
\label{adv-on-int}
\added{In recommender systems, adversarial training on interaction graphs has gained increasing attention.} These methods introduce adversarial perturbations within interaction data to enhance the system’s robustness against potential threats targeting user-item interactions. Recent studies, such as Conv-GCF~\cite{7_yuan2020exploring} and AdvGraph~\cite{42_chen2022adversarial}, embed adversarial perturbations directly into the interaction matrix. This approach can be formulated as the following optimization problem:
\begin{equation}
    \begin{aligned}
        & \Theta^* = \arg \min_{\Theta} \max_{\Delta, ||\Delta|| \leq \epsilon} \mathcal{L}(\mathcal{D}|\Theta) + \lambda \mathcal{L}(\mathcal{D}+\Delta|\Theta), \\
        & \text{where} ~~\mathcal{D}+\Delta \coloneqq (\mathcal{U}, \mathcal{I}, \mathcal{R}+\Delta),
    \end{aligned}
\end{equation}
where $\mathcal{R}+\Delta$ represents the perturbed interactions.

Beyond static interaction data, AdvTrain~\cite{30_yue2022defending} explores adversarial training in sequential recommendation settings. Their approach perturbs interaction sequences, which is particularly relevant for sequential recommendation systems, where user temporal behavior significantly influences recommendation outcomes.

\subsubsection{Discussion of Adversarial Training}
\added{Adversarial training has introduced various strategies to enhance the robustness of recommender systems.} Among these, \textit{Adversarial Perturbation on Parameters} has gained significant attention as a prevalent approach. By introducing perturbations at the parameter level, this method equips models to counteract disruptive manipulations from malicious users~\cite{4_he2018adversarial, 24_chen2019adversarial}. 
\textit{Adversarial Perturbation on User Profile} adopts a defensive strategy similar to that of attackers. By injecting users into the system, it strengthens the model’s resistance against adversarial manipulation~\cite{17_wu2021fight}. In contrast, \textit{Adversarial Perturbation on Item Profile} aligns more closely with content-based recommendation frameworks, focusing on perturbations related to item-specific attributes~\cite{3_anelli2021study, 2_tang2019adversarial}. Finally, \textit{Adversarial Perturbation on Interaction} introduces discrete perturbations in user-item interactions to enhance model resilience against interaction-level attacks~\cite{30_yue2022defending, 42_chen2022adversarial}. However, given the limited adversarial capabilities in practice, interaction-level attacks are relatively rare in real-world recommendation scenarios.

Future research could explore synergies between different adversarial perturbation strategies, such as those targeting user profiles, item profiles, and system parameters, to develop comprehensive defense mechanisms against diverse threats. Additionally, designing novel adversarial training methodologies that adapt to the dynamic nature of modern recommender systems presents an exciting avenue for further investigation.

%% file: Tex_Graph/2-Adversarial.tex
\begin{tikzpicture}[node distance=2cm, auto]
\footnotesize

    \draw[->] (0.82, -7) -- (15.5, -7);
    \foreach \x/\year in {2.5/2018, 4.5/2019, 6.8/2020, 8.8/2021, 10.2/2022, 12.2/2023, 13.75/2024}
    \draw[dashed, lightgray] (\x, -7) -- (\x, 1.7) node[below=8.7cm, black] {\year};
    
    \node[text width=1.5cm, align=center] (a1) {Adversarial Training \\ on Parameters};
    \node[text width=1.5cm, align=center, below=1cm of a1] (a2) {Adversarial Training \\ on User};
    \node[text width=1.5cm, align=center, below=0.3cm of a2] (a3) {Adversarial Training \\ on Item};
    \node[text width=1.5cm, align=center, below=0.7cm of a3] (a4) {Adversarial Training \\ on Interaction};

    \node (4) [right=2cm of a1, yshift=0.8cm]{APR\cite{4_he2018adversarial}};
    \node (1) [below=1.2cm of 4]{FNCF\cite{1_du2018enhancing}};
    \node (24) [right=1cm of 4, align=center, yshift=0.8cm]{ATF\cite{24_chen2019adversarial} \\ APC\cite{66_tran2019adversarial}};
    \node (5) [right=0.8cm of 1, align=center]{ACAE\cite{5_yuan2019adversarial} \\ FG-ACAE\cite{6_yuan2019adversarial}};
    \node (32) [right=0.8cm of 24, yshift=-1.6cm]{SACRA\cite{32_li2020adversarial}};
    \node (40) [right=3.5cm of 32]{RocSE\cite{40_ye2023towards}};
    \node (new1) [right=3.5cm of 32, yshift=1.6cm]{\added{SharpCF\cite{chen2023adversarial}}};
    \node (new2) [right=5.1cm of 32, yshift=0.8cm, align=center]{\added{VAT\cite{zhang2024improving}} \\ \added{PamaCF\cite{zhangunderstanding}}};
    \node (new2_description) [below=0.5cm of new2, align=center, font=\scriptsize, text=gray] {\added{Theoretical Proof}};

    \node (ept1) [right=1cm of a1] {};
    \draw[line width=0.25mm] (a1) to (ept1);
    \node (ept2) [right=12cm of 4] {};
    \draw[line width=0.25mm] (4) to (new2);
    \draw[line width=0.25mm] (new2) to (ept2);

    \node (17) [right=8cm of a2, yshift=0.5cm]{APT\cite{17_wu2021fight}};
    
    \node (ept3) [right=10cm of a2] {};
    \draw[line width=0.25mm] (a2) to (ept3);

    \node (2) [right=4.3cm of a3, yshift=0.5cm]{AMR\cite{2_tang2019adversarial}};
    \node (3) [right=2.2cm of 2, yshift=-1cm]{VAR\cite{3_anelli2021study}};

    \node (ept4) [right=10cm of a3] {};
    \draw[line width=0.25mm] (a3) to (ept4);

    \node (7) [right=6cm of a4, align=center, yshift=0.6cm]{Conv-GCF\cite{7_yuan2020exploring}};
    \node (42) [right=1.5cm of 7]{AdvGraph\cite{42_chen2022adversarial}};
    \node (30) [below=1.2cm of 42]{AdvTrain\cite{30_yue2022defending}};

    \node (ept5) [right=10cm of a4] {};
    \draw[line width=0.25mm] (a4) to (ept5);

    \draw[->, >=latex, line width=0.25mm] ($(a1)!.95!(ept1)$) .. controls +(up:0.6cm) and +(left:1.0cm).. node[pos=0.6, above=0.2cm, font=\scriptsize, text=gray] {on Embedding} (4);
    \draw[dashed, ->, >=latex, line width=0.25mm] ($(4)!.1!(ept2)$) .. controls +(up:0.6cm) and +(left:1.0cm).. (24);
    \draw[dashed, ->, >=latex, line width=0.25mm] ($(4)!.25!(ept2)$) .. controls +(down:0.6cm) and +(left:1.0cm).. (32);
    \draw[dashed, ->, >=latex, line width=0.25mm] ($(4)!.65!(new2)$) .. controls +(down:0.6cm) and +(left:1.0cm).. (40);
    \draw[dashed, ->, >=latex, line width=0.25mm] ($(4)!.7!(new2)$) .. controls +(up:0.6cm) and +(left:1.0cm).. node[pos=0.6, above=0.2cm, font=\scriptsize, text=gray] {\added{Accelerations}} (new1);
    \draw[dashed, ->, >=latex, line width=0.25mm] ($(4)!.06!(ept2)$) .. controls +(down:1.6cm) and +(left:1.0cm).. (5);
    \draw[->, >=latex, line width=0.25mm] ($(a1)!.95!(ept1)$) .. controls +(down:0.6cm) and +(left:1.0cm).. node[pos=0.6, below=0.2cm, font=\scriptsize, text=gray] {on Deep Layer} (1);
    \draw[dashed, ->, >=latex, line width=0.25mm] (1) to (5);
    \draw[dashed, ->, >=latex, line width=0.25mm, color=gray] (new2_description) to (new2);

    % \draw[->, >=latex, line width=0.25mm] (a1) to node[pos=0.6, above, font=\scriptsize, text=gray] {on Embedding} (4);
    % \draw[->, >=latex, line width=0.25mm] ($(a1)!.35!(4)$) .. controls +(down:0.9cm) .. node[pos=0.8, above, font=\scriptsize, text=gray] {on Deep Layer} (1);
    % \draw[->, >=latex, line width=0.25mm] (4) to (24);
    % \draw[->, >=latex, line width=0.25mm] (24) to (32);
    % \draw[->, >=latex, line width=0.25mm] (32) to (40);
    % \draw[->, >=latex, line width=0.25mm] ($(4)!.35!(24)$) .. controls +(down:0.9cm) .. (5);
    % \draw[->, >=latex, line width=0.25mm] (1) to (5);

    \draw[dashed, ->, >=latex, line width=0.25mm] ($(a2)!.5!(ept3)$) .. controls +(up:0.6cm) and +(left:1.0cm).. (17);
    % \draw[->, >=latex, line width=0.25mm] (a2) to (17);

    \draw[dashed, ->, >=latex, line width=0.25mm] ($(a3)!.3!(ept4)$) .. controls +(up:0.6cm) and +(left:1.0cm).. node[pos=0.6, above=0.1cm, font=\scriptsize, text=gray] {\added{Feature-level}} (2);
    \draw[dashed, ->, >=latex, line width=0.25mm] ($(a3)!.6!(ept4)$) .. controls +(down:0.6cm) and +(left:1.0cm).. node[pos=0.6, above=0.0cm, font=\scriptsize, text=gray] {\added{Sample-level}} (3);
    % \draw[->, >=latex, line width=0.25mm] (a3) to (2);
    % \draw[->, >=latex, line width=0.25mm] (2) to (3);

    \draw[dashed, ->, >=latex, line width=0.25mm] ($(a4)!.5!(ept5)$) .. controls +(up:0.6cm) and +(left:1.3cm).. (7);
    \draw[dashed, ->, >=latex, line width=0.25mm] ($(a4)!.85!(ept5)$) .. controls +(up:0.6cm) and +(left:1.0cm).. (42);
    \draw[dashed, ->, >=latex, line width=0.25mm] ($(a4)!.8!(ept5)$) .. controls +(down:1cm) and +(left:1.0cm).. node[pos=0.9, above=0.2cm, font=\scriptsize, text=gray] {\added{Sequential Interaction}} (30);
    % \draw[->, >=latex, line width=0.25mm] (a4) to (7);
    % \draw[->, >=latex, line width=0.25mm] (7) to (42);
    % \draw[->, >=latex, line width=0.25mm] ($(7)!.35!(42)$) .. controls +(down:0.9cm) .. node[pos=0.8, above, font=\scriptsize, text=gray] {Sequential} (30);

\end{tikzpicture}

%% file: Section/3-subsection/3-Certifiable.tex
Certifiable robust training plays a crucial \added{role in ensuring the integrity and reliability of recommender systems against malicious attacks}. This approach focuses on designing algorithms that provide formal guarantees of robustness against adversarial perturbations, ensuring that recommendations remain accurate even in the presence of malicious users.

\added{Liu et al.~\cite{9_liu2020certifiable} exemplify this approach by defining the robust boundary of the Factorization Machine (FM) model}. Given a trained FM, denoted as $f$, and an input instance $\bm{x}$ with $d$-dimensional features in the binary space $\{0,1\}^{1 \times d}$, the FM-based recommendation task with second-order weights is expressed as:
\begin{equation}
    \begin{aligned}
        f(\bm{x}) = w_0 + \sum_{i=1}^dw_ix_i + \sum_{i=1}^d \sum_{j=i}^d<v_i,v_j>x_ix_j,
    \end{aligned}
    \label{eq:fm}
\end{equation}
where $w_0$ is the global bias, $w_i$ represents the weight of the $i$-th feature, $v_i \in \mathbb{R}^{1 \times k}$ is the embedding of the $i$-th feature, and the inner product $<v_i,v_j>$ captures interactions between features $i$ and $j$. The variable $x_i$ corresponds to the $i$-th dimension of instance $\bm{x}$.

With a perturbation budget $p$, which limits the maximum number of feature flips in $\bm{x}$, Liu et al.~\cite{9_liu2020certifiable} estimate the upper bound of prediction alteration, denoted as $b(x)$. If $f(\bm{x}) \leq 0$, then $b(x) = \max_{\bm{x}'} f(\bm{x}') - f(\bm{x})$; otherwise, $b(x) = \min_{\bm{x}'} f(\bm{x}') - f(\bm{x})$. Here, $\bm{x}'$ represents a perturbed version of $\bm{x}$, constrained by the condition $|\bm{x} \oplus \bm{x}'| \leq p$, where $\oplus$ denotes the XOR operation. This measures the model’s output shift under the maximum allowable perturbation. Evaluating whether the prediction changes significantly under this bound provides insight into the model’s robustness, as further discussed in Section~\ref{5-2-mertics}. Building on this foundation, \citet{9_liu2020certifiable} propose a robust training algorithm formulated as:
\begin{equation}
    \begin{aligned}
        \Theta^* = \arg \min_{\Theta} \sum_{\mathcal{D}} \log \left[ 1 + \mathrm{exp} 
        \left( \left(-y \right) \left(f \left( \bm{x} \right) +b \left( \bm{x} \right) \right) \right) 
        \right],
    \end{aligned}
\end{equation}
where $f(\bm{x}) + b(\bm{x})$ represents the prediction bound under the maximum perturbation shift.

Certifiable robust training provides a valuable framework for assessing and enhancing the robustness of recommender systems, allowing developers to better safeguard their models against adversarial attacks while maintaining accurate and reliable recommendations.

%% file: Section/4-against_Noise.tex
Recommender systems also encounter challenges from natural noise, which arises due to inconsistencies, inaccuracies, or missing information in input data. Such noise can stem from various sources, including human error, uncertainty, and vagueness~\citep{20_yera2016fuzzy}. Ensuring robustness against natural noise is essential for maintaining accurate and reliable recommendations. To mitigate this issue, researchers have explored various techniques, including regularization, purification, and self-supervised learning. This section provides a detailed discussion of these methods, highlighting their contributions to enhancing the robustness of recommender systems against natural noise.

\subsection{Regularization}
\label{sec:reg}
\input{Section/4-subsection/1-Regularization}

\subsection{Purification}
\label{sec:pur}
\input{Section/4-subsection/2-Purification}

\subsection{Self-supervised Learning}
\label{sec:ssl}
\input{Section/4-subsection/3-Self-supervised}

%% file: Section/4-subsection/1-Regularization.tex
Regularization is a widely used technique in machine learning to prevent overfitting by constraining model complexity. By limiting the model’s capacity to learn intricate and fluctuating noise patterns, regularization enhances generalization and improves robustness against natural noise in input data. Mathematically, the regularization objective is expressed as:
\begin{equation}
    \begin{aligned}
        \Theta^* = \arg \max_\Theta \mathcal{L}(\mathcal{D}|\Theta) + \lambda \left \Vert\Theta \right \Vert_k,
    \end{aligned}
\end{equation}
where $\mathcal{L}(\mathcal{D}|\Theta)$ denotes the model's loss function based on dataset $\mathcal{D}$ and parameters $\Theta$. The term $\left \Vert\Theta \right \Vert_k$ represents the regularization penalty, determined by the $L_k$ norm of the model parameters, and $\lambda$ is a hyperparameter that balances model fit and complexity.

\added{Beyond simple parameter constraints}, Zhang et al.~\cite{15_zhang2017robust} introduced $R_1$-norm\footnote{The $R_1$-norm of a matrix $X \in \mathbb{R}^{d \times l}$ is defined as $\left \Vert X \right \Vert_{R_1} = \sum_{j=0}^l \sqrt{\sum_{i=0}^d x_{i,j}^2}$} regularization on predictions to reduce sensitivity to noise. Their loss function is formulated as:
\begin{equation}
    \begin{aligned}
        \Theta^* = \arg \max_\Theta \sum_{i\in \mathcal{I}} \sqrt{\sum_{u \in \mathcal{U}}(r_{u,i}-\hat{r}_{u,i})^2} + \left \Vert \Theta \right \Vert_F^2,
    \end{aligned}
\end{equation}
where $r_{u,i}$ is the ground truth rating, and $\hat{r}_{u,i}$ is the predicted rating computed by $\Theta$. The $R_1$-norm provides greater robustness against outliers compared to the Euclidean distance (i.e., $L_2$-norm). By integrating $R_1$-norm regularization, the model prioritizes essential features over noise, leading to more refined recommendations. \added{Recent research continues to advance regularization strategies for enhancing robustness in noisy recommender systems}. For instance, Chen et al.~\cite{39_chen2022denoising} combine Jacobian regularization~\cite{kim2021lipschitz} with transformer blocks in sequential recommendation models. This approach significantly reduces the model’s susceptibility to noisy sequences, resulting in more stable and reliable recommendations.

While regularization serves as a versatile method to improve noise robustness across different recommender systems, it may lack efficacy against specific noise types. Although effective in preventing overfitting and improving generalization, regularization does not directly address the root cause of noise. Therefore, combining regularization with other noise-mitigation strategies is often necessary for optimal performance.

%% file: Section/4-subsection/2-Purification.tex
\begin{figure}[t]
    \centering
    \resizebox{0.95\textwidth}{!}{\input{Tex_Graph/3-Purification}}
    \caption{\added{Main development trajectory of purification methods.}}
    \label{fig:pur_his}
\end{figure}
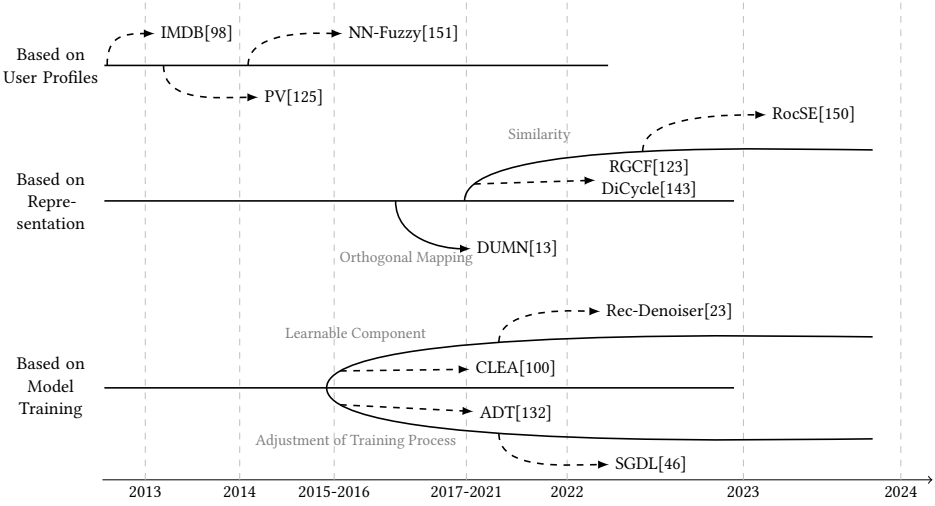

Purification is an effective technique targeted for identifying and rectifying noise in user-item interactions, thereby enhancing the performance and robustness of recommender systems. The focus of this approach is to detect and eliminate noise from the input data during the training process to account for the presence of noise. The primary objective of purification can be mathematically formulated as follows:
\begin{equation}
    \begin{aligned}
        \Theta^*= \arg \min_{\Theta} \mathcal{L}(\mathcal{D}|\Theta, W),
    \end{aligned}
\end{equation}
where $W$ is the weight matrix for all interactions. More specifically:
\begin{equation}
    \begin{aligned}
        \mathcal{L}(\mathcal{D}|\Theta, W) = \mathbb{E}_{(u,i,r)}\left [w_{u,i}\psi \left (f \left (u,i \right ), r_{u,i} \right ) \right ],
    \end{aligned}
\end{equation}
where $w_{u,i}$ is the weight of interaction $r_{u,i}$ in $W$.
To provide a comprehensive understanding of the development of the methods based on purification, we illustrate their trajectories in Figure~\ref{fig:pur_his}.

\textit{Based on User Profiles.} Historically, various works have proposed strategies to identify noisy interactions based solely on user profiles. 
IMDB~\cite{71_pham2013preference} integrates interactions and item-related information to formulate a user preference model, marking deviations from anticipated user preferences as anomalies. Moving forward, PV~\cite{41_toledo2015correcting} proposes an innovative preference model that obviates the need for supplementary information. In a different approach, NN-Fuzzy~\cite{20_yera2016fuzzy} \added{introduces a fuzzy inference system} to detect noisy ratings, leveraging uncertainty modeling to refine preference predictions.

\textit{Based on Representations.} More recently, methods have been proposed that leverage learned representations to re-weight noisy interactions.
DUMN~\cite{44_bian2021denoising} leverages the representations of explicit feedback (devoid of noise) to refine the representations of implicit feedback (potentially tainted with noise) through orthogonal mapping. Both RGCF~\cite{27_tian2022learning} and RocSE~\cite{40_ye2023towards} measure the congruence between user-item pairs to identify noise, considering interactions with minimal similarity as potential noise. DiCycle~\cite{70_xie2022denoising} further considers the dynamics of interaction. By employing a continuous translation-invariant kernel~\cite{xu2019self}, it translates timestamps with item representation into embeddings. The resulting inner product of the embeddings of timestamps $t_i$ and $t_j$ reflects the consistency of user interactions across these timestamps, positing that inconsistent interactions could be tinged with noise. \added{Additionally, NeuFilter~\cite{xia2024neural} introduces Kalman Gain to estimate user representations, thereby mitigating the unreliability caused by noisy interactions. These adaptive estimates enhance robustness against varying noise patterns.}

\textit{Based on Model Training.} Another paradigm in noise purification focuses on model training. 
Several methodologies integrate learnable components into the recommendation model to spotlight noise~\cite{54_qin2021world, 39_chen2022denoising}.
CLEA~\cite{54_qin2021world} \added{and GDMSR~\cite{quan2023robust}} use a multi-layer perceptron (MLP) to detect the noisy items in historical interactions during training. \added{GDMSR~\cite{quan2023robust} further integrates Dunbar’s number theory to adaptively adjust noise levels for each user based on the upper limit of social relationships, thereby incorporating social constraints into the denoising process.} Meanwhile, Rec-Denoiser~\cite{39_chen2022denoising} integrates a differentiable mask in its attention mechanism, detecting potential noisy interactions. \added{Additionally, BOD~\cite{wang2023efficient} directly learns interaction-specific weights to refine the denoising process, offering a more granular noise-adjustment mechanism.} 

Concurrently, other researches~\citep{43_wang2021denoising, 31_gao2022self} detect noise by observing the data pattern in the training phase. Both ADT~\cite{43_wang2021denoising} and SGDL~\cite{31_gao2022self} posit that clean and noisy interactions reveal distinct patterns throughout training. In particular, ADT utilizes a truncated loss to eliminate noisy interactions, while SGDL employs meta-learning to fine-tune the denoising process. 
\added{Moreover, DCF~\cite{he2024double} extends ADT by preserving hard examples with relatively high losses.}
\added{Furthermore, PLD~\cite{zhang2025personalized} innovatively leverages users' personalized loss distributions to adjust the training sampling strategy, enhancing both efficiency and performance in denoising. By tailoring the denoising process to individual users, PLD adapts to diverse user behaviors, reducing over-filtering risks.}

\textit{Conclusion.} The purification technique is a powerful method for enhancing the robustness of recommender systems by identifying and correcting noise in user-item interactions. However, it is a challenging task that requires careful design and selection of appropriate models and strategies, considering the specific characteristics of the recommendation scenarios and noise types.

%% file: Tex_Graph/3-Purification.tex
\begin{tikzpicture}[node distance=2cm, auto]
\footnotesize

    \draw[->] (0.82, -6.5) -- (14.5, -6.5);
    \foreach \x/\year in {1.5/2013, 3/2014, 4.5/2015-2016, 6.6/2017-2021, 8.2/2022, 10.2/2023, 12/2024, 13.5/\added{2025}}
    \draw[dashed, lightgray] (\x, -6.5) -- (\x, 1) node[below=7.5cm, black] {\year};

    \node[text width=1.5cm, align=center] (a1) {Based on User Profiles};
    \node[text width=1.5cm, align=center, below=1.2cm of a1] (a2) {Based on Representation};
    \node[text width=1.5cm, align=center, below=1.8cm of a2] (a3) {Based on Model Training};

    \node (71) [right=0.8cm of a1, yshift=0.5cm]{IMDB\cite{71_pham2013preference}};
    \node (41) [right=0.3cm of 71, yshift=-1cm]{PV\cite{41_toledo2015correcting}};
    \node (20) [right=0.2cm of 41, yshift=1cm]{NN-Fuzzy\cite{20_yera2016fuzzy}};

    \node (ept1) [right=8cm of a1] {};
    \draw[line width=0.25mm] (a1) to (ept1);

    \node (27) [right=7.8cm of a2, yshift=0.35cm, align=center]{RGCF\cite{27_tian2022learning} \\ DiCycle\cite{70_xie2022denoising}};
    \node (40) [right=0.1cm of 27, yshift=1cm]{RocSE\cite{40_ye2023towards}};
    \node (44) [left=0.5cm of 27, yshift=-1.1cm]{DUMN\cite{44_bian2021denoising}};
    \node (new1) [right=1.5cm of 27, yshift=-0.1cm]{\added{NeuFilter\cite{xia2024neural}}};

    \node (ept2) [right=10cm of a2] {};
    \draw[line width=0.25mm] (a2) to (ept2);
    \node (ept3) [right=2cm of ept2, yshift=0.8cm] {};
    \draw[line width=0.25mm] ($(a2)!.6!(ept2)$) .. controls +(up:1.0cm) and +(left:1.2cm).. node[pos=0.3, above=0.2cm, font=\scriptsize, text=gray]{Similarity} (ept3) coordinate[pos=0.1] (EPT3_M1) coordinate[pos=0.5] (EPT3_M2) coordinate[pos=0.65] (EPT3_M3);

    \node (54) [right=5.8cm of a3, yshift=0.3cm]{CLEA\cite{54_qin2021world}};
    \node (39) [right=0.5cm of 54, yshift=0.9cm, align=center]{Rec-\\Denoiser\cite{39_chen2022denoising}};
    \node (43) [below=0.22cm of 54]{ADT\cite{43_wang2021denoising}};
    \node (31) [right=0.6cm of 43, yshift=-0.8cm]{SGDL\cite{31_gao2022self}};
    \node (new2) [right=0.0cm of 39, yshift=-0.8cm, align=center]{\added{GDMSR\cite{quan2023robust}} \\ \added{BOD\cite{wang2023efficient}}};
    \node (new3) [right=4cm of 43]{\added{DCF\cite{he2024double}}};

    \node (ept4) [right=9cm of a3] {};
    \draw[line width=0.25mm] (a3) to (ept4);
    \node (ept5) [right=2cm of ept4, yshift=0.8cm] {};
    \draw[line width=0.25mm] ($(a3)!.4!(ept4)$) .. controls +(up:1cm) and +(left:1.2cm).. node[pos=0.15, above=0.25cm, font=\scriptsize, text=gray]{Learnable Component} (ept5) coordinate[pos=0.1] (EPT5_M1) coordinate[pos=0.4] (EPT5_M2) coordinate[pos=0.65] (EPT5_M3);
    \node (new4) [right=3.3cm of ept4, yshift=-0.8cm] {\added{PLD\cite{zhang2025personalized}}};
    \node (new4_description) [above=0.5cm of new4, align=center, font=\scriptsize, text=gray] {\added{Probabilistic Resampling}};
    \draw[dashed, ->, >=latex, line width=0.25mm] (new4_description) to (new4);
    \node (ept6) [right=0.5cm of new4] {};
    \draw[line width=0.25mm] ($(a3)!.4!(ept4)$) .. controls +(down:1cm) and +(left:1.2cm).. node[pos=0.15, below=0.25cm, font=\scriptsize, text=gray]{Adjustment of Training Process} (new4) coordinate[pos=0.1] (EPT6_M1) coordinate[pos=0.4] (EPT6_M2);
    \draw[line width=0.25mm] (new4) to (ept6);

    \draw[dashed, ->, >=latex, line width=0.25mm] ($(a1)!.1!(ept1)$) .. controls +(up:0.45cm) and +(left:1.2cm).. (71);
    \draw[dashed, ->, >=latex, line width=0.25mm] ($(a1)!.2!(ept1)$) .. controls +(down:0.6cm) and +(left:1.2cm).. (41);
    \draw[dashed, ->, >=latex, line width=0.25mm] ($(a1)!.35!(ept1)$) .. controls +(up:0.6cm) and +(left:1.2cm).. (20);
    % \draw[->, >=latex, line width=0.25mm] (a1) to (71);
    % \draw[->, >=latex, line width=0.25mm] (71) to (41);
    % \draw[->, >=latex, line width=0.25mm] (41) to (20);

    \draw[dashed, ->, >=latex, line width=0.25mm] (EPT3_M1) to (27);
    \draw[dashed, ->, >=latex, line width=0.25mm] (EPT3_M2) .. controls +(up:0.6cm) and +(left:1.2cm).. (40);
    \draw[dashed, ->, >=latex, line width=0.25mm] (EPT3_M3) .. controls +(down:0.6cm) and +(left:1.2cm).. (new1);
    \draw[->, >=latex, line width=0.25mm] ($(a2)!.5!(ept2)$) .. controls +(down:0.6cm) and +(left:1.2cm).. node[pos=0.3, below=0.25cm, font=\scriptsize, text=gray] {Orthogonal Mapping} (44);
    % \draw[->, >=latex, line width=0.25mm] (a2) to node[pos=0.62, above, font=\scriptsize, text=gray]{Similarity} (27);
    % \draw[->, >=latex, line width=0.25mm] ($(a2)!.5!(27)$) .. controls +(down:0.6cm) .. node[pos=0.8, above, font=\scriptsize, text=gray] {Mapping} (44);
    % \draw[->, >=latex, line width=0.25mm] (27) to (40);

    \draw[dashed, ->, >=latex, line width=0.25mm] (EPT5_M1) to (54);
    \draw[dashed, ->, >=latex, line width=0.25mm] (EPT6_M1) to node[pos=0.5, above=-0.05cm, font=\scriptsize, text=gray]{\added{Loss-guided Reweighting}} (43);
    \draw[dashed, ->, >=latex, line width=0.25mm] (43) to (new3);
    \draw[dashed, ->, >=latex, line width=0.25mm] (EPT5_M2) .. controls +(up:0.6cm) and +(left:1.3cm).. (39);
    \draw[dashed, ->, >=latex, line width=0.25mm] (EPT5_M3) .. controls +(down:0.3cm) and +(left:1.3cm).. (new2);
    \draw[dashed, ->, >=latex, line width=0.25mm] (EPT6_M2) .. controls +(down:0.6cm) and +(left:1.3cm).. (31);
    
    % \draw[->, >=latex, line width=0.25mm] (a3) to node[pos=0.75, above, font=\scriptsize, text=gray]{on Model Component} (54);
    % \draw[->, >=latex, line width=0.25mm] ($(a3)!.5!(54)$) .. controls +(down:0.8cm) .. node[pos=0.8, above, font=\scriptsize, text=gray] {on Training Stage} (43);
    % \draw[->, >=latex, line width=0.25mm] (54) to (39);
    % \draw[->, >=latex, line width=0.25mm] (43) to (31);

\end{tikzpicture}

%% file: Section/4-subsection/3-Self-supervised.tex
Self-supervised learning has \added{emerged as a powerful machine learning paradigm}, extensively applied in scenarios where explicit labels are scarce or unavailable~\cite{liu2021self, schiappa2023self}. This approach exploits the inherent structure of the data~\cite{liu2021self} or generates data variants~\cite{jaiswal2020survey} to enhance generalization capabilities. \added{In recommender systems, self-supervised strategies} can be broadly classified into two categories: generative-based~\cite{8_li2015deep, 21_wu2016collaborative, 38_zheng2021multi} and contrastive-based self-supervised learning~\cite{67_zhou2020s3, 36_wu2021self, 68_xie2022contrastive}. \added{Generative-based methods corrupt input data and use the restoration process as a learning signal, whereas contrastive-based methods generate multiple data views through augmentation}. The model is trained to extract essential features across these views while minimizing noise by maximizing mutual information between different views. These methods play a crucial role in mitigating the impact of noise in user-item interactions and auxiliary information~\cite{yu2023self, hendrycks2019using}.

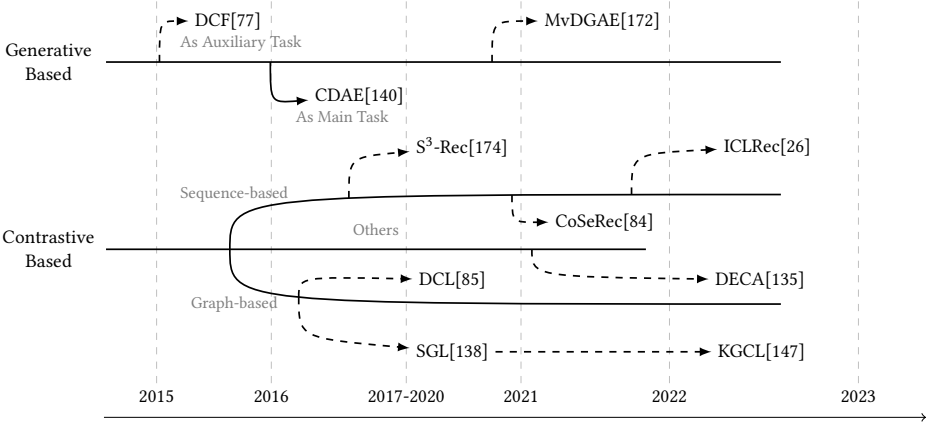
\begin{figure}[t]
    \centering
    \resizebox{0.95\textwidth}{!}{\input{Tex_Graph/4-Self-supervised}}
    \caption{\added{Main development trajectory of self-supervised methods.}}
    \label{fig:sp_his}
\end{figure}

To provide a comprehensive overview of the development of self-supervised learning techniques in robust recommender systems, we illustrate their progression in Figure~\ref{fig:sp_his}. \added{Given the extensive research on} self-supervised learning in recommendation, we focus on methods explicitly designed for denoising. For a broader view of self-supervised methods, we refer readers to the survey~\cite{yu2023self}.

\subsubsection{Generative-based Self-supervised Learning}
\label{sec:4-3-1-gen}
Denoising auto-encoders (DAEs)~\cite{vincent2008extracting} serve as a cornerstone of generative-based self-supervised learning in recommender systems. These methods operate on \textbf{masked} user or item side information~\citep{8_li2015deep}, user-item interaction matrices~\citep{21_wu2016collaborative}, or other attributes~\cite{38_zheng2021multi}, aiming to reconstruct the original information from corrupted inputs.

Consider a matrix $X \in \{0, 1\}^{m \times n}$, \added{which could represent attributes~\cite{8_li2015deep}, interactions~\cite{21_wu2016collaborative}, or other features~\cite{38_zheng2021multi}}. By applying a mask $M \in \{0, 1\}^{m \times n}$, the corrupted matrix is generated as $\tilde{X} = X \odot M$, where $\odot$ denotes element-wise multiplication. Certain techniques leverage the restoration task as a supplementary objective. For instance, in DCF~\cite{8_li2015deep}, a DAE, symbolized as $g$, is integrated into recommender systems to revive the original attributes matrix $X$ from its corrupted version, $\tilde{X}$. Expanding upon this, MvDGAE~\cite{38_zheng2021multi} delves into both user and item interconnections. Specifically, MvDGAE is dedicated to reconstructing both the intricate relationships among users and the coexistence patterns between items. These methods can be mathematically articulated as:
\begin{equation}
    \begin{aligned}
        \Theta^*, g^* = \arg \min_{\Theta, g} \mathcal{L}(\mathcal{D}|\Theta) + \underbrace{\lambda \left \Vert g(\tilde{X}) - X \right \Vert^2_F}_{\mathrm{Reconstruction}},
    \end{aligned}
\end{equation}
where $\Theta$ is composed of the latent features in $g$.

\added{Additionally, with the rapid advancement of diffusion models, recent studies such as RecDiff~\cite{li2024recdiff} and DiffKG~\cite{jiang2024diffkg} have begun leveraging diffusion-based denoising capabilities to enhance recommender system robustness. These methods iteratively refine representations through a progressive noise removal process, leading to improved denoising capability.}

Conversely, an alternative avenue focuses on deploying DAEs to generate recommendations, aligning with the principles of auto-encoder-based recommender systems~\cite{sedhain2015autorec}. In CDAE~\cite{21_wu2016collaborative}, the term $\mathcal{L}(\mathcal{D}|\Theta)$ can be defined as:
\begin{equation}
    \begin{aligned}
        \mathcal{L}(\mathcal{D}|\Theta) = \frac{1}{|\mathcal{U}|} \sum_{u \in \mathcal{U}} \underbrace{\left \Vert g_{\Theta}(\tilde{X}_u)-X_u\right \Vert^2_F}_{\mathrm{Reconstruction}},
    \end{aligned}
\end{equation}
where $X_u$ denotes the interactions of user $u$, and $g_{\Theta}$ denotes the DAE-based recommendation model.

\subsubsection{Contrastive-based Self-supervised Learning}
\label{sec:4-3-2-con}
Contrastive learning maximizes mutual information between positive pairs---different augmented views of the same sample. Applied to recommender systems, this approach is formalized as:
\begin{equation}
    \begin{aligned}
        \Theta^* = \arg \min_{\Theta} \mathcal{L}(\mathcal{D}|\Theta) - \underbrace{\lambda \mathbb{E}_{u \in \mathcal{U}}\left[I(v_{u}, v_{u}')|\Theta\right]
        ~\mathrm{or}~~\lambda \mathbb{E}_{i \in \mathcal{I}}\left[I(v_{i}, v_{i}')|\Theta\right]
        }_{\mathrm{Contrastive}},
    \end{aligned}
\label{eq:contrast}
\end{equation}
where $I(\cdot, \cdot)$ denotes mutual information. \added{Common loss functions include InfoNCE~\citep{oord2018representation} and Cross-Entropy. Positive pairs $(v, v')$ originate from different augmentation strategies, primarily categorized as sequence-based or graph-based.}

\textit{Sequence-based.} In the domain of sequence-based contrastive learning, pioneering methods like S$^3$-Rec~\cite{67_zhou2020s3} emphasize aligning items with their intrinsic attributes. This is achieved by masking either attributes, items, or specific segments of sequences, thus engendering diverse views tailored for contrastive learning. Furthering this approach, CoSeRec~\cite{23_liu2021contrastive} innovatively substitutes and inserts items into historical item sequences based on item co-appearances in interaction sequences. Additionally, ICLRec~\cite{37_chen2022intent} delves into the relationship between historical interactions and a user's shopping intent, guaranteeing a congruous connection between sequence views and their intent.

\textit{Graph-based.} Contrastive learning techniques based on interaction graphs, such as SGL~\cite{36_wu2021self}, employ diverse graph augmentations, including node dropping, edge masking, and sampling distinct subgraphs via random walks. Similarly, DCL~\cite{68_liu2021contrastive} utilizes stochastic edge masking to perturb specific network structures, leading to dual augmented neighborhood subgraphs. However, random augmentations in SGL may compromise the preservation of useful interactions for contrastive learning, as pointed out by KGCL~\cite{34_yang2022knowledge}. To address this, KGCL integrates supplementary knowledge graph to improve the masking approach. \added{Furthermore, KACL~\cite{wang2023knowledge} refines this approach by introducing an adaptive view generator that dynamically selects important edges, effectively filtering out irrelevant information and mitigating knowledge overload in recommender systems.}

\added{Beyond knowledge-based enhancements, various methods leverage different assumptions to construct more robust augmented views. For instance, AdaGCL~\cite{jiang2023adaptive} incorporates a graph variational autoencoder along with a constrained denoising matrix to generate diverse views, thereby improving robustness. DSL~\cite{wang2023denoised} constructs denoised views by aligning user interaction similarity with social relationship similarity, reducing noise from both perspectives. Additionally, SSD-ICGA~\cite{sun2024self} applies the Independent Cascade Graph model to simulate influence diffusion, generating enhanced graph views by filtering out inactive edges, which improves noise reduction.}

Additionally, the seminal work by Wang et al.~\cite{35_wang2022learning} highlights the consistency in clean sample predictions across models, contrasting it with the variability observed in noisy samples. Building on this observation, DeCA~\cite{35_wang2022learning} employs multiple models as a means of data augmentation, resulting in enhanced contrastive learning effectiveness.

In conclusion, the incorporation of self-supervised learning into recommender systems stands as a potent remedy for inaccuracies induced by noise. Apart from the aforementioned approaches, several self-supervised strategies specifically designed for graphs, be they static~\cite{zhu2020deep, mo2022simple} or dynamic~\cite{rossi2020temporal, zhang2023dyted}, can be aptly adapted to recommendation scenarios. These techniques bolster the model's ability to extract resilient and flexible representations, which are paramount for delivering superior recommendations in real-world settings.

%% file: Tex_Graph/4-Self-supervised.tex
\begin{tikzpicture}[node distance=2cm, auto]
\footnotesize

    \draw[->] (0.82, -4.7) -- (14, -4.7);
    \foreach \x/\year in {1.6/2015, 3.3/2016, 5.3/2017-2020, 7/2021, 9.2/2022, 11.2/2023, 13/\added{2024}}
    \draw[dashed, lightgray] (\x, -4.7) -- (\x, 1) node[below=5.7cm, black] {\year};
    
    \node[text width=1.5cm, align=center] (a1) {Generative Based};
    \node[text width=1.5cm, align=center, below=2cm of a1] (a2) {Contrastive Based};

    \node (21) [right=3cm of a1, yshift=-0.5cm]{CDAE\cite{21_wu2016collaborative}};
    \node (8) [left=0.6cm of 21, yshift=1.1cm]{DCF\cite{8_li2015deep}};
    \node (38) [right=4cm of 8]{MvDGAE\cite{38_zheng2021multi}};
    \node (new1) [right=3.8cm of 38, align=center] {\added{RecDiff\cite{li2024recdiff}} \\ \added{DiffKG\cite{jiang2024diffkg}}};

    \node (ept1) [right=10cm of a1]{};
    \draw[line width=0.25mm] (a1) to node[pos=0.15, below=0.6cm, font=\scriptsize, text=gray] {As Main Task} node[pos=0.15, below=0.02cm, font=\scriptsize, text=gray] {As Auxiliary Task} (ept1);

    \node (67) [right=4.5cm of a2, yshift=1.5cm]{$\rm{S^3}$-Rec\cite{67_zhou2020s3}};
    \node (36) [right=4.5cm of a2, yshift=-1.5cm]{SGL\cite{36_wu2021self}};
    \node (68) [above=0.6cm of 36]{DCL\cite{68_liu2021contrastive}};
    \node (23) [right=0.5cm of 67, yshift=-1.1cm]{CoSeRec\cite{23_liu2021contrastive}};
    \node (37) [right=3cm of 67]{ICLRec\cite{37_chen2022intent}};
    \node (34) [right=3.2cm of 36]{KGCL\cite{34_yang2022knowledge}};
    \node (35) [right=3.2cm of 68, yshift=0.3cm]{DECA\cite{35_wang2022learning}};
    \node (new2) [right=0.1cm of 34]{\added{KACL\cite{wang2023knowledge}}};
    \node (new3) [above=0.5cm of new2, align=center]{\added{AdaGCL\cite{jiang2023adaptive}} \\ \added{DSL\cite{wang2023denoised}}};
    \node (new4) [right=0.1cm of new3, align=center]{\added{SSD-} \\ \added{ICGA\cite{sun2024self}}};

    \node (ept2) [right=8cm of a2]{};
    \draw[line width=0.25mm] (a2) to node[pos=0.3, above=0.0cm, font=\scriptsize, text=gray] {Others} (ept2);
    \node (ept3) [above=0.6cm of ept2, xshift=4.3cm]{};
    \draw[line width=0.25mm] ($(a2)!.3!(ept2)$) .. controls +(up:0.8cm) .. node[pos=0.2, above=0.2cm, font=\scriptsize, text=gray] {Sequence-based} (ept3) coordinate[pos=0.5] (EPT3_M1) coordinate[pos=0.6] (EPT3_M2) coordinate[pos=0.8] (EPT3_M3);
    \node (ept4) [below=0.6cm of ept2, xshift=4.3cm]{};
    \draw[line width=0.25mm] ($(a2)!.3!(ept2)$) .. controls +(down:0.8cm) .. node[pos=0.2, below=0.2cm, font=\scriptsize, text=gray] {Graph-based} (ept4) coordinate[pos=0.5] (EPT4_M1) coordinate[pos=0.9] (EPT4_M2) coordinate[pos=0.99] (EPT4_M3);
    
    \draw[->, >=latex, line width=0.25mm] ($(a1)!.1!(ept1)$)  .. controls +(down:0.6cm) .. (21);
    \draw[dashed ,->, >=latex, line width=0.25mm] ($(a1)!.6!(ept1)$)  .. controls +(up:0.6cm) .. (38);
    \draw[dashed ,->, >=latex, line width=0.25mm] ($(a1)!.15!(ept1)$) .. controls +(up:0.6cm) ..  (8);
    \draw[dashed ,->, >=latex, line width=0.25mm] ($(a1)!.9!(ept1)$) .. controls +(up:0.6cm) .. node[pos=0.8, below=0.1cm, font=\scriptsize, text=gray] {\added{Diffusion-based}}  (new1);
    % \draw[->, >=latex, line width=0.25mm] (a1) to node[pos=0.62, above, font=\scriptsize, text=gray]{on Interaction}  node[pos=0.62, below=0.2cm, font=\scriptsize, text=gray] {on Side-information} (21);
    % \draw[->, >=latex, line width=0.25mm] ($(a1)!.35!(21)$) .. controls +(down:0.6cm) ..  (8);
    % \draw[->, >=latex, line width=0.25mm] (21) to (38);1

    % \draw[->, >=latex, line width=0.25mm] (a2) to (67);
    % \draw[dashed ,->, >=latex, line width=0.25mm] ($(67)!.2!(ept2)$)  .. controls +(up:0.6cm) .. (36);
    % \draw[dashed ,->, >=latex, line width=0.25mm] ($(67)!.5!(ept2)$)  .. controls +(up:0.6cm) .. (68);
    \draw[dashed ,->, >=latex, line width=0.25mm] (EPT3_M1)  .. controls +(up:0.6cm) .. (67);
    \draw[dashed ,->, >=latex, line width=0.25mm] (EPT3_M2)  .. controls +(down:0.4cm) .. (23);
    \draw[dashed ,->, >=latex, line width=0.25mm] (EPT3_M3)  .. controls +(up:0.6cm) .. (37);

    \draw[dashed ,->, >=latex, line width=0.25mm] (EPT4_M1)  .. controls +(down:0.6cm) .. (36);
    \draw[dashed ,->, >=latex, line width=0.25mm] (EPT4_M1)  .. controls +(up:0.3cm) .. (68);
    \draw[dashed ,->, >=latex, line width=0.25mm] (EPT4_M2)  .. controls +(up:0.3cm) .. (new3);
    \draw[dashed ,->, >=latex, line width=0.25mm] (EPT4_M3)  .. controls +(up:0.3cm) .. (new4);
    % \draw[dashed ,->, >=latex, line width=0.25mm] (EPT3_M2)  .. controls +(down:0.8cm) .. (34);
    \draw[dashed, ->, >=latex, line width=0.25mm] (36) to node[pos=0.6, above=0.0cm, font=\scriptsize, text=gray] {\added{w/ Knowledge Graph}} (34);
    \draw[dashed, ->, >=latex, line width=0.25mm] (34) to (new2);

    \draw[dashed ,->, >=latex, line width=0.25mm] ($(a2)!.8!(ept2)$) .. controls +(down:0.2cm) .. (35);
    
    % \draw[->, >=latex, line width=0.25mm] (36) to (68);
    % \draw[->, >=latex, line width=0.25mm] ($(67)!.35!(36)$) .. controls +(down:1.0cm) .. node[pos=0.62, below=0.2cm, font=\scriptsize, text=gray] {with Non-random Augmentation} (23);
    % \draw[->, >=latex, line width=0.25mm] (23) to (34);

\end{tikzpicture}

%% file: Section/5-Evaluation.tex
% In this section, we discuss the evaluation of recommender systems, focusing on three aspects: the evaluation metrics of recommender systems, the evaluation methods of robustness in recommender systems, and the common datasets for evaluation. These three aspects help researchers measure the performance of their recommender systems and understand their robustness under various conditions and perturbations, thus facilitating the development of more effective and reliable models.

% \subsection{Evaluation Metrics of Recommender Systems}
% \input{Section/5-subsection/1-Eva_RS}

% In this section, we discuss the evaluation of recommender systems, focusing on two aspects: the evaluation methods of robustness in recommender systems, and the common datasets for evaluation. % 随后，我们针对一些代表性的鲁棒性方法进行了实验对比并分析，展示了在不同场景下的鲁棒性方法适用性

% These two aspects help researchers measure the performance of their recommender systems and understand their robustness under various conditions and perturbations, thus facilitating the development of more effective and reliable models.

% In this section, we discuss the evaluation of recommender systems, focusing on two aspects: the evaluation methods of robustness in recommender systems and the common datasets used for evaluation. Subsequently, we conduct experimental comparisons and analyses of representative robustness methods, demonstrating their applicability across different recommendation scenarios.

In this section, we discuss the evaluation of recommender systems, focusing on two key aspects: the evaluation methods for robustness in recommender systems and the commonly used datasets for assessment. \added{Furthermore, we perform experimental comparisons and analyses of representative robustness methods, illustrating their effectiveness across various recommendation scenarios.}

\subsection{Evaluation Methods of Robustness in Recommender Systems}
\label{5-2-mertics}
\input{Section/5-subsection/2-Eva_Robustness}

\subsection{Common Datasets}
\label{5-3-dataset}
\input{Section/5-subsection/3-Eva_dataset}

\subsection{\added{Comparative Evaluation of Robustness Methods}}
\label{5-4-eva}

\input{Section/5-subsection/4-Evaluation}

%% file: Section/5-subsection/2-Eva_Robustness.tex
The average prediction shift~\cite{69_burke2015robust} between the clean training set and the perturbed training set is commonly used to evaluate the robustness of a given recommender system. In this section, we introduce three types of commonly used evaluation methods for assessing recommender systems' robustness, as listed in Table~\ref{tab:eva_method}. Note that we only include publications that directly employ one of these three methods in the table. For instance, some approaches analyze the variation in model performance as noise levels or attack intensity change. While this approach is a variant of \textit{Offset on Metrics}, it does not directly present quantitative evaluation indicators but rather trends. Therefore, we do not list such publications in Table~\ref{tab:eva_method}.

\input{Table/Evaluation_method}

\subsubsection{Offset on Metrics}
To efficiently evaluate the robustness of recommender systems, some methods measure the offset on performance metrics as an indirect representation of prediction shift, formulated as follows:
\begin{equation}
    \begin{aligned}
        \Delta M = \frac{\vert M' - M\vert}{M},
    \end{aligned}
\end{equation}
where $M$ represents the performance of the recommender system trained on clean data, and $M'$ represents the performance of the system trained on perturbed data. For example, several studies~\cite{48_deldjoo2020dataset, 59_shrestha2021empirical, 6_yuan2019adversarial, 7_yuan2020exploring} utilize incremental HR$@k$\footnote{HR$@k$ measures the fraction of test items appearing in the top-$k$ recommendation list relative to all test items.} to compute the shift as
$\Delta_{\mathrm{HR}@k} = |\mathrm{HR}@k' - \mathrm{HR}@k|.$
Similarly, other works~\cite{23_liu2021contrastive, 37_chen2022intent, 40_ye2023towards} assess robustness using the offset on NDCG$@k$\footnote{NDCG$@k$ measures ranking quality by considering both item relevance and ranking position.}, computed as
$ \Delta_{\mathrm{NDCG}@k} = |\mathrm{NDCG}@k' - \mathrm{NDCG}@k|.$
Furthermore, Wu et al.~\cite{17_wu2021fight} propose a refined method, Robustness Improvement (RI), which better evaluates defense mechanisms:
\begin{equation}
    \begin{aligned}
        \rm{RI} = 1 - \frac{\mathrm{HR}@\mathit{k}'-\mathrm{HR}@\mathit{k}}{\mathrm{HR}@\mathit{k}^*-\mathrm{HR}@\mathit{k}},
    \end{aligned}
\end{equation}
where $\mathrm{HR}@k^*$ is the top-$k$ HR of the model trained on perturbed data without any defense.

\added{In the context of adversarial attacks, numerous studies~\cite{zhang2024lorec, zhang2024llm4dec, zhang2024improving, zhangunderstanding, 17_wu2021fight} measure defense effectiveness by evaluating the change in recommendation metrics of the target item, defined as:}
\added{\begin{equation}
    \begin{aligned}
        \Delta T\text{-}M = \vert T\text{-}M' - T\text{-}M \vert,
    \end{aligned}
\end{equation}}
\added{where $T\text{-}M$ denotes the recommendation metric for the target item. In practice, it is common to choose unpopular items as target items during evaluation. As a result, the value of $T\text{-}M$ in the absence of attacks is typically close to 0.0. Consequently, some studies~\cite{zhang2024improving, zhang2024lorec, zhangunderstanding} approximate $\Delta T\text{-}M$ directly using $T\text{-}M'$.}

% \added{where $T\text{-}M$ denotes the recommendation metric for the target item. Additionally, it is common to select unpopular items as the target items in evaluation. As a result, the value of $T\text{-}M$ in the absence of attacks is generally close to 0.0, and some works~\cite{zhang2024improving, zhang2024lorec, zhangunderstanding} approximate $\Delta T\text{-}M$ directly using $T\text{-}M'$.}

Compared to alternative approaches, offset-based metrics are straightforward to compute. However, they do not always provide an accurate measure of robustness. For example, consider two top-$k$ recommendation lists $\hat{L}_u@k$ and $\hat{L}_v@k$ for users $u$ and $v$ in a system trained on clean data. Without loss of generality, assume $k=3$, ideal recommendation lists $L_u = \{i_1, i_2, \dots, i_8\}, L_v = \{i_5, i_6, \dots, i_{12}\}$, and $\hat{L}_u@k=\{i_8, i_9, i_{10}\}, \hat{L}_v@k=\{i_1, i_5, i_6\}$. If, after perturbation, these lists change to $\hat{L}_u'@k=\{i_7, i_8, i_{10}\}$ and $\hat{L}_v'@k=\{i_1, i_3, i_6\}$, we obtain $\Delta_{\mathrm{HR}@k} = 0$ \added{despite noticeable shifts in recommendations}. \added{In such cases, offset-based metrics may fail to reflect robustness accurately.}

\subsubsection{Offset on Output}
To address the challenge of accurately measuring recommender system robustness, some methods evaluate the offset on system output to quantify prediction shifts. This can be formulated as follows:
\begin{equation}
    \begin{aligned}
        \Delta O = \frac{1}{|\mathcal{U}|} \sum_{u \in \mathcal{U}} \text{sim}(\hat{L}_u@k, \hat{L}_u^{'}@k),
    \end{aligned}
    \label{eq:delta_o}
\end{equation}
where $\hat{L}_u@k$ represents the top-$k$ recommendation list from a system trained on clean data, $\hat{L}_u^{'}@k$ denotes the top-$k$ list from a system trained on perturbed data, and $\rm{sim}(\cdot)$ is a similarity function that quantifies differences between the two lists.

Shriver et al.~\cite{57_shriver2019evaluating} propose a metric called \textit{Top Output} (TO), which is sensitive only to the top-ranked item for a user. This item typically has the highest likelihood of being preferred:
\begin{equation}
    \begin{aligned}
        TO = \frac{1}{|\mathcal{U}|} \sum_{u \in \mathcal{U}} \mathbb{I}[\hat{L}_u^{'}@k[0] \in \hat{L}_u],
    \end{aligned}
\end{equation}
where $\mathbb{I}[\cdot]$ is an indicator function that returns 1 if the condition holds and 0 otherwise.

Oh et al.~\cite{47_oh2022rank} employ two similarity metrics in Eq.~\ref{eq:delta_o} to compute $\Delta O$: \textit{Rank-Biased Overlap} (RBO)~\cite{kendall1948rank} and \textit{Jaccard Similarity} (Jaccard)~\cite{jaccard1912distribution}. RBO captures similarity in item ordering, whereas Jaccard measures the overlap in the top-$k$ items regardless of order. Given two lists $L_1$ and $L_2$, these metrics are formulated as:
\begin{equation}
    \begin{aligned}
        \mathrm{RBO}@k(L_1, L_2) = (1-p)\sum_{d=1}^{k}p^{d-1}\dfrac{|L_1[0:d] \cap L_2[0:d]|}{d},
    \end{aligned}
\end{equation}
where $p$ is a tunable parameter (recommended value: 0.9), and
\begin{equation}
    \begin{aligned}
        \mathrm{Jaccard}@k(L_1, L_2) = \dfrac{|L_1[0:k] \cap L_2[0:k]|}{|L_1[0:k] \cup L_2[0:k]|}.
    \end{aligned}
\end{equation}

The top-$k$ Jaccard metric is particularly valuable in industry due to its computational efficiency compared to RBO. Conversely, RBO enables a more detailed robustness analysis by focusing on full-ranked lists, offering a more comprehensive evaluation of recommender system stability. By leveraging these diverse metrics, researchers can gain deeper insights into system performance under various perturbations, facilitating the development of more robust and reliable models.

\subsubsection{Certifiable Robustness}
Certifiable robustness focuses on finding the robustness boundary for a given instance in the recommender systems model. Traditional methods for certifiable robustness~\cite{cohen2019certified, li2019certified} can be categorized into two approaches: randomized smoothing and directly finding the worst perturbation. Randomized smoothing is a technique that smoothes the input by applying random noise, aiming to find an adversarial boundary that causes the model to produce incorrect outputs. However, in the recommender systems scenario, it is difficult to smooth the input of the model due to the semantics and discreteness of the features.

Directly finding the worst perturbation involves searching for the worst perturbation that can lead to an incorrect prediction for a given input. 
\citet{9_liu2020certifiable} provide both non-robust certification and robust certification by approximately calculating the worst perturbation for the FM model. For a given FM model $f$, input sample $\bm{x}$, which includes historical interaction and other features, and the perturbation budget $q$, let $\bm{x}'$ denote the perturbed instance corresponding to $\bm{x}$.
Recall the formulation of the FM model $f(\bm{x})$ in Eq.~\ref{eq:fm}, 
Liu et al.~\cite{9_liu2020certifiable} formulate the prediction shift $\delta$ as:
\begin{equation}
    \begin{aligned}
        \delta =  f(\bm{x}') - f(\bm{x})
        = & \sum_{j=1}^d w_j(x_j-x_j') + \sum_{f=1}^k \sum_{i=1}^d \sum_{j=1}^d v_{i,f} v_{j,f} x_i (x_j-x_j') \\ 
        & + \frac{1}{2} \sum_{f=1}^k\left(\sum_{j=1}^d v_{j,f} (x_j-x_j')\right)^2 + \frac{1}{2} \sum_{f=1}^k \sum_{j=1}^d v_{j,f}^2(x_j-x_j')^2.
    \end{aligned}
    \label{eq:certi_dealta}
\end{equation}
\added{Due to space constraints, Appendix~\ref{sec:ap_certi} details the approximation process for maximizing the prediction shift $\delta$ to achieve robustness certification. In summary, certifiable robustness is a key metric for evaluating robustness, as it establishes robustness bounds for individual instances.}

\input{Table/Dataset}

% Due to space limitations, Appendix~\ref{sec:ap_certi} details the process of approximating the maximization of the prediction shift $\delta$ to obtain robustness certification. In conclusion, certifiable robustness is a crucial dimension for assessing recommender systems' robustness, as it is focused on identifying the robustness boundary for a specific instance within the recommender systems.

\added{Additionally, numerous publicly available toolkits assist researchers in evaluating the robustness of recommender systems. For example, RecBole~\cite{zhao2022recbole} provides a comprehensive collection of up-to-date recommender system implementations, while RecAD~\cite{wang2023recad} offers a suite of existing attack and defense methods for evaluating robustness.}

%% file: Table/Evaluation_method.tex
\begin{table}[t]
    \centering
    \resizebox{0.95\linewidth}{!}{
    \begin{threeparttable}[b]
    \caption{Comparison of Different Evaluation Methods}
        \begin{tabular}{lccccp{0.25\textwidth}}
            \toprule
            \textbf{Evaluation Method} &\added{\textbf{Formula}}       &\added{\textbf{Efficient}}              &\added{\textbf{Certifiable}}      &\added{\textbf{Model-agnostic}}             &\multirow{1}{0.3\textwidth}{\centering \textbf{Publication}}\\
            \midrule
            \multirow{1}*[-1.5em]{Offset on Metrics}   & \multirow{1}*[-1.5em]{\added{$\Delta M = \frac{\vert M' - M\vert}{M}$}}            & \multirow{1}*[-1.5em]{\checkmark}                                      & \multirow{1}*[-1.5em]{}                            & \multirow{1}*[-1.5em]{\checkmark}   
            & \textcolor{\myColor}{\cite{48_deldjoo2020dataset, 59_shrestha2021empirical, 1_du2018enhancing, 2_tang2019adversarial, 3_anelli2021study, 4_he2018adversarial, 5_yuan2019adversarial, 6_yuan2019adversarial, 7_yuan2020exploring, 8_li2015deep, 16_zhang2020gcn, 17_wu2021fight, 18_vinagre2019statistically, 23_liu2021contrastive, 27_tian2022learning, 30_yue2022defending,35_wang2022learning, 37_chen2022intent, 39_chen2022denoising, 40_ye2023towards, 55_strub2015collaborative, zhang2024lorec, zhang2024llm4dec, zhang2024improving, zhangunderstanding, zhang2025personalized}}\\
            Offset on Output          & \added{$\Delta O = \mathbb{E}_{u} \left[ \text{sim}(\hat{L}_u@k, \hat{L}_u^{'}@k)\right]$}      & \checkmark                                &                        & \checkmark   
            & \cite{17_wu2021fight, 47_oh2022rank, 57_shriver2019evaluating}\\     
            Certifiable Robustness     & \added{$\delta =  f(\bm{x}') - f(\bm{x})$}     &                                     & \checkmark                           &     
            & \cite{9_liu2020certifiable}\\
            \bottomrule
        \end{tabular}
        % \begin{tablenotes}
        %     \item[1] Computational cost of the evaluation method.
        %     \item[2] Agreement level between computed results and the model's true robustness.
        %     \item[3] Applicability of the evaluation method to diverse models.
        % \end{tablenotes}
        \label{tab:eva_method}
    \end{threeparttable}
    }
\end{table}

%% file: Table/Dataset.tex
\begin{table}[t]
    \centering
    \caption{Common datasets}
    \resizebox{0.9\linewidth}{!}{
    \begin{tabular}{lrrrp{0.55\textwidth}}
         \toprule
         \textbf{Dataset}       &\textbf{\# User}       &\textbf{\# Item}     &\textbf{\# Interaction}   &\multirow{1}{0.3\textwidth}{\centering \textbf{Publication}}\\
         \midrule
         \multirow{1}*[-0.5em]{MovieLens-100K~\citep{harper2015movielens}}           & \multirow{1}*[-0.5em]{943}           & \multirow{1}*[-0.5em]{1682}          & \multirow{1}*[-0.5em]{100,000}
         & \textcolor{\myColor}{\cite{1_du2018enhancing,8_li2015deep,11_yang2016re,12_yang2017spotting,13_aktukmak2019quick,17_wu2021fight,20_yera2016fuzzy,26_zhou2014detection,31_gao2022self,33_chung2013betap,35_wang2022learning,41_toledo2015correcting,45_zou2013belief,51_cao2013shilling,52_zhou2016svm,56_wang2021implicit,57_shriver2019evaluating, xia2024neural, he2024double}}\\
         
         \multirow{1}*[-0.5em]{MovieLens-1M}           & \multirow{1}*[-0.5em]{6040}                 & \multirow{1}*[-0.5em]{3900}              & \multirow{1}*[-0.5em]{1,000,209} 
         &\textcolor{\myColor}{\cite{1_du2018enhancing,5_yuan2019adversarial,6_yuan2019adversarial,7_yuan2020exploring,8_li2015deep,17_wu2021fight,18_vinagre2019statistically,25_zhang2014hht,26_zhou2014detection,27_tian2022learning,28_wu2012hysad,29_beutel2014cobafi,30_yue2022defending,40_ye2023towards,55_strub2015collaborative,56_wang2021implicit, xia2024neural}}\\
         MovieLens-20M          & 138,493              & 27,278             & 20,000,263 
         &\cite{30_yue2022defending,39_chen2022denoising,46_shenbin2020recvae,48_deldjoo2020dataset}\\
         \midrule
         Amazon-Beauty~\citep{ni2019justifying}          & 1,210,271            & 249,274            & 2,023,070 
         &\multirow{1}*{\cite{14_cao2020adversarial,23_liu2021contrastive,30_yue2022defending,37_chen2022intent,39_chen2022denoising,59_shrestha2021empirical,67_zhou2020s3}}\\
         Amazon-Book            & 8,026,324            & 2,330,066          & 22,507,155 
         &\multirow{1}*{\textcolor{\myColor}{\cite{27_tian2022learning,34_yang2022knowledge,36_wu2021self,40_ye2023towards,42_chen2022adversarial,43_wang2021denoising,68_xie2022contrastive, wang2023knowledge}}}\\
         \midrule
         LastFM~\citep{Celma:Springer2010}                 & 980                  & 1000              & 1,293,103 
         &\textcolor{\myColor}{\cite{24_chen2019adversarial,30_yue2022defending,47_oh2022rank,48_deldjoo2020dataset,67_zhou2020s3, xia2024neural, jiang2023adaptive, wang2023knowledge, jiang2024diffkg}}\\
         Netflix                & 480,189              & 17,770             & 100,480,507 
         &\cite{15_zhang2017robust,20_yera2016fuzzy,21_wu2016collaborative,26_zhou2014detection,28_wu2012hysad,29_beutel2014cobafi,46_shenbin2020recvae,56_wang2021implicit}\\
         \multirow{1}*[-0.5em]{Yelp}                   & \multirow{1}*[-0.5em]{1,987,897}            & \multirow{1}*[-0.5em]{150,346}            & \multirow{1}*[-0.5em]{6,990,280}
         &\textcolor{\myColor}{\cite{4_he2018adversarial,9_liu2020certifiable,16_zhang2020gcn,17_wu2021fight,21_wu2016collaborative,23_liu2021contrastive,27_tian2022learning,31_gao2022self,34_yang2022knowledge,36_wu2021self,37_chen2022intent,38_zheng2021multi,40_ye2023towards,42_chen2022adversarial,43_wang2021denoising,48_deldjoo2020dataset,59_shrestha2021empirical,67_zhou2020s3,68_xie2022contrastive, quan2023robust, li2024recdiff, jiang2023adaptive, wang2023denoised, he2024double}}\\
         \added{MIND~\cite{wu2020mind}}                         & \added{1,000,000} &\added{161,013} &\added{24,155,470} &\textcolor{\myColor}{\cite{zhang2024improving, zhang2024llm4dec, zhang2024lorec, zhangunderstanding, zhang2025personalized, jiang2024diffkg}} \\
         \bottomrule
    \end{tabular}
    }
    \label{tab:dataset}
\end{table}

%% file: Section/5-subsection/3-Eva_dataset.tex
% In this survey, we have analyzed the evaluation datasets used in papers on recommender systems' robustness and found that over 50 common datasets have been utilized. However, \added{nearly} 50\% of these datasets appear in only one paper, and merely 20\% have been used in more than three papers. Here, we introduce the statistics of \added{most frequently used 20\% of datasets} in Table~\ref{tab:dataset}. The primary datasets include MovieLens\footnote{https://grouplens.org/datasets/movielens/}, Amazon\footnote{https://nijianmo.github.io/amazon/index.html}, LastFM\footnote{http://ocelma.net/MusicRecommendationDataset/lastfm-1K.html}, Netflix\footnote{https://www.kaggle.com/datasets/netflix-inc/netflix-prize-data}, Yelp\footnote{https://www.yelp.com/dataset}, and \added{MIND}\footnote{https://msnews.github.io/} datasets. Due to space limitations, detailed descriptions of these datasets are provided in Appendix~\ref{sec:ap_dataset}.

% \input{Table/Dataset}

% During our statistical analysis, we observed that the datasets used in most papers are diverse and inconsistent. Furthermore, some papers sample from large datasets for testing, which complicates the replication of experiments and may lead to unfair method comparisons in follow-up research. We encourage researchers to select more widely used datasets when conducting studies and to adopt a cautious and rigorous approach when sampling from large datasets for experiments. \added{Establishing standardized benchmarks in this field would further facilitate its development and enhance reproducibility.}

In this survey, we have analyzed the evaluation datasets used in papers on recommender systems' robustness and found that over 50 common datasets have been utilized. However, \added{nearly} 50\% of these datasets appear in only one paper, and merely 20\% have been used in more than three papers.

Here, we introduce the statistics of \added{the 20\% most frequently used datasets} in Table~\ref{tab:dataset}. The primary datasets include MovieLens, Amazon, LastFM, Netflix, Yelp, and \added{MIND}. \added{Due to space limitations, detailed descriptions of these datasets are provided in Appendix~\ref{sec:ap_dataset}.} During our statistical analysis, we found that the datasets used across many papers are inconsistent. In addition, some papers sample subsets from large datasets for testing, which makes it difficult to replicate experiments and can lead to unfair comparisons between methods in later studies. We encourage researchers to use more widely adopted datasets and to take a careful and rigorous approach when sampling from large datasets. \added{Establishing standardized benchmarks would further support the field’s progress.}

% During our statistical analysis, we observed that the datasets used in most papers are inconsistent. Furthermore, some papers sample from large datasets for testing, which complicates the replication of experiments and may lead to unfair method comparisons in follow-up research. We encourage researchers to select more widely used datasets when conducting studies and to adopt a cautious and rigorous approach when sampling from large datasets for experiments. \added{Establishing standardized benchmarks in this field would further facilitate its development and enhance reproducibility.}

% In addition to the data sets shown above, some data sets have also been used in more than two papers, 
% such as content-based image recommendation dataset Pinterest\footnote{https://www.pinterest.com/; https://github.com/hexiangnan/neural\_collaborative\_filtering}~\cite{geng2015learning}, 
% check-in dataset Gowalla\footnote{http://dawenl.github.io/data/gowalla\_pro.zip}~\cite{liang2016modeling, cho2011friendship},
% Film rating dataset FilmTrust\footnote{https://www.librec.net/datasets.html}~\cite{guo2013novel},
% and so on.
% DVD rating dataset CiaoDVD\footnote{https://www.librec.net/datasets.html}~\cite{guo2014etaf},
% news reading dataset Adressa\footnote{https://www.adressa.no/},}

%% file: Section/5-subsection/4-Evaluation.tex
\added{In this section, we experimentally evaluate representative robustness methods across different categories. Some methods are inherently dependent on specific models or scenarios, making direct comparisons with other categories infeasible. For instance, certifiable robustness methods~\cite{9_liu2020certifiable} are restricted to FM models and, therefore, are excluded from our comparative analysis.}

\subsubsection{\added{Experimental Setup}}

\added{Due to space constraints, we briefly introduce the experimental setup here. More detailed settings for robustness against malicious attacks can be found in~\cite{zhang2024lorec, zhang2024improving, 17_wu2021fight}, while details on robustness against natural noise are available in~\cite{zhang2025personalized, 43_wang2021denoising}.}

\input{Table/Attack_eval}

\added{We conduct experiments on the \textbf{MIND} dataset~\cite{wu2020mind}, which includes item side information. Following~\cite{zhang2024lorec, zhang2024improving}, we sample users and filter out those with fewer than five interactions. To ensure comparability across model types, we adopt two representative backbones: \textbf{SASRec}~\cite{kang2018selfattentive} (sequential) and \textbf{MF}~\cite{koren2009matrix} (collaborative filtering). For robustness evaluation, we implement the \textbf{Bandwagon Attack}~\cite{mobasher2007toward} and simulate 10\% \textbf{random misclicks} as natural noise. We use $\mathrm{NDCG}@20$ to assess performance, $\mathrm{T}$-$\mathrm{NDCG}@20$ for attack defense, and further evaluate robustness via metric offset ($\Delta \mathrm{NDCG}@20$, $\Delta \mathrm{T}$-$\mathrm{NDCG}@20$) and output offset (TO@50).}

% \added{\textbf{Evaluation Metrics.} To assess recommendation performance, we employ Normalized Discounted Cumulative Gain at 20 ($\mathrm{NDCG}@20$). To evaluate attack defending, we measure the ranking performance of target items using $\mathrm{T}\text{-}\mathrm{NDCG}@20$. Furthermore, we assess robustness using both metric deviation analysis ($\Delta \mathrm{NDCG}@20$, $\Delta \mathrm{T}\text{-}\mathrm{NDCG}@20$) and output perturbation evaluation (TO@50).}

\subsubsection{\added{Robustness Methods against Malicious Attacks}}

\added{Table~\ref{tab:eva_attack} presents the performance of various robustness methods against malicious attacks. In the Collaborative Filtering (CF) setting}:
\begin{itemize}[leftmargin=*]
    \item \added{\textbf{Adversarial training} on parameters not only reduces attack success rates but also enhances recommendation performance, aligning with the theorem provided by PamaCF~\cite{zhangunderstanding}.}
    \item \added{\textbf{Adversarial training on parameters better preserves output stability.} In contrast, fraudster detection or adversarial training on user profiles inherently alters user profiles by adding or removing users, leading to greater output fluctuations.}
\end{itemize}
\added{In the Sequential Recommendation setting:}
\begin{itemize}[leftmargin=*]
    \item \added{\textbf{Fraudster detection methods} show better performance in mitigating attacks, while traditional adversarial training methods (e.g., APR) tend to significantly degrade recommendation quality. This may be due to the fact that parameter perturbations are more likely to disrupt the reasoning process in sequential models, resulting in inaccurate recommendations.}
    \item \added{\textbf{Leveraging LLM-based open-world knowledge} substantially enhances defense effectiveness. Both \textbf{LLM4Dec} and \textbf{LoRec} markedly reduce attack success rates while} \added{offering greater output stability compared to other methods}.
\end{itemize}
\added{In summary, in CF-based recommender systems, parameter-level adversarial training is a preferred defense strategy due to its dual benefits of robustness and performance enhancement~\cite{zhangunderstanding}. However, detection methods must accurately identify fraudsters while minimizing modifications to user profiles to maintain output stability. In Sequential Recommendation, Fraudster Detection approaches are more suitable for defense.} \added{Furthermore, integrating} \added{external knowledge in the detection phase} \added{has the potential to} \added{enhance detection accuracy and overall robustness.}

\input{Table/Noise_eval}

\subsubsection{\added{Robustness Methods against Natural Noise}}

\added{Table~\ref{tab:eva_noise} presents the performance of various robustness methods against natural noise. Our key observations are:}

\begin{itemize}[leftmargin=*]
    \item \added{\textbf{Regularization} is not universally suitable for all types of recommender systems. While improving output stability to some extent, they significantly degrade recommendation performance.}
    \item \added{\textbf{Purification and Self-Supervised Learning (SSL) methods} both effectively enhance recommendation performance.} \added{Compared to SSL, Purification exhibits superior output stability, particularly \textbf{PLD}. PLD modifies the sampling strategy for positive samples during training, which increases the likelihood of selecting clean samples rather than directly discarding potential noise.}
\end{itemize}

\added{In summary, both Purification and SSL serve as strong defenses against natural noise. Future research should explore hybrid strategies that combine SSL with purification techniques, leveraging their complementary strengths to enhance both robustness and recommendation performance.}

% \subsubsection{Robustness Methods against Natural Noise}

% Table~\ref{tab:eva_noise} presents the performance of various robustness methods against natural noise. Our key observations are:

% \begin{enumerate}[leftmargin=*]
%     \item \textbf{Regularization-based methods} are not universally suitable for all recommendation scenarios. While they improve output stability to some extent, they significantly degrade recommendation performance.
%     \item \textbf{Purification and Self-Supervised Learning (SSL) methods} both effectively enhance recommendation performance.
%     \item Compared to \textbf{SSL}, \textbf{Purification methods achieve better output stability}, particularly \textbf{PLD~\cite{zhang2025personalized}}. This is because \textbf{PLD modifies only the sampling strategy for positive samples during training}, increasing the likelihood of selecting clean samples rather than directly discarding noisy ones.
% \end{enumerate}

% In summary, both \textbf{Purification} and \textbf{Self-Supervised Learning} serve as strong defenses against natural noise. Future research could explore \textbf{hybrid strategies} that combine \textbf{self-supervised learning with purification techniques}, leveraging their complementary strengths to enhance both robustness and recommendation performance.

%% file: Table/Attack_eval.tex
\begin{table}[t]
    \centering
    \caption{\added{Performance Comparison Against Malicious Attacks}}
    \resizebox{0.95\linewidth}{!}{
    \begin{threeparttable}[b]
    
    \begin{tabular}{llcccccc}
         \toprule
         \multirow{1}*{\added{\textbf{Category}}} &\multirow{1}*{\added{\textbf{Method}}} &\added{\textbf{NDCG@20}($\uparrow$)} &\added{\textbf{NDCG@20'}($\uparrow$)} &\added{$\bm{\Delta}$\textbf{NDCG@20}($\downarrow$)} &\added{$\bm{\Delta}$\textbf{T-NDCG@20}($\downarrow$)} & \added{\textbf{T0@50}($\uparrow$)} & \added{\textbf{Time/epoch}($\downarrow$)}\\
         \midrule
         \added{\textbf{Backbone}} &\added{\textbf{MF}\tnote{1}}  & \added{0.69$\pm$0.01}  & \added{0.67$\pm$0.01}  & \added{2.90$\pm$1.02}  & \added{0.055$\pm$0.005}  & \added{70.59$\pm$2.62} &  \added{20.28$\pm$2.39} \\
         \cmidrule(lr){2-8}
         \multirow{2}*{\added{\textbf{Fraudster Detection}}} &~~+\added{\textbf{GraphRfi}~\cite{16_zhang2020gcn}}  & \added{0.67$\pm$0.00}  & \added{0.65$\pm$0.01}  & \added{2.99$\pm$1.25}  & \added{0.050$\pm$0.005}  & \added{61.03$\pm$1.90}&  \added{47.50$\pm$10.32} \\
         &~~+\added{\textbf{LLM4Dec}~\cite{zhang2024llm4dec}}  & \added{0.68$\pm$0.00}  & \added{0.68$\pm$0.01}  & \added{\textbf{0.32$\pm$0.91}}  & \added{\underline{0.025$\pm$0.003}}  & \added{65.10$\pm$1.71}&  \added{-} \\
         \cmidrule(lr){2-8}
         \multirow{3}*{\added{\textbf{Adversarial Training}}} &~~+\added{\textbf{APR}~\cite{4_he2018adversarial}}  & \added{\underline{0.70$\pm$0.01}}  & \added{\underline{0.71$\pm$0.01}}  & \added{\underline{1.43$\pm$1.62}}  & \added{0.026$\pm$0.003}  & \added{\underline{90.22$\pm$2.91}}& \added{26.19$\pm$3.03}  \\
         &~~+\added{\textbf{APT}~\cite{17_wu2021fight}}  & \added{0.65$\pm$0.02}  & \added{0.63$\pm$0.02}  & \added{3.08$\pm$2.96}  & \added{0.046$\pm$0.002}  & \added{66.52$\pm$5.07}& \added{-}  \\
         &~~+\added{\textbf{PamaCF}~\cite{zhangunderstanding}}  & \added{\textbf{0.71$\pm$0.00}}  & \added{\textbf{0.73$\pm$0.00}}  & \added{2.82$\pm$0.27}  & \added{\textbf{0.006$\pm$0.001}}  & \added{\textbf{96.31$\pm$1.66}} & \added{26.78$\pm$2.96}  \\
         \midrule
         \added{\textbf{Backbone}} &\added{\textbf{SASrec}}  & \added{6.50$\pm$0.33}  & \added{6.34$\pm$0.50}  & \added{2.46$\pm$1.51}  & \added{0.083$\pm$0.001}  & \added{60.59$\pm$2.02}& \added{197.04$\pm$4.66}  \\
         \cmidrule(lr){2-8}
         \multirow{3}*{\added{\textbf{Fraudster Detection}}} &~~+\added{\textbf{GraphRfi}~\cite{16_zhang2020gcn}}  & \added{6.35$\pm$0.39}  & \added{6.29$\pm$0.21}  & \added{0.94$\pm$1.35}  & \added{0.067$\pm$0.001}  & \added{69.72$\pm$3.50}&  \added{310.57$\pm$26.72} \\
         &~~+\added{\textbf{LLM4Dec}~\cite{zhang2024llm4dec}}  & \added{\underline{6.39$\pm$0.27}}  & \added{6.31$\pm$0.16}  & \added{1.25$\pm$0.90}  & \added{\underline{0.060$\pm$0.002}}  & \added{\textbf{81.27$\pm$4.63}}&  \added{-} \\
         &~~+\added{\textbf{LoRec}~\cite{zhang2024lorec}}  & \added{\textbf{6.52$\pm$0.20}}  & \added{\textbf{6.72$\pm$0.13}}  & \added{\textbf{3.07$\pm$0.68}}  & \added{\textbf{0.003$\pm$0.001}}  & \added{\underline{76.26$\pm$3.91}}& \added{260.86$\pm$13.01}  \\
         \cmidrule(lr){2-8}
         \multirow{2}*{\added{\textbf{Adversarial Training}}} &~~+\added{\textbf{APR}~\cite{4_he2018adversarial}}  & \added{6.14$\pm$0.39}  & \added{6.09$\pm$0.33}  & \added{0.81$\pm$1.68}  & \added{0.088$\pm$0.001}  & \added{70.34$\pm$2.10}&  \added{245.82$\pm$15.18} \\
         &~~+\added{\textbf{ADVTrain}~\cite{30_yue2022defending}}  & \added{6.33$\pm$0.41}  & \added{\underline{6.36$\pm$0.38}}  & \added{\underline{0.47$\pm$1.31}}  & \added{0.104$\pm$0.001}  & \added{71.01$\pm$3.66}&  \added{226.27$\pm$6.02} \\
         \bottomrule
    \end{tabular}
    \begin{tablenotes}
        \item[1] \added{When the number of epochs in the MF model is too high, the T-NDCG@20 scores across different defense methods become excessively low, making comparisons difficult~\cite{17_wu2021fight}. To ensure a fair comparison, we follow the epoch settings as proposed in~\cite{17_wu2021fight}.}
        
    \end{tablenotes}
    \end{threeparttable}
    }
    
    \label{tab:eva_attack}
\end{table}

%% file: Table/Noise_eval.tex
\begin{table}[t]
    \centering
    \caption{\added{Performance Comparison Against against Natural Noise}}
    \resizebox{0.85\linewidth}{!}{
    \begin{tabular}{llccccc}
         \toprule
         \multirow{1}*{\added{\textbf{Category}}} &\multirow{1}*{\added{\textbf{Method}}} &\added{\textbf{NDCG@20}($\uparrow$)} &\added{\textbf{NDCG@20'}($\uparrow$)} &\added{$\bm{\Delta}$\textbf{NDCG@20}($\downarrow$)} & \added{\textbf{T0@50}($\uparrow$)} & \added{\textbf{Time/epoch}($\downarrow$)}\\
         \midrule
         \added{\textbf{Backbone}} &\added{\textbf{MF}}  & \added{4.57$\pm$0.02}  & \added{3.55$\pm$0.05}  & \added{22.32$\pm$1.05}  & \added{64.70$\pm$2.30}& \added{20.36$\pm$1.98}  \\
         \cmidrule(lr){2-7}
         \added{\textbf{Regularization}} &~~+\added{\textbf{R1}~\cite{15_zhang2017robust}}  & \added{4.01$\pm$0.03}  & \added{3.29$\pm$0.02}  & \added{17.96$\pm$4.03}  & \added{70.39$\pm$2.12}& \added{21.20$\pm$1.88}  \\
         \cmidrule(lr){2-7}
         \multirow{3}*{\added{\textbf{Purification}}} &~~+\added{\textbf{ADT}~\cite{43_wang2021denoising}}  & \added{4.90$\pm$0.12}  & \added{4.04$\pm$0.09}  & \added{17.55$\pm$2.05}  & \added{63.95$\pm$6.97}& \added{22.89$\pm$1.88}  \\
         &~~+\added{\textbf{BOD}~\cite{wang2023efficient}}  & \added{5.16$\pm$0.10}  & \added{4.60$\pm$0.12}  & \added{\underline{10.85$\pm$1.59}}  & \added{\textbf{87.47$\pm$3.27}}&  \added{48.59$\pm$1.02} \\
         &~~+\added{\textbf{PLD}~\cite{zhang2025personalized}}  & \added{\textbf{5.24$\pm$0.07}}  & \added{\textbf{4.76$\pm$0.10}}  & \added{\textbf{9.16$\pm$1.10}}  & \added{\underline{86.85$\pm$1.59}}&  \added{21.95$\pm$1.81} \\
         \cmidrule(lr){2-7}
         \added{\textbf{Self-supervised Learning}} &~~+\added{\textbf{DeCA}~\cite{35_wang2022learning}}  & \added{5.19$\pm$0.06}  & \added{4.21$\pm$0.09}  & \added{18.88$\pm$2.09}  & \added{74.66$\pm$5.80}&  \added{29.32$\pm$4.17} \\
         \midrule
         \added{\textbf{Backbone}} &\added{\textbf{SASrec}}  & \added{6.50$\pm$0.33}  & \added{5.46$\pm$0.51}  & \added{16.00$\pm$2.48}  & \added{25.03$\pm$5.31}&  \added{197.04$\pm$4.66} \\
         \cmidrule(lr){2-7}
         \added{\textbf{Regularization}} &~~+\added{\textbf{R1}~\cite{15_zhang2017robust}}  & \added{5.78$\pm$0.50}  & \added{5.19$\pm$0.29}  & \added{10.21$\pm$3.73}  & \added{26.79$\pm$1.55}&  \added{209.11$\pm$5.00} \\
         \cmidrule(lr){2-7}
         \multirow{2}*{\added{\textbf{Purification}}} &~~+\added{\textbf{ADT}~\cite{43_wang2021denoising}}  & \added{6.00$\pm$0.52}  & \added{5.49$\pm$0.30}  & \added{8.50$\pm$1.78}  & \added{26.79$\pm$2.81}&  \added{212.73$\pm$3.17} \\
         &~~+\added{\textbf{PLD}~\cite{zhang2025personalized}}  & \added{6.60$\pm$0.23}  & \added{\textbf{6.08$\pm$0.28}}  & \added{\textbf{7.88$\pm$1.45}}  & \added{\textbf{30.78$\pm$1.39}}& \added{207.59$\pm$ 2.88}  \\
         \cmidrule(lr){2-7}
         \added{\textbf{Self-supervised Learning}} &~~+\added{\textbf{S$^3$-Rec}~\cite{67_zhou2020s3}}  & \added{\textbf{6.61$\pm$0.29}}  & \added{\underline{6.06$\pm$0.39}}  & \added{\underline{8.32$\pm$1.61}}  & \added{\underline{28.60$\pm$3.76}}&  \added{251.39$\pm$ 4.39} \\ 
         \bottomrule
    \end{tabular}
    }
    \label{tab:eva_noise}
\end{table}

%% file: Section/6-Discussion.tex
This section explores the considerations of robustness in recommender systems from two perspectives: the nature of the recommendation task and the context of the application. In the first subsection, we delve into the four main types of recommendation tasks---content-based, collaborative filtering-based, sequential, and hybrid. Challenges in each of these tasks primarily arise from their distinct operational methodologies and \added{their reliance} on various data types.

In the second subsection, we shift our focus to the application contexts of these recommender systems. Specifically, we discuss e-commerce, media, news, and social recommender systems. We aim to highlight \added{how robustness requirements vary based on} their application context.

\subsection{From the Perspective of Recommendation Tasks}

\textit{Content-based Recommender Systems:} In these systems~\cite{lops2011content}, robustness \added{depends heavily on} the quality of item metadata, such as tags, descriptions, and other side information. They recommend items \added{similar to those the user has liked before, making them especially vulnerable to errors or noise in the item metadata.} Such issues can significantly affect the system’s ability to measure item similarity accurately, leading to poor recommendations. Therefore, a key research goal in these systems is to design algorithms that improve the reliability of side information (Section~\ref{sec:adv_item}).

\textit{Collaborative Filtering (CF)-based Recommender Systems:} 
CF-based recommender systems generate recommendations based on the premise that users with similar behaviors tend to share similar preferences. Since these systems rely on user behaviors, they are vulnerable to interference from malicious users, particularly shilling attacks~\cite{tang2020revisiting, christakopoulou2019adversarial}, which can distort user behavior patterns, leading to biased and inaccurate recommendations~\cite{lam2004shilling}. Given these vulnerabilities, \added{research efforts in CF-based systems primarily focus on detecting and mitigating such attacks (Section~\ref{3-Attack}).} Specifically, researchers are developing methods to identify these attacks and devising training methods that bolster the system's defenses, \added{ensuring recommendation accuracy and reliability~\citep{zhang2024llm4dec}}.

\textit{Sequential Recommender Systems:} Sequential recommender systems generate recommendations based on users' historical interaction sequences~\cite{quadrana2018sequence}. Consequently, they are highly sensitive to temporal fluctuations, such as evolving user preferences, seasonal variations, and emerging trends~\cite{30_yue2022defending}. The primary challenge in these systems is to maintain reliable recommendations despite such changes in user behavior. To address this, researchers are focusing on designing models that \added{are more resilient to deviations in users' historical behavior (Section~\ref{4-Noise} and Section~\ref{adv-on-int}).}

\textit{Hybrid Recommender Systems:} Hybrid recommender systems integrate multiple recommendation strategies to leverage their respective strengths~\citep{burke2002hybrid}. Consequently, they face a diverse set of robustness challenges, each stemming from the methods they incorporate. Researchers in this domain aim to develop an optimal combination of these techniques, ensuring that the strengths of one approach compensate for the weaknesses of another. \added{The goal is to create a robust framework that delivers consistently accurate recommendations across varying scenarios.}

\subsection{From the Perspective of Applications}

\textit{E-commerce Recommender Systems:} 
E-commerce recommender systems are a \added{key feature of platforms like Amazon and eBay, helping customers discover products more easily}~\cite{smith2017two}. \added{These systems typically rely on} historical browsing and purchase data, using both user-item interactions and item-side information. Despite their effectiveness, they face several robustness challenges. One significant issue is the presence of natural noise in user data, such as unintentional clicks or purchases that may not accurately reflect genuine user preferences~\cite{20_yera2016fuzzy}. Additionally, attackers can manipulate item-side information to promote specific products or fabricate user profiles to artificially boost a product's exposure~\cite{cohen2021black}. \added{In an e-commerce context, ensuring recommender system robustness requires a multifaceted approach.} Beyond the strategies outlined in Sections~\ref{3-Attack} and ~\ref{4-Noise}, real-world applications demand that these systems adapt to the dynamic nature of product trends, handle extensive product catalogs, and maintain the long-term relevance of products. \added{These factors collectively introduce new challenges in sustaining robustness.}

\textit{Media Recommender Systems:} 
Media recommender systems are widely used on streaming platforms like Netflix and TikTok. \added{They suggest content---such as movies or songs---based on a user's watch or listen history and stated preferences~\cite{gomez2015netflix}}. Like e-commerce systems, short-media recommendations \added{face robustness issues}, including accidental likes, misleading video tags, and orchestrated efforts to inflate likes or views \added{(Section~\ref{4-Noise} and Section~\ref{3-Attack})}. These challenges are further intensified by the rapid shift in trends and the short lifespan of content. Conversely, long-media recommendations—such as those for movies and music—better capture user preferences through indicators like viewing or listening duration, which mitigates the impact of natural noise. However, they are still vulnerable to fake user interactions~\cite{tang2020revisiting}, and \added{the manipulation of item information, such as fake tags, remains common due to the promotional nature of media content~\cite{cohen2021black} (Section~\ref{3-Attack}).}

\textit{News Recommender Systems:} 
News recommender systems, \added{commonly used on news websites and apps, suggest articles based on a user's reading history and stated interests~\cite{karimi2018news}.} These systems face unique challenges due to the short lifespan of \added{news content and their core role in spreading information, which makes them attractive targets for attackers aiming to spread misinformation.}  \added{In this context, the primary concern is not just the presence of noise but whether attackers have manipulated news content to introduce false information, as discussed in Section~\ref{sec:adv_item}.}

\textit{Social Recommender Systems:} 
Social recommender systems, common on platforms such as Facebook and Twitter, suggest connections, pages, or groups based on user interactions and existing social ties~\cite{yang2014survey}. A distinguishing characteristic of these systems is that every participant functions as both a user and an item. This dual role heightens the significance of side information, including user-generated content and social interactions, such as following activities~\cite{cresci2020decade}. Given the prevalence of biased content and artificial followers on social platforms, ensuring robustness is paramount. A key challenge is detecting and mitigating the influence of harmful content, fake users, and social bots. \added{To address these challenges, researchers are developing methods for detecting fraudulent activity and safeguarding platform authenticity, as outlined in Section~\ref{sec:detetct}.}

%% file: Section/7-Relationship.tex
\input{Table/Tradeoffs}

% Robustness is a key property in trustworthy recommender systems. In this section, we \added{examine} the interplay between robustness and other key properties of trustworthy recommender systems. The four critical aspects are accuracy, explainability, privacy, and fairness. We discuss the potential trade-offs and synergies that may arise when pursuing robustness alongside these performance goals, as shown in Table~\ref{tab:robustness_relationship}.

Robustness is a key property of trustworthy recommender systems. In this section, we \added{examine how robustness interacts with other important aspects of trustworthiness.} These four aspects are accuracy, interpretability, privacy, and fairness. We explore the possible trade-offs and synergies that may occur when improving robustness alongside these goals, as illustrated in Table~\ref{tab:robustness_relationship}.

\subsection{Robustness vs. Accuracy}
Several studies have shown that \added{improving model robustness often reduces accuracy, especially on clean data}~\cite{schmidt2018adversarially, wang2020once}. This trade-off has also been theoretically established in specific scenarios~\cite{tsipras2018robustness, yu2021understanding}. The trade-off often arises due to the limitations of defense methods, such as the difficulty in identifying adversarial examples that subtly alter label semantics. \added{Another reason is the model’s struggle to learn an optimal decision boundary after adversarial perturbations are introduced.}

Within recommender systems, this trade-off remains evident. When facing malicious data---whether due to natural noise or attacks---researchers often introduce assumptions to build more robust models. Some approaches emphasize distinguishing between clean and adversarial data during training, aiming to detect and minimize the impact of ``adversarial-like'' data by reducing their weight in training~\cite{43_wang2021denoising, 31_gao2022self}. While these methods effectively mitigate the effects of noise or attacks, they often lead to fundamental divergences in feature representations between optimal standard and robust models~\cite{tsipras2018robustness}, thereby reducing recommendation accuracy on clean data. 

In real-world scenarios, reducing the trade-off between robustness and accuracy is critically important. Since most recommender systems are designed to serve platform goals, significant computational resources are invested in improving robustness. \added{Notably, in certain settings such as collaborative filtering, \citep{zhangunderstanding} theoretically demonstrate that adversarial training can improve both robustness and recommendation performance. This finding offers new insights into mitigating the trade-off between accuracy and robustness in recommender systems. However, how to generalize these benefits to other recommendation paradigms remains an open question for future research.}

% In practical recommendation scenarios, mitigating the trade-off between robustness and accuracy is of paramount importance. Since most recommender systems are optimized to serve platform interests, significant computational investments are made in improving robustness. \added{Notably, in specific settings such as collaborative filtering (CF), \citep{zhangunderstanding} theoretically demonstrate that adversarial training can simultaneously enhance both robustness and recommendation performance. This opens up new perspectives for addressing the trade-off between accuracy and robustness in recommender systems. However, how to generalize such benefits to other recommendation paradigms remains an open research problem requiring further investigation.}

% In practical recommendation scenarios, mitigating the trade-off between robustness and accuracy is of paramount importance. Since most recommender systems are optimized to serve platform interests, significant computational investments are made in improving robustness. However, if prioritizing robustness leads to a substantial decline in recommendation quality for legitimate users, the overall impact may be detrimental. While some studies suggest that increased robustness inevitably compromises accuracy~\cite{tsipras2018robustness}, others indicate that a balance can be achieved~\cite{pang2022robustness}. The trade-off between accuracy and robustness in recommender systems remains an open research problem requiring further investigation.

\subsection{Robustness vs. Interpretability}
Interpretability in machine learning and deep learning refers to \added{how clearly a model’s decision-making process can be understood, either locally or globally~\cite{zhang2020explainable}}. A highly interpretable model not only reveals the basis for its decisions but also explains the key factors influencing its predictions. Recent studies highlight a \added{strong correlation} between them~\cite{tomsett2018why, liu2018adversarial}, suggesting that understanding \added{how a model behaves on adversarial samples can help improve robustness.}

In recommender systems, there is a growing emphasis on interpretability~\cite{seo2017interpretable, zhang2022neuro}. For instance, \cite{zhang2022neuro} formulates recommendation as neuro-symbolic reasoning, where symbolic reasoning not only explains the model’s logic but also helps identify noise or attacks by tracing the reasoning behind incorrect recommendations.
In real-world systems, robustness and interpretability \added{often support each other}. Users are more likely to trust recommendations \added{when they are both robust and interpretable}, fostering greater economic value. From a technical standpoint, these objectives reinforce each other, making their joint pursuit beneficial.
% \added{Ensuring both robustness and interpretability enhances confidence in recommendations, creating an ecosystem where both users and providers benefit.}

\subsection{Robustness vs. Privacy}
Privacy refers to protecting individuals’ or organizations’ information, actions, and identities from unauthorized access, use, or disclosure~\cite{de2020overview}. There is a complex interplay between privacy and robustness, with each potentially enhancing the other. Prior research suggests that privacy algorithms can inherently provide robustness, and some general methods exist for converting robust algorithms into privacy-preserving ones~\cite{dwork2009differential}. A widely used approach to protecting privacy is \textbf{Differential Privacy} (DP), which reduces the risk of exposing individual data by adding noise to data or model parameters~\cite{dwork2006differential, abadi2016deep}. Interestingly, incorporating differential privacy has been shown to enhance model robustness, making them less sensitive to small perturbations~\cite{lecuyer2019certified, naseri2022local}. 

In recommender systems, differential privacy is especially relevant in federated learning scenarios. When users send their data to central servers for gradient computation, they often add perturbations to protect privacy~\cite{shin2018privacy, minto2021stronger}. Similar perturbation techniques are applied in adversarial training for recommender systems~\cite{4_he2018adversarial, 42_chen2022adversarial}. \added{As real-world recommender systems increasingly operate under privacy regulations, ensuring both privacy and robustness through their mutual reinforcement offers a promising solution for secure recommendations.}

\subsection{Robustness vs. Fairness}
Fairness in machine learning refers to generating impartial and unbiased predictions across different groups or individuals~\cite{wang2023survey, pessach2022review}. Many approaches define fairness as ensuring model predictions remain unchanged under variations in sensitive attributes (e.g., gender)~\cite{kim2019multiaccuracy, pessach2022review}. This idea resembles adversarial training, where small perturbations in input data should not alter predictions~\cite{yuan2019adversarial}. However, a key distinction remains: fairness focuses on changes to specific attributes \added{regardless of their size}, whereas robustness enforces that changes remain small to preserve original labels. Recent studies suggest that enhancing robustness can indirectly improve fairness~\cite{pruksachatkun2021does}.

In recommender systems, fairness on the user side closely follows traditional definitions, ensuring that recommendations remain consistent when users’ sensitive attributes change~\cite{wang2023survey}. However, robustness in recommender systems mainly aims to reduce \added{the impact of noisy interactions or malicious behavior, which involves a different set of concerns.} \added{That said, in real-world scenarios, the scope of noise might extend beyond just user behaviors, and fairness issues may begin to overlap with robustness objectives. This possible convergence points to a promising direction for future work on using robustness techniques to support fairness in recommender systems.}

%% file: Table/Tradeoffs.tex
\begin{table}[t]
    \centering
    \caption{\added{Relationship Between Robustness and Other Key Properties in Recommender Systems}}
    \resizebox{0.9\linewidth}{!}{
    \begin{tabular}{ll}
         \toprule
         \textcolor{\myColor}{\textbf{Property}} & \multicolumn{1}{c}{ \textcolor{\myColor}{ \textbf{Relationship with Robustness}}} \\
         \midrule
         \multirow{1}*{\textcolor{\myColor}{\textbf{Accuracy}}} & \textcolor{\myColor}{\textbf{Often negatively correlated} (\(\downarrow\)); robustness may reduce accuracy on clean data} \\
         \multirow{1}*{\textcolor{\myColor}{\textbf{Interpretability}}} & \textcolor{\myColor}{\textbf{Positively correlated} (\(\uparrow\)); interpretable models facilitate adversarial detection} \\
         \multirow{1}*{\textcolor{\myColor}{\textbf{Privacy}}} & \textcolor{\myColor}{\textbf{Positively correlated} (\(\uparrow\)); privacy-preserving techniques often enhance robustness} \\
         \multirow{1}*{\textcolor{\myColor}{\textbf{Fairness}}} & \textcolor{\myColor}{\textbf{Positively correlated} (\(\uparrow\)); robust models can mitigate bias, but fairness operates under different constraints} \\
         \bottomrule
    \end{tabular}
    }
    \label{tab:robustness_relationship}
\end{table}

%% file: Section/8-Future_Direction.tex
As recommender systems evolve, they present numerous challenges to researchers \added{seeking} to develop robust models that maintain \added{their} performance under various conditions and perturbations. This section explores open issues and future directions in the field of robust recommender systems.

\subsection{Mitigating Gap between Defense Assumption and Attack Goal}
\added{A key challenge in robust recommender systems is the mismatch between defense assumptions and the actual goals of attacks.} Attacks, especially shilling attacks, often aim to promote or diminish product exposure rather than merely undermining recommendation performance. However, many existing defense \added{methods are built on the assumption that} attackers primarily seek to disrupt the functionality of recommender systems. For instance, \citet{4_he2018adversarial} incorporate adversarial perturbations \added{into} model parameters during \added{the} training phase. Yet, these perturbations often \added{fail to match} real-world attack patterns, meaning some adversarial examples optimized during training might be ineffective for defense. In another method, \citet{17_wu2021fight} \added{propose} adding empirical risk-minimizing users to defend fraudsters who are treated as a threat to recommendation quality.

\added{Such approaches often struggle to address attacks driven by different motivations, leading to fragmented and less effective defenses.} To advance the robustness of recommender systems, \added{future research should aim to bridge this gap by aligning defense assumptions more closely with real-world attack goals. Promising directions include:}(1)~imposing more rigorous constraints on adversarial samples during training---potentially by adopting knowledge---enhanced adversarial perturbations or similar techniques that infuse prior knowledge into perturbation generation. And (2)~considering new adversarial training paradigms could be valuable. For instance, shifting from the established ``min-max'' framework to a ``min-min'' perspective~\cite{17_wu2021fight} might pave the way for a more specific defense against attacks like shilling.

\subsection{Improving Generalization of Defense Methods}
The generalization of defense methods in recommender systems presents a considerable challenge. Many existing strategies \added{focus} on specific assumptions or model constraints. For example, \citet{9_liu2020certifiable} is tailored exclusively for models based on the Factorization Machine, while \citet{14_cao2020adversarial} is designed for models grounded in Reinforcement Learning. Moreover, numerous pre-processing detection techniques hinge on predefined features for detecting malicious users~\cite{29_beutel2014cobafi, 50_lee2012shilling}, and some resort to specific attack methods for generating supervised data~\cite{25_zhang2014hht, 11_yang2016re}. \added{These model- or attack-specific designs} limit the applicability of defenses across different scenarios, making it difficult to handle adversarial threats in large-scale, real-world systems.

Given these limitations, there is a growing need to develop defense methods that are scalable and generalizable across diverse models and environments. Future research should focus on shared characteristics of poisoned data and general patterns in model training or loss functions, rather than overly tailoring defenses to specific models, datasets, or attack types. Promising directions include:  (1)~The development of feature-free detection methods, ensuring they neither depend on pre-selected features nor limit dataset adaptability. (2)~In-process detection methodologies that can discern patterns exhibited by malicious users during the model's training phase. \added{(3)~Leveraging external knowledge to enhance generalization, such as incorporating LLM-based semantic understanding.}

\subsection{Standard Evaluation and Benchmark}
A main challenge in robust recommender system research is the absence of consistent evaluation methods and standardized benchmark datasets. As discussed in Section~\ref{5-2-mertics}, most methods compute the offset on metrics to indirectly show the robustness of the recommender system. However, the offset on the metrics sometimes does not effectively measure the robustness. Furthermore, as pointed out in Section~\ref{5-3-dataset}, more than 50 datasets are commonly used, but only about 20\% of them are referenced in more than three studies. \added{Notably,} only three \added{datasets} are used in over 10 papers.

\subsection{Trustworthy Recommender Systems}
As discussed in Section~\ref{7-Rel}, the relationship between robustness and other trustworthiness properties---such as interpretability, privacy, and fairness---is complex. These properties can either \added{reinforce one another or conflict, especially in relation to accuracy.} One of the main challenges in current research is managing these trade-offs with accuracy. Several studies have shown that improving a model’s robustness~\cite{tsipras2018robustness}, interpretability~\cite{bell2022s}, privacy~\cite{rezaei2023accuracy}, or fairness~\cite{cooper2021emergent} often leads to reduced accuracy. While some strides have been made to diminish these trade-offs—by establishing the upper bounds of these trade-offs in specific instances~\cite{tsipras2018robustness, yu2021understanding, zhangunderstanding}---there remains ample scope for advancement. Future avenues of exploration include: (1)~The theoretical examination of upper bounds in more intricate scenarios. (2)~Investigating the interplay between various trustworthiness properties and accuracy, with the intent of deciphering their shared traits and distinctions, and thereby identifying more optimal balancing techniques.

Another pressing research challenge entails the synergistic enhancement of robustness, interpretability, privacy, and fairness. Existing literature has suggested that the enhancement techniques for these properties often share similar optimization objectives~\cite{lecuyer2019certified, naseri2022local, pruksachatkun2021does}. Moreover, certain methodologies can be adapted from one property to another~\cite{dwork2009differential}. Forward-looking research should revisit the formal definitions of each property, aiming to delineate unified optimization objectives or devise methods to translate strategies across properties.

\subsection{LLM for Robust Recommender Systems}
Large language models (LLMs), like ChatGPT~\cite{openai_chat}, have brought significant advancements in natural language processing (NLP) in recent years. These models, trained on vast amounts of text data, are capable of generating human-like text, answering questions, and performing various language tasks with high accuracy. Lately, there's been a growing interest in utilizing LLMs for a range of tasks beyond their usual scope~\cite{fan2023recommender, gao2023chat}. Moreover, some studies indicate that LLMs can be employed to improve model robustness~\cite{chen2023graph}. The broad knowledge base within LLMs can improve defense mechanisms. Future research avenues in this area could include: (1)~Leveraging LLMs in adversarial training, for instance, in generating adversarial examples. (2)~Using LLMs to enhance techniques for detecting fraudsters. (3)~Investigating recommendation systems based on LLMs, taking advantage of their built-in robustness for delivering more reliable recommendations.

\subsection{\added{Defending Adaptive Attackers}}
\added{Most existing defense methods assume that attackers are unaware of the defense mechanisms, effectively treating the attack process as fully black-box. While this assumption simplifies evaluation, it limits the ability to assess the true robustness of defense strategies. In real-world scenarios, however, the interaction between attackers and defenders is a dynamic and continuous adversarial process. Attackers actively adapt their strategies based on the current robustness of the recommender system. For instance, when detection mechanisms are deployed, attackers may adopt more stealthy behaviors to evade detection. Likewise, as adversarial training techniques become more common, attackers may shift toward more destructive or targeted strategies. Robustness can be more meaningfully evaluated in such adaptive attack settings, where the attacker deliberately responds to the defense. From the defender’s perspective, it is essential to develop dynamic defense strategies that can continuously learn and respond to evolving attacks. Promising directions include reinforcement learning-based defenses, adversarial game-theoretic frameworks, and self-evolving training methods that proactively counter adaptive attacks.}

%% file: Section/9-Conclusion.tex
In this survey, we offer a comprehensive review of seminal contributions in the field of robust recommender systems. We propose a taxonomy to systematically organize the vast array of publications in this domain. Specifically, we examine the robustness of recommender systems along two primary dimensions: malicious attacks and natural noise. Regarding techniques, we delve into state-of-the-art methods for building robust recommender systems, including methods tailored for malicious attacks---fraudster detection, adversarial training, and certifiable robust training---as well as methods designed to mitigate natural noise---regularization, purification, and self-supervised learning. Additionally, we discuss evaluation metrics and prominent datasets, shedding light on \added{how robustness in recommender systems is assessed}. Our analysis explores the nuanced dimensions of robustness across various scenarios and its interplay with other trustworthy properties, such as accuracy, interpretability, privacy, and fairness. We also \added{highlight open challenges and envision potential future directions} in robust recommender systems research. \added{Our goal with this survey is to} equip researchers with a thorough understanding of the key aspects of robust recommender systems, elucidate significant advancements, and inspire further exploration of this critical area.

%% file: Section/10-Appendix.tex
\input{Table/symbol}

\subsection{Evaluation with Certifiable Robustness}
\label{sec:ap_certi}

For simplicity and without loss of generality, we assume $f(\bm{x}) \le 0$. Consequently, in subsequent discussions, we solely show the process of maximizing the value of $\delta$. In the non-robust certification approach, a greedy strategy is utilized to identify the dimension $j$ of $\bm{x}$ that maximizes the impact on the output. This dimension is given by
\begin{equation}
    \begin{aligned}
        j^* = \arg \max_j w_j + \sum_{f=1}^k \sum_{i=1}^d v_{i,f} v_{j,f} x_i + \sum_{f=1}^k v_{i,f}^2.
    \end{aligned}
\end{equation}
To compute $\bm{x}'$, they iteratively flip the $j$ dimension of $\bm{x}$ until the total number of flipped dimensions is equal to budget $p$. If the prediction of the perturbed $\bm{x}'$ differs from that of $\bm{x}$, i.e., $\mathrm{sign}\left(f(\bm{x})\right) \neq \mathrm{sign}\left(f(\bm{x}')\right)$,the model $f$ is considered certifiably non-robust under the budget $q$ for the instance $\bm{x}$. It's worth noting that the calculation of $\delta$ in the non-robust certification through the greedy strategy might not yield the maximum value. Therefore, even if we cannot identify a perturbed $\bm{x}'$ with a prediction different from that of $\bm{x}$, we cannot definitively assert the model's $f$ robustness.

In contrast, robust certification seeks to ascertain the model's $f$ robustness by approximating the upper bound $\delta_{\max}$ of the prediction change under the perturbation budget $q$. \citet{9_liu2020certifiable} split $\delta$ into two problems, i.e., $\delta = \delta_1(\bm{x}) + \delta_2(\bm{x}),$ that are easy to solve the upper bound $\delta_{\max}$:
\begin{equation}
    \begin{aligned}
        \delta_1(\bm{x}) = \sum_{j=1}^dw_jx_j'+\sum_{f=1}^k \sum_{i=1}^d \sum_{j=1}^d v_{i,f} v_{j,f} x_i x_j', ~~~\delta_2(\bm{x}) = \frac{1}{2} \sum_{f=1}^k\left( \sum_{j=1}^d v_{j,f} x_j' \right) + \frac{1}{2} \sum_{f=1}^k \sum_{j=1} ^d v_{j,f}^2 {x_j'}^2,
    \end{aligned}
\end{equation}

The approximated upper bound $\Bar{\delta}$ can be computed by $\Bar{\delta} = \max_{\bm{x}'} \delta_1(\bm{x}) + \delta_2(\bm{x}),$
where $\Bar{\delta} \ge \delta_{\rm{max}}$. If the approximated upper bound $\Bar{\delta}$ does not alter the prediction, i.e., $\mathrm{sign}\left(f(\bm{x})\right) = \mathrm{sign}\left(f(\bm{x})+\Bar{\delta}\right),$
the model $f$ can be declared certifiably robust. However, if the upper bound $\Bar{\delta}$ does change the prediction, we cannot claim that the model $f$ is non-robust because $\Bar{\delta} \ge \delta_{\rm{max}}$.

\subsection{Datasets}
\label{sec:ap_dataset}

\textit{MovieLens.}\footnote{https://grouplens.org/datasets/movielens/}
GroupLens Research has collected and provided rating datasets from the MovieLens website. These datasets come in different sizes, with the most commonly used versions being MovieLens-100K, MovieLens-1M, and MovieLens-20M. MovieLens-100K was collected over a seven-month period (September 19, 1997 – April 22, 1998) and has been cleaned by removing users with fewer than 20 ratings or incomplete demographic information (age, gender, occupation, zip code). MovieLens-1M was collected in 2000, comprising users who joined MovieLens during that year. MovieLens-20M (January 9, 1995 – March 31, 2015) includes randomly selected users who had rated at least 20 movies but does not contain demographic information.

\textit{Amazon.}\footnote{https://nijianmo.github.io/amazon/index.html}
This dataset encompasses Amazon reviews (ratings, text, helpfulness votes), product metadata (descriptions, category information, price, brand, and image features), and links (also viewed/also bought graphs). It consists of 29 subcategories, such as Beauty, Books, Movies, and Electronics. The most frequently used subsets are Amazon-Beauty and Amazon-Book. Additionally, each dataset has meta, rating-only, and k-score versions. The meta version contains all available information, including reviews, product metadata, and links. The rating-only version includes user ratings for items, while the $k$-score is a dense subset where each remaining user and item has at least $k$ reviews.

\textit{LastFM.}\footnote{http://ocelma.net/MusicRecommendationDataset/lastfm-1K.html}
LastFM 1K is a dataset released by LastFM, collecting the entire listening history (approximately 20 million records) of 992 users from July 2005 to May 2009. Additionally, it includes user information such as gender, age, country, and registration time.

\textit{Netflix.}\footnote{https://www.kaggle.com/datasets/netflix-inc/netflix-prize-data}
The Netflix movie rating dataset contains over 100 million ratings from 480 thousand randomly chosen, anonymous Netflix customers across 17 thousand movie titles. The data, collected between October 1998 and December 2005, reflects the distribution of all ratings received during this period. Ratings are on a scale from 1 to 5 (integral) stars.

\textit{Yelp.}\footnote{https://www.yelp.com/dataset}
The Yelp dataset is a subset of Yelp businesses and reviews, featuring over 1.2 million business attributes such as hours, parking availability, and ambiance. It covers 11 metropolitan areas.

\added{\textit{MIND.}}\footnote{https://msnews.github.io/}
\added{The MIND dataset is a large-scale dataset for news recommendation research. It is collected from anonymized behavior logs of the Microsoft News website.}

% \subsection{\added{Experimental Setup}}
% \label{sec:ap_set}

% \added{\textbf{Dataset.} Considering that some approaches require item side-information, we employ the \textbf{MIND} news recommendation dataset~\cite{wu2020mind}, as summarized in Table~\ref{tab:dataset}. Following~\cite{zhang2024lorec, zhang2024improving}, we sample a subset of users and filter out those with fewer than five interactions.}

% \added{\textbf{Backbone Models.} Since different methods are designed for distinct model types, we select two widely used architectures as backbones to ensure broad comparability: (1) A sequential recommender system: \textbf{SASRec}~\cite{kang2018selfattentive}, and (2) A collaborative filtering recommender system: \textbf{MF}~\cite{koren2009matrix}.} 

% \added{\textbf{Robustness Settings.} For malicious attacks, we implement the widely used heuristic attack, Bandwagon Attack~\cite{mobasher2007toward}. For natural noise, we simulate noise by introducing 10\% randomly misclick.}

% \added{\textbf{Evaluation Metrics.} To assess recommendation performance, we employ $\mathrm{NDCG}@20$. To evaluate the effectiveness of attack defenses, we measure the ranking performance of target items using $\mathrm{T}\text{-}\mathrm{NDCG}@20$. Furthermore, we assess robustness through metric deviation analysis ($\Delta \mathrm{NDCG}@20$, $\Delta \mathrm{T}\text{-}\mathrm{NDCG}@20$) and output perturbation evaluation (TO@50).}

%% file: Table/symbol.tex
\begin{table}[t]
    \centering
    \caption{Nomenclature}
    \resizebox{0.8\linewidth}{!}{
    \begin{tabular}{cp{0.7\textwidth}}
        \toprule
         \textbf{Abbreviation}                     & \textbf{Description}\\
         \midrule
         $(u,i,r_{u,i})$            & User-item-rating triplets \\
         $\mathcal{U}$              & User set, $u \in \mathcal{U}$ \\
         $\mathcal{I}$              & Item set, $i \in \mathcal{I}$ \\
         $\mathcal{R}$              & Interaction set, $r_{u,i} \in \mathcal{R}$ \\
         $\mathcal{I}_u$ & The items related to user $u$\\
         $\mathcal{R}_u/\mathcal{R}_i$ & The ratings given by user $u$/ The ratings related to item $i$\\
         % $\mathcal{R}_{u, \mathrm{max}}$ & The highest ratings given by user $u$\\
         $\mathcal{D}$              & Collected dataset, $(u,i,r_{u,i}) \in \mathcal{D}$ \\
         $\mathcal{D}^+$            & Training dataset \\
         $\mathcal{D}^-$            & Test dataset \\
         $\Delta$                   & Perturbations of data \\
         $f_*$                      & recommender systems trained on $*$ \\
         $M$ /  $M'$                         & Performance of recommender systems trained on clean data / perturbed data\\
                              % & Performance of recommender trained on perturbed data\\
         $\psi(\cdot, \cdot)$       & Error function \\
         $\delta(\cdot, \cdot)$     & Difference function \\
         $\mathcal{L}$              & Loss function \\
         $\Theta$                   & Model parameters \\
         $L_u$                      & Ground truth preferred list for user $u$ \\
         $\hat{L}_u@k$ / $\hat{L}_u^{'}@k$              & Top-k recommendation list for user $u$ given by recommender systems trained on clean data / perturbed data\\
         % $\hat{L}_u^{'}@k$          & Top-k recommendation list for user $u$ given by recommender systems trained on perturbed data\\
         \bottomrule
    \end{tabular} 
    }
    \label{tab:symbol}
\end{table}